\documentclass[a4paper,11pt]{article}
\usepackage{bm, upgreek}
\usepackage{multicol}
\usepackage{graphicx,color}
\usepackage{amsmath}
\usepackage{amssymb}
\usepackage{epsfig}
\usepackage{mathrsfs}
\usepackage{amsbsy}
\usepackage{bbm}
\usepackage{textcomp}
\usepackage{enumerate}
\usepackage{marvosym}
\usepackage{pifont}
\usepackage{upgreek}
\usepackage[hidelinks]{hyperref}
\usepackage[font=small,labelfont=bf]{caption}
\usepackage[makeroom]{cancel}
\usepackage[symbol]{footmisc}
\usepackage{mathtools}
\usepackage{kotex}
\usepackage[table]{xcolor}
\usepackage{multirow}

\usepackage[many]{tcolorbox}
\usepackage{setspace}
\usepackage{amsfonts}
\usepackage[bbgreekl]{mathbbol}
\usepackage{authblk}
\usepackage{orcidlink}
\usepackage{lmodern}

\DeclareSymbolFontAlphabet{\mathbbm}{bbold}
\DeclareSymbolFontAlphabet{\mathbb}{AMSb}%

\newcommand{\nEYE}{{\boldsymbol{\nu}\mathbf{EYE}}}

\numberwithin{equation}{section}
\numberwithin{figure}{section}

\counterwithout{figure}{section}

\usepackage{tikz}
\usetikzlibrary{arrows}
\usetikzlibrary{shapes}

\usepackage{enumitem}
\definecolor{mycolor1}{rgb}{0.50, 1.00, 0.00}   
\hypersetup{
    colorlinks=true,
    linkcolor=blue,
    filecolor=magenta,      
    urlcolor=cyan!60!gray,
    pdftitle={Overleaf Example},
    pdfpagemode=FullScreen,
    }

\author[1]{Shaomin CHEN\,\orcidlink{0000-0002-2376-8413}} 
\affil[1]{Center for High Energy Physics
Department of Engineering Physics
Tsinghua University, Beijing, China}   
\author[2]{YangHwan Ahn} 
\affil[2]{Institute of Particle and Nuclear 
Physics, Henan Normal University, Xinxiang, Henan, China} 
\author[3]{Davide Franco\,\orcidlink{0000-0001-5604-2531}} 
\affil[3]{APC, Universit\'e Paris Cit\'e, CNRS, Astroparticule et Cosmologie, Paris, France}
\author[4]{Fabio Mantovani}  
\affil[4]{Department of Physics and Earth Sciences
University of Ferrara
Polo Scientifico e Tecnologico,
Ferrara, Italy}
\author[5]{Aldo Ianni} 
\affil[5]{I.N.F.N. Laboratori Nazionali del Gran Sasso,
Assergi (AQ), Italy}
\author[6]{Jiyong Choi\,\orcidlink{0000-0001-7009-4346}} 
\author[6]{S. Gwon\,\orcidlink{0000-0002-2384-632X}}
\author[7]{K. K. Joo\,\orcidlink{0000-0002-5515-0087}}
\affil[6]{Center for Precision Neutrino Research, Chonnam National University, Republic of Korea}
\affil[7]{Physics Department, Chonnam National University, Republic of Korea}
\author[8]{Chang Hyon Ha\,\orcidlink{0000-0002-9598-8589}}
\author[8]{Kim Siyeon\,\orcidlink{0000-0003-1871-9972}}
\affil[8]{Department of Physics, Chung-Ang University, Seoul, Republic of Korea}
\author[9]{Jong-Chul Park\,\orcidlink{ 0000-0002-1276-875X}} 
\affil[9]{Department of Physics, Chungnam National University, Korea}
\author[10]{M. Pac,\orcidlink{0000-0001-6085-1462}} 
\affil[10]{Department of Radiology, Dongshin University,
Naju-si, Korea}
\author[11]{Pouya Bakhti}   
\author[11]{Meshkat Rajaee} %
\author[11,12]{Seodong Shin\,\orcidlink{0000-0001-6956-841X}} 
\affil[11]{Laboratory for Symmetry and Structure of the Universe, Department of Physics, Jeonbuk National University, Jeonju, Korea}
\affil[12]{Center for Theoretical Physics 
of the Universe, Institute for Basic Science, Daejeon, Korea}
\author[13]{Young Ju Ko\,\orcidlink{0000-0002-5055-8745}}
\affil[13]{Department of Physics, Jeju National University, Jeju, Korea}
\author[14]{Bo-Young Han\,\orcidlink{0000-0002-3939-6271}} 
\affil[14]{Korea Atomic Energy Research Institute
Daejeon, Republic of Korea}
\author[15]{Jihoon Choi \,\orcidlink{0000-0001-6244-6284}} 
\affil[15]{Korea Astronomy and Space Science 
Institute, Yuseong-gu, Daejeon, Republic of Korea}
\author[16]{B. R. Ko\,\orcidlink{0000-0002-4591-5269)}}     
\author[16]{HyangKyu Park\,\orcidlink{0000-0002-6966-1689}} 
\affil[16]{Accelerator Science Department, Korea University
Sejong Campus, Republic of Korea}
\author[17]{Yunseo Choi\,\orcidlink{0009-0003-4022-0139}}    
\author[17]{Gihan Hong\,\orcidlink{0000-0002-2247-3728}}    
\author[17]{Eunwoo Jang\,\orcidlink{0009-0003-3927-4527}}   
\author[17]{Jaebak Kim\,\orcidlink{0000-0002-2072-6082}}    
\author[17]{Minseo Kim}    
\author[17]{Kyungmin Lee\,\orcidlink{0000-0002-2247-3728}}  
\author[17,18]{Heegyu Park\,\orcidlink{0009-0004-9535-1651}}  
\author[17]{E. Won\,\orcidlink{0000-0002-4245-7442}}        
\author[17]{Jae Hyeok Yoo\,\orcidlink{0000-0003-0463-3043}} 
\affil[17]{Korea University, Physics Department, Republic of Korea}
\affil[18]{Korea University, Department of Integrative Energy Engineering, Republic of Korea}
\author[19]{J.Y. Cho\,\orcidlink{0000-0002-4199-7068}}   
\author[19]{J.Y. Lee\,\orcidlink{0000-0003-4444-6496}}   
\author[19]{D.W. Jeong\,\orcidlink{0000-0003-4764-3671}} 
\author[19]{H.J. Kim\,\orcidlink{0000-0001-9787-4684}}   
\affil[19]{Kyungpook National University, Republic of Korea}
\author[20]{Sin Kyu Kang\,\orcidlink{0000-0001-7508-3881}}  
\affil[20]{School of Liberal Arts, Seoul National University of Science and Technology, Republic of Korea}
\author[21]{Myung-Ki Cheoun\,\orcidlink{0000-0001-7810-5134}}  
\affil[21]{Soongsil University, Physics Department, Republic of Korea}
\author[22]{V. Kornoukhov\,\orcidlink{0000-0003-4891-4322}}    
\affil[22]{National Research Nuclear University MEPhI: Moscow, Russia}
\author[23]{V. Kobychev\,\orcidlink{0000-0003-0030-7451}}   
\author[23,24]{V.I.~Tretyak\,\orcidlink{0000-0002-2369-0679}} 
\affil[23]{Institute for Nuclear Research of NASU, Kyiv, Ukraine}
\affil[24]{Institute of Experimental and Applied Physics, CTU Prague, Prague, Czech Republic}
\author[25]{Steve Elliott\,\orcidlink{0000-0001-9361-9870}}
\affil[25]{Los Alamos National Laboratory,
Los Alamos, United States}

\author[26]{Jose R. Alonso\,\orcidlink{0000-0002-0220-4846}}

\author[26]{Janet M. Conrad\,\orcidlink{0000-0002-6393-0438}}

\author[27]{Michael H. Shaevitz\,\orcidlink{0000-0002-7436-8655}}

\affil[26]{Laboratory for Nuclear Science, Massachusetts Institute of Technology, Cambridge, MA, USA}
\affil[27]{Department of Physics, Columbia University, New York, NY, United States}
\affil[28]{Physics Department, University of Michigan, Ann
Arbor, MI, United States}

\author[28]{Joshua Spitz\,\orcidlink{0000-0002-6288-7028}}

\author[26]{Daniel Winklehner\,\orcidlink{0000-0002-0715-6310}}

\begin{document}
\vspace{0.1in}
\noindent

\vspace{0.05in}
\setstretch{1.5}
\title{\Huge \bf{The $\nEYE$ Neutrino Telescope}: Conceptual Design Report}
\maketitle
\begin{abstract}
The {$\bf\nu EYE$} (``new eye'', Neutrino Experiment at YEmilab, \href{https://sites.google.com/korea.ac.kr/the-nueye-telescope} {\tt nuEYE.korea.ac.kr}) neutrino project leverages the existing large pit at Yemilab located in South Korea, to reveal the existence of sterile neutrino, the up-turn of the neutrinos from the Sun, and the first minimum of the neutrino oscillation over distances on the order of tens of kilometers for the first time. This initiative is expected to facilitate a wide range of significant scientific and technological advancements within both South Korean and international communities engaged in neutrino science and technology. The {$\bf\nu EYE$} aims to investigate the largely unexplored sector of almost-massless lepton in the elementary particle physics in detail. The emphasis will be placed on the study of real time nuclear processes and reactions involving possible sterile neutrinos on timescales down to nanoseconds in ultra-high intense or radioactive neutrino beams for the first time in the world; the {$\bf\nu EYE$} looks at to-be universal oscillation (``up-turn'' in the electron neutrino survival probability) of neutrinos predicted by the three neutrino oscillation paradigm. This will confirm or deny our current understanding on the particle interactions of the lepton sector; and measurement of the first oscillation minimum between the first and second neutrinos in mass.
\end{abstract}
\hfill \break
\vspace*{\fill}
\hfill \break
\textbullet ~Prepared by the $\nEYE$ proto-collaboration.
\setstretch{1.0}
\newpage 
\newpage

\thispagestyle{empty}
\tableofcontents
\thispagestyle{empty}
\newpage

{\setstretch{1.5}
  \section{The $\nEYE$ Project Executive Summary}

 The $\nEYE$ (``new eye'', Neutrino Experiment at YEmilab,
\href{https://sites.google.com/korea.ac.kr/the-nueye-telescope}
{\tt nuEYE.korea.ac.kr})
neutrino project leverages the existing
large pit at Yemilab located in South Korea, to 
reveal the existence of sterile neutrino, the up-turn of the neutrinos from the Sun,
and the first minimum of the neutrino oscillation  
over distances on the order of
tens of kilometers 
for the first time.
This initiative is expected to facilitate
a wide range of significant scientific and
technological advancements within both
South Korean and international communities engaged
in neutrino science and technology.
\hfill \break
\hfill \break
The $\nEYE$ aims to investigate the largely
unexplored sector of almost-massless
lepton in 
the elementary particle physics in
detail. The emphasis will be placed on
\begin{itemize}
\item the study of real time nuclear
processes and reactions involving possible sterile neutrinos on
timescales down to nanoseconds in ultra-high intense
or radioactive neutrino beams for the first time in the
world; 
\item the $\nEYE$ looks at to-be universal oscillation (``up-turn''
in the electron neutrino survival probability)
of neutrinos predicted by the three
neutrino oscillation paradigm.
This will confirm
or deny our current understanding on the particle
interactions of the lepton sector; and 
\item Measurement of the first oscillation minimum
between the first and second neutrinos in mass.
\end{itemize}
The major scientific themes outlined above 
represent significant questions that warrant
further exploration.
\hfill \break
\hfill \break
The $\nEYE$ facility is proposed to be constructed in the 
``Liquid Scintillator Counter'' (LSC) hall with a 
ultra high-intensity accelerator based neutrino beam
facility and/or high radioactive source. 
Adjacent to the LSC hall, a 
a highly sophisticated radio-purification facility
will be built to provide the
radio-purified scintillator and water to the $\nEYE$
experiment.
\hfill \break
\hfill \break
Key international collaborations have been developed with 
the United States IsoDAR team (MIT, Michigan), and European
team for the detector construction (INFN, APC).
\newpage

  %
%
\section{Introduction}
Curiosity about nature and the universe has led to 
enormous advancements in science and technology throughout 
human history. Today, fields such as physics, 
mathematics, chemistry, and life science are highly advanced and 
complex, providing significant technological benefits to humanity. 
From particle physics to 
cosmology, we study objects in the universe on scales ranging from the smallest 
particles to the largest cosmic structures, all based on 
first principles.
\hfill \break
Understanding the nature of interactions 
among fundamental particles at the smallest scale has been
the primary interest in particle physics. 
This interest has resulted in what we now call the Standard Model
(SM) of particle physics, which has been experimentally 
and theoretically developed and validated since the last century.
Nevertheless, there are compelling reasons
to believe that the SM is not the
ultimate model of nature at the most miniature scale. One very 
relevant example of this proposal is the non-zero neutrino mass. 
\hfill \break
The existence of a hypothetical new neutrino may explain
the finite neutrino mass. Studies of the Sun's metallicity 
with neutrinos are crucial in astrophysics.
The ordering of the neutrino mass, and 
the nature of neutrinos as leptons (Dirac vs. Majorana), and $CP$ 
violation are
of great interest. This proposal will discuss
some of the most outstanding physics programs.
\begin{center}
\includegraphics[width=0.8\linewidth]{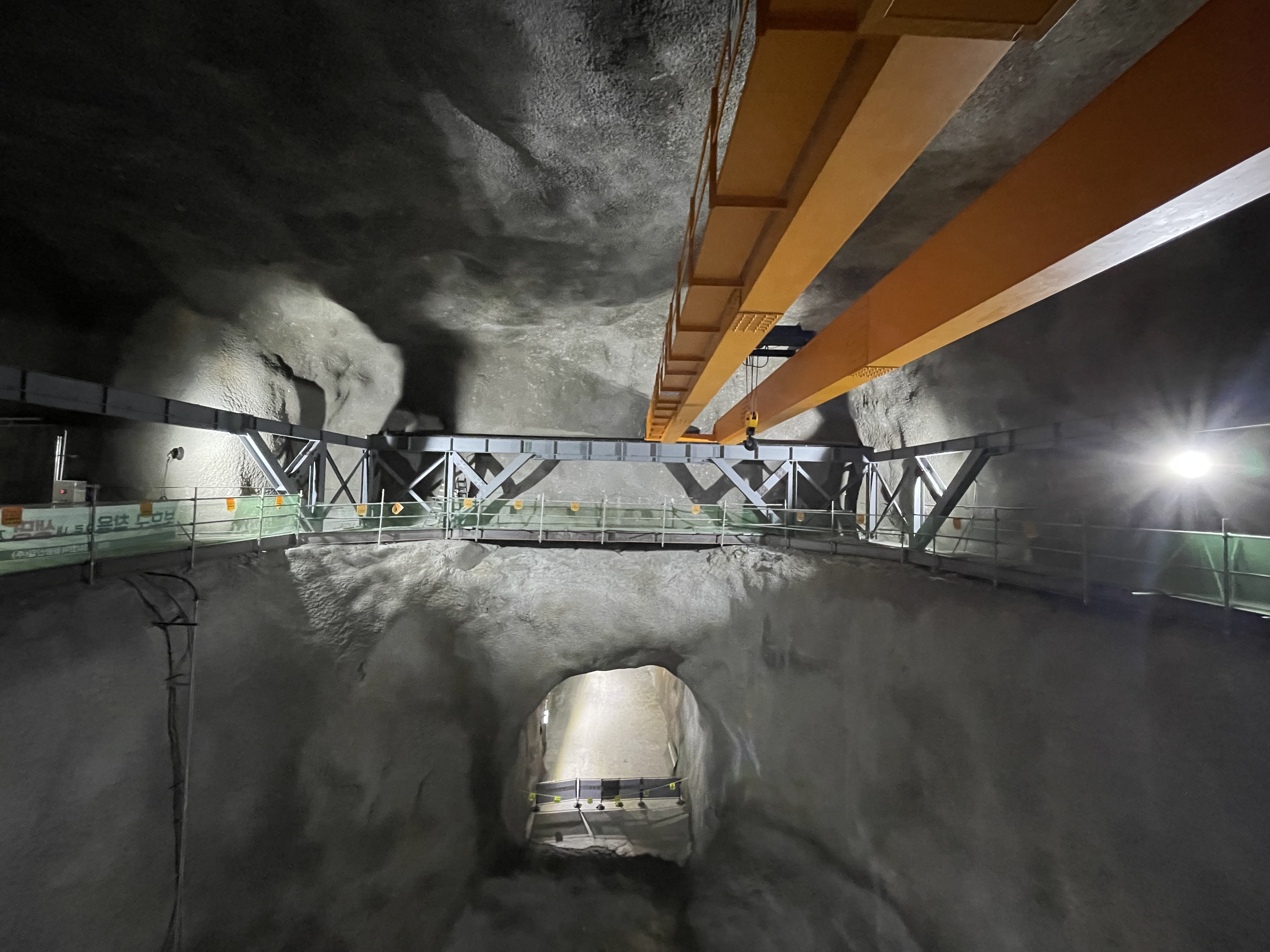}
\captionof{figure}
{A picture of a large Yemilab hall, courtesy of Dr.\ K. S. Park. Similar
one can also be seen
from \href{https://www.nature.com/articles/d41586-024-01347-3} 
{ Nature 30 May, 2024}.
}
\label{fig:yemilab_nature}
\end{center}
Yemilab is a new underground laboratory located in Korea 
at a depth of 1,000 m 
(\url{https://arxiv.org/abs/2402.13708}).
Yemilab includes a cylindrical pit with 19.5 m in diameter and 22 m in height,
designed as a multipurpose laboratory for 
next-generation experiments, as shown in Fig.~\ref{fig:yemilab_nature}. 
At this site, our proposed $\nEYE$ ({\bf N}eutrino 
{\bf E}xperiment at {\bf YE}milab) detector 
is planned to be installed through this proposal.
Here the $\nEYE$ detector refers to a 
proposed $\mathcal{O}(2)$-kilo tonne liquid scintillator detector that will
be described in detail in the following sections. 
First, we discuss the physics programs with the $\nEYE$ detector, 
and later, we discuss the status as well as the plan for 
the international collaboration.

Further information about $\nEYE$ can be found on its website at
\href{https://sites.google.com/korea.ac.kr/the-nueye-telescope}
{\tt nuEYE.korea.ac.kr}.

\section{Detailed Plan of Research and International Collaboration} 
\subsection{\fontsize{11}{15}\selectfont Sterile neutrino search program}
\label{sec:sterile}
In the neutrino sector, the SM of particle physics requires
extension to account for  
the observed neutrino oscillations, which 
imply nonzero neutrino masses. Further extensions may involve 
one or more sterile neutrinos ($\nu_s$) to explain experimental 
anomalies observed over the past few decades. 
\subsubsection{\fontsize{11}{15}\selectfont Current status of sterile $\nu$ search}
The first experimental anomaly, indicating a possible
sterile neutrino signal was reported by the LSND 
experiment more than 20 years ago
\href{https://journals.aps.org/prd/abstract/10.1103/PhysRevD.64.112007}
{PRD 64 112007 (2001)}, reporting an excess of
87.9 $\pm$ 22.4 $\pm$ 6.0 events in $\bar{\nu}_\mu \rightarrow
\bar{\nu}_e$ appearance search.
Utilizing a neutron spallation facility, 
the KARMEN experiment did not confirm 
\href{https://journals.aps.org/prd/abstract/10.1103/PhysRevD.65.112001}
{PRD 65 112001, (2002)} this observation, but couldn't entirely exclude it.
The MiniBooNE $\bar{\nu_e} + \nu_e$ combined 
\href{https://journals.aps.org/prl/abstract/10.1103/PhysRevLett.129.201801}
{PRL 129 201801 (2022)} is compatible with LSND but the
low energy spectrum has
excess. However, the $\mu$BooNE result
\href{https://journals.aps.org/prl/abstract/10.1103/PhysRevLett.130.011801}
{PRL 130 011801 (2023)}
disfavors
MiniBooNE and LSND but does not rule out these results  completely.
Recent $\mu$BooNE result \href{https://www.nature.com/articles/s41586-025-09757-7}
{Nature volume 648, 64–69 (2025)} disfavors a single sterile neutrino state.

\hfill \break
In nuclear-reactor-based $\bar{\nu}_e$ disappearance
experiments, there have been extensive searches for 
sterile neutrinos. The Daya Bay and RENO experiments found
a rate deficit 
but it is more likely that the flux calculation needs to be revised.
NEOS, STEREO, PROSPECT, DANSS, SOLID, and
Neutrino4 studied the neutrino spectrum with near and far
detectors for the flux-independent oscillation study.
Most experiments saw no evidence, except that Neutrino4
reported an arguable $3\sigma$ signal with $\Delta m^2 = 7~\textrm{eV}^2$
\href{https://www.sciencedirect.com/science/article/pii/S0370269322001885?via%3Dihub}
{PLB 829, 137054, (2022)}.
\hfill \break
The SAGE and GALLEX experiments, solar $\nu_e$ disappearance
experiments with the radiochemical extraction of $^{71}\textrm{Ge}$ 
indicated so-called the Gallium anomaly 
\href{https://journals.aps.org/prd/abstract/10.1103/PhysRevD.97.073001}
{PRD 97, 073001 (2018)}. 
Most recently, the BEST experiment, which used a 3.4 MCi 
$^{51}\textrm{Cr}$ radioactive neutrino source with the 
radiochemical extraction method, 
revealed the $4\sigma$ deviation consistent with
$\nu_e \rightarrow \nu_s$ oscillations 
\href{https://journals.aps.org/prc/abstract/10.1103/PhysRevC.105.065502}
{PRC 105, 065502 (2022)}.
\hfill \break
Having a consistent picture in current
sterile neutrino search
experiments is somewhat tricky. 
In particular, it is increasingly important
to carefully examine the BEST result, which uses the same  
neutrino source but real-time neutrino detection 
as an independent cross-check. 
\subsubsection{\fontsize{11}{15}\selectfont Sterile 
$\nu$ search in the $\nEYE$ experiment}
\tikz\draw[black,fill=black] (0,0) circle (.5ex); 
\colorbox{mycolor1!30}{\bf High-intensity accelerator (IsoDAR)}
\hfill \break
The $\nEYE$ detector's combination of low energy threshold and large-size is unmatched by any above-ground detector for electroweak physics.    Events in a detector such as $\nEYE$ would be swamped by cosmic-ray backgrounds, hence must be constructed underground.     Experiments at accelerators, which are constructed on the Earth’s surface, must make use of detectors with smaller footprints and higher energy thresholds to reduce the cosmic-ray background.    They also make use of ``beam compression" ---  delivering high intensity beam in short bursts --- which adds expense to an accelerator.
\hfill \break Most underground laboratories cannot easily be retrofitted to allow an accelerator to be constructed underground.   But because $\nEYE$ is being constructed in an entirely new laboratory, this is the opportunity to re-think the paradigm.    This is the concept behind the isotope decay-at-rest (IsoDAR) system – a novel compact cyclotron accelerator and target system that is proposed for installation in an existing tunnel adjacent to $\nEYE$ (\href{https://arxiv.org/abs/2404.06281}{arxiv:2404.06281},
\href{https://journals.aps.org/prd/abstract/10.1103/PhysRevD.105.052009}
{PRD 105 052009, 2022}).   With this integration,
$\nEYE$ will become the first underground accelerator-based 
experiment, thus advancing the
frontiers of rare event searches.
\hfill \break In the US, the IsoDAR antineutrino source is being designed and proposed to the US government. 
The 60 MeV protons from IsoDAR impinge on 
a $^9\textrm{Be}$ target, resulting in a high flux neutrons.
The neutrons then enter a surrounding $^7\textrm{Li}$ sleeve and produce $^8\textrm{Li}$ 
via neutron capture, resulting in $^8\textrm{Li} \rightarrow$ $^8\textrm{Be}
+ e^- + \bar{\nu}_e$.  
The IsoDAR source can produce $1.67\times 10^6$
 IBD events and 7000 $\bar{\nu}_e e^-
\rightarrow \bar{\nu}_e e^-$ elastic scattering (ES) events
with the $\nEYE$ detector 
during 5 calendar years of operation.
This allows one to search 
for oscillations associated with one or more sterile neutrinos 
with higher sensitivity than that with
the radioactive sources 
in the higher $\Delta m^2$ region if the design current of 10 mA is 
reached. This can be seen in Fig.~\ref{fig:IsoDAR_CL}.
The US group intends to 
place the IsoDAR accelerator in Yemilab and a proposal is planned to
be submitted along with this proposal. A horizontally long tunnel
next to the $\nEYE$ hall is already placed for the IsoDAR 
accelerator complex. Given that Yemilab already has an accelerator tunnel to house the IsoDAR
accelerator system, 
we have regularly discussions with a part of the IsoDAR collaboration from MIT and
U. of Michigan about future possibilities. A sensitivity figure 
provided from IsoDAR collaboration is shown in Fig.~\ref{fig:IsoDAR_CL}
where five year running is assumed.
\begin{center}
\includegraphics[width=0.8\linewidth]{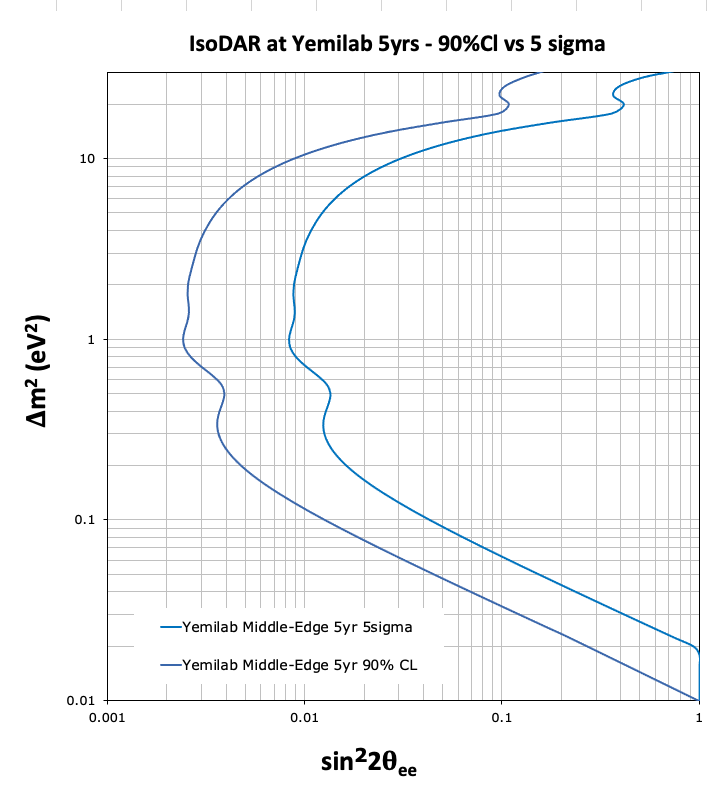}
\captionof{figure}
{
Estimated sensitivities for five year running of IsoDAR are shown.
Both 5$\sigma$ and 90\% CL sensitivities are shown.
}
\label{fig:IsoDAR_CL}
\end{center}
\hfill \break
\hfill \break
\tikz\draw[black,fill=black] (0,0) circle (.5ex); 
\colorbox{mycolor1!30}{\bf Radioactive source}
\hfill \break
Here, we propose a sterile neutrino search program with
radioactive neutrino sources, directly measuring the
inverse beta decay (IBD) or $\nu_e e^-$ scattering with the $\nEYE$ 
detector. This approach avoids
complicated radiochemical processes and instead detects
neutrinos in real time. At this stage, we 
consider two sources; $^{144}$Ce and/or $^{51}$Cr. The
$^{144}$Ce produces $\bar{\nu}_e$ via radioactive beta
decays and $^{51}$Cr produces $\nu_e$ via nuclear electron
capture. 
Note that $^{144}$Ce is obtained from
the nuclear spent fuel, whereas $^{51}$Cr is obtained by
activating enriched $^{50}$Cr with thermal
neutrons inside reactors. Therefore, it is relatively easier to
prepare $^{51}$Cr than $^{144}$Ce source.
Additionally, 3.4 MCi $^{51}$Cr has been used in the
BEST experiment but $\mathcal{O}$(100) kCi $^{144}$Ce source has never
been produced yet, with an attempt by the SOX 
experiment, \href{https://journals.aps.org/prl/abstract/10.1103/PhysRevLett.107.201801}
{PRL 107 201801 (2011)}.  Note that in fact the neutrinos above the
IBD threshold come from the daughter product of $^{144}$Ce as
indicated in Fig.~\ref{fig:144ce}. Also, neutrino energies from
$^{51}$Cr are also indicated (from
\href{https://www-nds.iaea.org/relnsd/vcharthtml/VChartHTML.html}
{IAEA}).
\begin{center}
\includegraphics[width=0.45\linewidth]{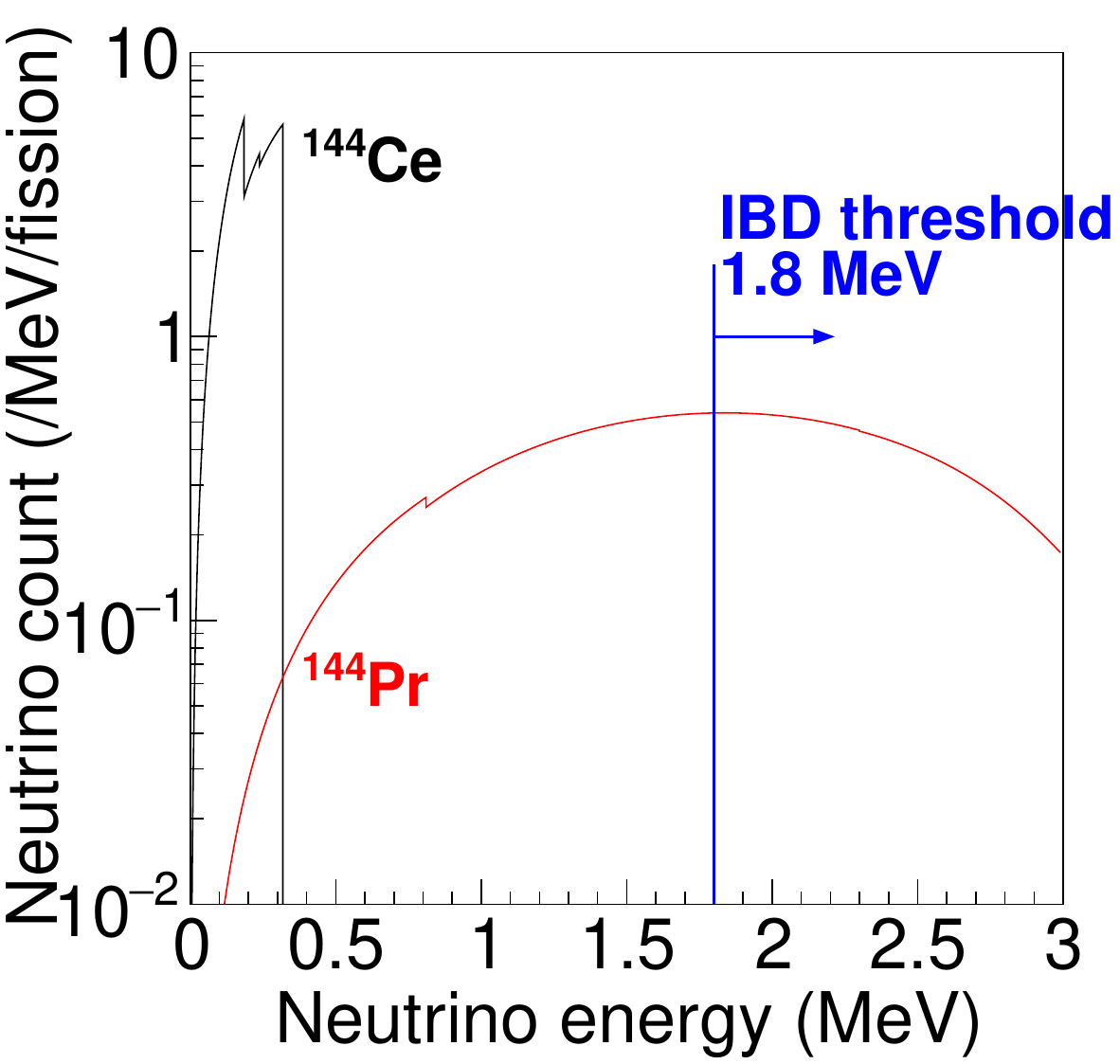}
\includegraphics[width=0.43\linewidth]{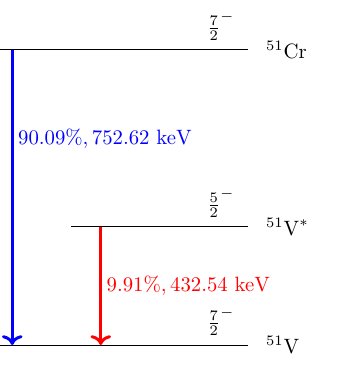}
\captionof{figure}
{
Left: energy distributions of $\bar{\nu}_e$ from $\beta^-$ decays of
$^{144}$Ce and $^{144}$Pr.
Right: a simplified decay transitions of $^{51}$Cr with 
released energies of $\nu_e$.
}
\label{fig:144ce}
\end{center}
\begin{center}
\begin{tabular}{lcc}
\hline
    	& $^{144}$Ce & $^{51}$Cr {\rule{0pt}{2.9ex}}\\
\hline
Assumed Activity  & 100 kCi & 10 MCi {\rule{0pt}{2.9ex}}\\
Detection Methods & IBD & $\nu_e  e^-$ scattering \\
Half-life & 285 days & 27.7 days \\
Expected events (1 yr)   & 34,000 & 28,000 \\
\hline
\end{tabular}
\captionof{table}{
Comparison between two sources. For expected events,
a two-kilo-tonne liquid scintillator detector is assumed.
}
\label{table:source}
\end{center}
We estimate the sensitivity with $^{144}$Ce in
$\Delta m^2$ and $\sin^2 2\theta$ for the sterile neutrino search.
For that purpose, the $\nEYE$ detector
is assumed. The detailed design parameters of the detector
will be addressed later. Sensitivities for source position
are shown in Fig.~\ref{fig:sensitivity_000VS001}.
We emphasize here that the primary physics goal in this
case is to confirm or rule out the BEST positive signal,
which utilized a radioactive source with the
radiochemical extraction of the signal. 
The black curve in Fig.~\ref{fig:sensitivity_000VS001} represents
the combined limits from experiments with different
systematic effect.  The key point here is
that we propose the IBD 
or elastic scattering processes for the detection,
allowing real-time measurements.
Note also that $^{51}$Cr with parameters in Table~\ref{table:source}
provides essentially similar sensitivity.
\hfill \break
\begin{center}
\includegraphics[width=0.8\linewidth]{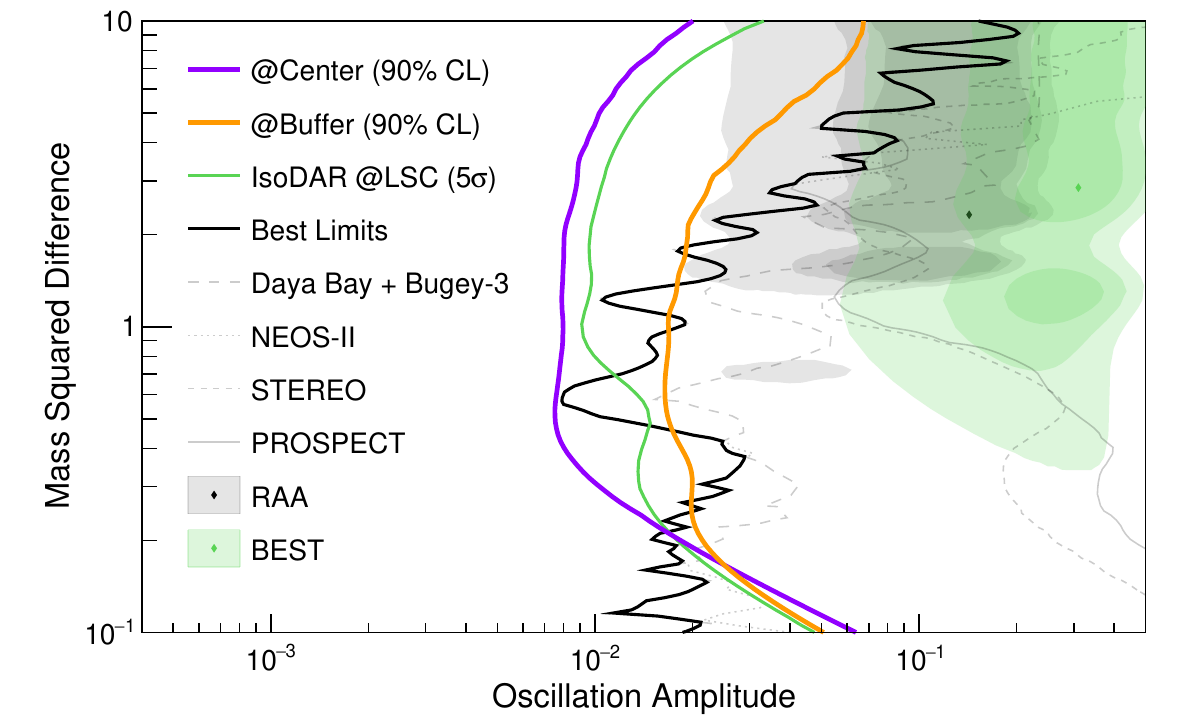}
\captionof{figure}
{
Comparison of sensitivity for source position. The purple 
and orange
curves represent the case where the source is located 
at the center (0, 0, 0) m and
side (9, 0, 0) m of the detector, respectively, with the 
source activity of 200 kCi
and the uncertainty in the radioactivity of 2\%. The thick black curve 
shows the best limit
by combining four results (gray curves) shown together.
Limits from other experiments are also indicated.
}
\label{fig:sensitivity_000VS001}
\end{center}

\subsection{BSM searches with the IsoDAR accelerator}
Since IsoDAR@$\nEYE$ is a first-of-its-kind, the physics 
program is still under development, but is clearly already rich.    
The areas that can explored are identified as international priorities, 
including Neutrino Properties, Cosmic Connections, and 
Rare Events.  The searches for new physics in each area 
are expected to have order of magnitude jumps in sensitivity. This is 
what can be achieved with the incredibly powerful and 
unprecedented combination of a kiloton-scale, precision 
and low-background detector and nearly $10^{25}$ protons on 
target with an accelerator nearby.  
The investigation into this physics utilizes two types of particle sources:
the ``direct flux" of electron antineutrinos from 
$^8$Li decay in the IsoDAR source that traverse the $\nEYE$ detector,
and ``indirect sources," which consist of gammas and neutrons 
contained within the IsoDAR target but that may subsequently 
produce Beyond SM particles capable of passing through the source 
shielding, enter $\nEYE$, and either interact or decay.    
With its extensive size and uniquely protected environment, $\nEYE$ 
can reconstruct the electrons, photons, and 
protons that are predicted to result from such signal particle creation 
inside the detector.
\hfill \break 
IsoDAR@$\nEYE$ will deliver an extensive program 
of neutrino property studies. A discussion of the 
very high sensitivity to oscillations to sterile neutrinos 
using $\bar \nu_e +p \rightarrow e^+ + n$, the inverse beta 
decay (IBD) interaction, can be found in Sec.~\ref{sec:sterile}. As another 
example of compelling new physics searches,  IsoDAR@$\nEYE$ 
will acquire 7000 electron antineutrino elastic 
scattering (ES) events---the world's largest sample. Both 
rate and energy dependence of ES are precisely predicted 
within the Standard Model (SM) from LEP and LHC data 
(\href{https://pdg.lbl.gov/2020/tables/contents_tables.html}{PTEP 083C01 (2020)}). 
The 1.67 million IBD events determine the normalization, reducing uncertainties, 
allowing IsoDAR$@$$\nEYE$ to measure $\sin^2\theta_W$ with a
precision of 1.9\% 
at low Q$^2$, more than five times better than reactor experiments.   
Because of the precise predicted energy dependence of the ES 
sample, $\nEYE$ can also perform a precise phenomenological search for 
non-standard neutrino interactions (NSIs) involving  lepton-lepton 
couplings, complementing the lepton-quark coupling 
searches by the active, global coherent neutrino-nucleus scattering program. 
\hfill \break
With $\bar{\nu}_e e^-$ ES, one can measure the weak mixing angle
at around $Q = 6$ MeV a region where no prior measurements exist. This
allows us to look for beyond the standard model physics for the first time in this 
kinematic range,
via so called non-standard neutrino interactions
(\href{https://journals.aps.org/prd/abstract/10.1103/PhysRevD.89.072010}
{PRD 89 072010 2014}).
Additionally, IsoDAR facilitates searches for new particles produced in the
target and sleeve that decay to $\nu_e \bar{\nu}_e$. The theoretical
motivation arises from generally called {\it light particles} which
refer to low-mass mediators of new interactions in the BSM.
The $\nEYE$ experiment will be most sensitive in the mass
range up to 10 MeV/$c^2$ 
(\href{https://journals.aps.org/prd/abstract/10.1103/PhysRevD.105.052009}
{PRD 105 052009, 2022}).
Once realized, this will be the first facility 
combining ultra-low background large detector with a 
high-power accelerator underground in the world.
\hfill \break $\nEYE$'s antineutrino program also allows searches for 
Dark Matter, which remains unexplained 
despite extensive search efforts over the years.   
A flaw in the approach may be that experiments 
have predominantly focused on a single type of dark matter, while the truth may 
be a combination of sources. For example, a fraction of the dark matter may 
consist of particles exhibit stronger interactions within SM than those
predicted by Weekly Interacting Massive Particles (WIMPs) of comparable mass.
Such a particle would be slowed down by interactions with 
the material in the atmosphere and the Earth before reaching our WIMP 
detectors underground.   However, IsoDAR’s intense 
underground antineutrino beam can 
enhance this ambient dark matter population through upscattering processes,
allowing potential observation in $\nEYE$ 
(\href{https://doi.org/10.1103/PhysRevD.106.035011}{PRD 106 035011 (2022)}).   
Additionally, it is plausible that dark matter includes axion-like
particles as part of its composition.
Axions can be produced through mixing 
with monoenergetic photons produced by nuclear excitations 
in the IsoDAR target, traverse the shielding, and decay or 
interact in the $\nEYE$ detector 
(\href{https://journals.aps.org/prd/abstract/10.1103/PhysRevD.107.095010}{PRD 107 095010 (2023)}). 
\hfill \break A wide range of 
cosmological questions are addressed by introducing a dark sector.  
Neutron--dark sector couplings are also motivated by the 
neutron beam/bottle experiment lifetime discrepancy 
(\href{https://www.annualreviews.org/content/journals/10.1146/annurev.nucl.55.090704.151611}{Ann. Rev. Nucl. Part. Sci. 55 27 (2005)}, \href{https://journals.aps.org/prl/abstract/10.1103/PhysRevLett.111.222501}{PRL 111 222501 (2013)}, \href{https://journals.aps.org/prc/abstract/10.1103/PhysRevC.97.055503}{PRC 97 055503 (2018)}, \href{https://journals.aps.org/prl/abstract/10.1103/PhysRevLett.127.162501}{PRL 127 162501 (2021)}). Such couplings would allow neutrons to disappear from the target and reemerge 
in $\nEYE$'s central and far side through 
$n \rightarrow n^\prime \rightarrow n$ transitions, and the underground, 
ultra-large design of $\nEYE$ allows for  much higher 
sensitivity than surface experiments 
(\href{https://journals.aps.org/prd/abstract/10.1103/PhysRevD.107.075034}{PRD 107 075034 (2023)}).   Alternatively, if the $n^\prime$ are Majorana, then 
$n \rightarrow n^\prime \rightarrow \bar n$ may occur, 
depositing $\sim$2 GeV of energy in the $\nEYE$ upon annihilation.  
This is a $B$ violating effect with implications for 
conditions in the early 
universe (\href{https://doi.org/10.1140/epjc/s10052-020-08824-9}{EPJC 81 33 (2021)}).
\hfill \break This is only a subset of the well-motivated theoretical questions that $\nEYE$ will be able to address.   However, because IsoDAR@$\nEYE$ is first-of-its-kind, it also opens search opportunities that are entirely motivated by the experimental design.   An example is a search for $X \rightarrow \nu_e \bar \nu_e$, where the new particle, $X$ is produced by mixing with the monoenergetic photons and the $\bar \nu_e$ interacts via IBD interactions, producing sharp peaks (\href{https://journals.aps.org/prd/abstract/10.1103/PhysRevD.105.052009}{PRD 105 052009 (2022)}).   No past detector has been able to perform such a search before because it requires the unique underground environment and enormous size of $\nEYE$, coupled with a neutrino source such as IsoDAR.   If such an effect were seen, it would transform theoretical physics.

\subsection{\fontsize{12}{15}\selectfont Solar science with neutrinos}
The Borexino experiment has significantly advanced recent measurements of 
solar neutrinos.
The most recent key findings  include
a measurement of neutrino flux from the CNO cycle of 
$6.7 ^{+1.2}_{-0.8} \times 10^8~\textrm{cm}^{-2}~\textrm{s}^{-1}$
(\href{https://journals.aps.org/prd/abstract/10.1103/PhysRevD.108.102005}
{PRD 108 102005 2023}) 
and
the measurement of $pp-$chain solar neutrinos at
$ 6.1 \pm 0.5 ^{+0.3}_{-0.5} \times 10^{10}~\textrm{cm}^{-2}~\textrm{s}^{-1}$
(\href{https://www.nature.com/articles/s41586-018-0624-y}
{Nature 562, 505 2018}). With this and the $^{7}$Be, $pep$, and $^8$B results,
the MSW-LMA prediction is favored; however, it still carries significant
uncertainties that necessitate more precise measurements to
investigate neutrino interactions beyond SM.
The $\nEYE$ experiment aims to achieve this challenging goal. 
\hfill \break
One important unresolved question in solar science pertains to
the metallicity of the Sun, which refers to  
the abundance of elements with $Z>2$. 
Various analyses of spectroscopic data yield divergent results
(``High or low'' metallicity,
\href{https://www.aanda.org/articles/aa/full_html/2022/05/aa42971-21/aa42971-21.html}{A\&A 661 A140 2022},
\href{https://www.aanda.org/articles/aa/full_html/2021/09/aa40445-21/aa40445-21.html}{A\&A 653 A141 2021}).
The solar neutrino fluxes, particularly those produced by 
reactions within the CNO cycles, 
can address this issue and the latest data prefer the high
metallicity model, but with significant uncertainty. The $\nEYE$ 
experiment will provide a clear resolution to this critical issue 
with high statistics
(see Fig.~\ref{fig:solar_background}) in an exceptionally radio-pure 
environment.
\hfill \break
We will examine the vacuum-MSW transition region of the
electron neutrino survival probability previously 
explored by Borexino 
(\href{https://www.nature.com/articles/s41586-018-0624-y}
{Nature 562, 505 2018}), 
by conducting 
precision measurements of $^7$Be, $pep$, and $^8$B solar neutrinos
to either confirm or refute the current understanding of the neutrino
oscillation phenomena. This will also allow for an investigation of
neutrino interactions within the context of Beyon SM physics.
\begin{center}
\includegraphics[width=\linewidth]{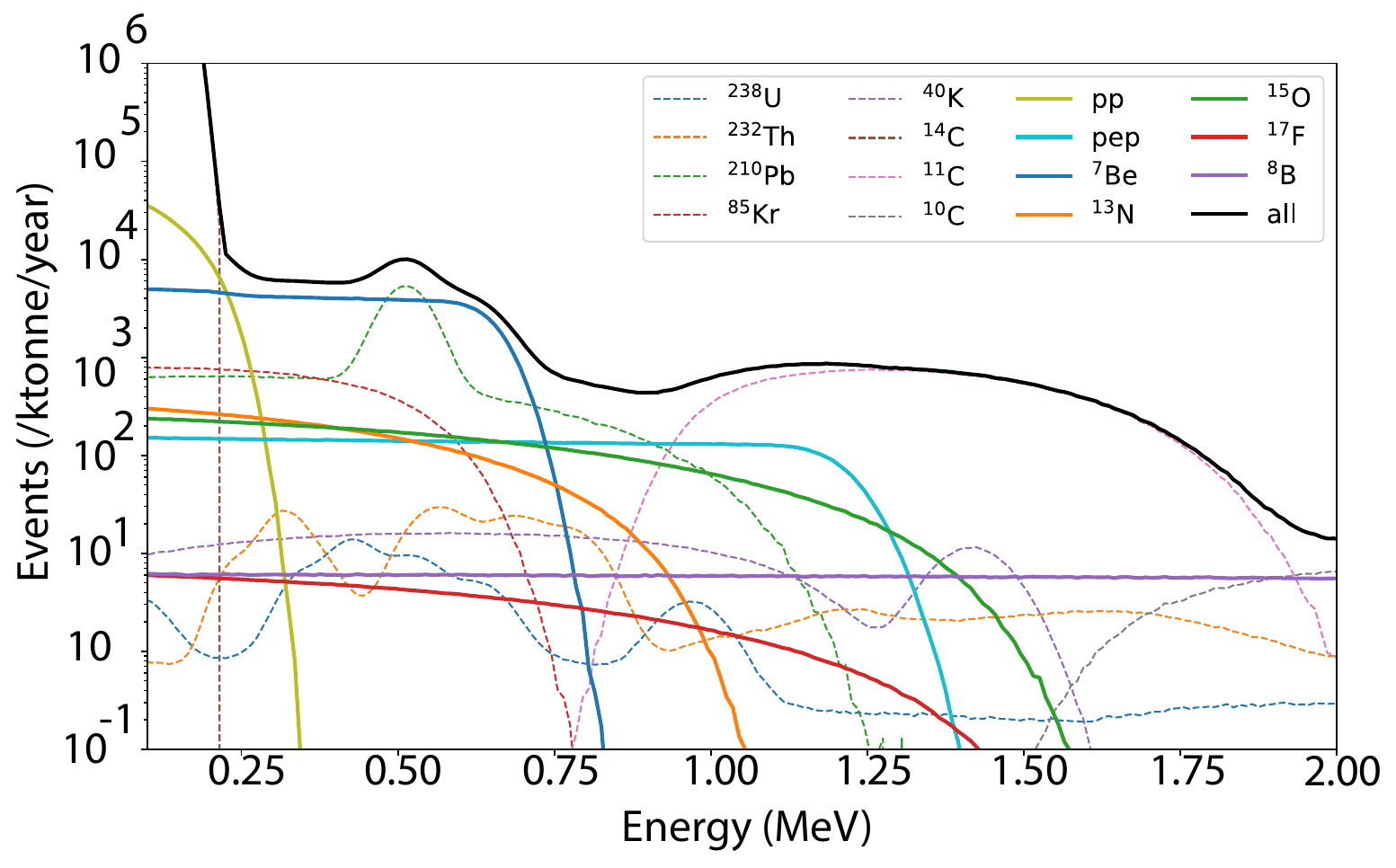}
\captionof{figure}
{
Expected neutrinos from the Sun and from background radioactive
sources as a function of the electron kinetic energy in
the $\nEYE$ 
detector are shown.
Here, we assume B16-GS98 model for the solar neutrino flux
calculation 
(\href{https://iopscience.iop.org/article/10.1088/1742-6596/1056/1/012058}
{J. Phys. Conf. Ser. 1056 012058 2018}),
$^{14}\textrm{C}/^{12}\textrm{C} = 2\times 10^{-18}$,
and the energy resolution of $5\%/\sqrt{E(\textrm{MeV})}$
at the Yemilab site.
}
\label{fig:solar_background}
\end{center}

Here we show the recent compilation of the $\nu_e$ survival
probability as a function of the neutrino energy measured
by Borexino (black dots with error bars,
\href{https://www.nature.com/articles/s41586-018-0624-y}
{Nature 562, 505 2018}), 
in Fig.~\ref{fig:MSW_LMA}, along with expectations from
the $\nEYE$ experiment, which assumes a similar period of data taking but
a target volume five times larger.
There is significant improvements in $^7$Be and $pep$ neutrino
measurements as shown in Fig.~\ref{fig:MSW_LMA}.
Note that
two points next to $^8$B represent the $\nEYE$ experiment's
expectations in the [3-5] MeV
and [5-10] MeV energy bins.
If we further increase the statistics, we may be able to see the
up-turn in the MSW-LMA solution.
\hfill \break
For this, we relied on the statistical uncertainty of the Borexino
results and scaled it down  according to the expected statistics
of the $\nEYE$ experiment.
These expectations may be over-simplified for $pp$ neutrino,
where pile-up from $^{14}$C is expected to be serious. The
frequency of $n$ pile-up in a certain time window $\Delta t$ for
an activity $f$ is
$e^{-f \Delta t} (f \Delta t)^{n-1}/(n-1)!$ from Poisson
statistics (\href{https://arxiv.org/abs/2303.08512}{arxiv:2303.08512}).
The expected rates for $pp$ solar neutrino, as well as the single and double
rates for $^{14}$C decays are summarized in Table~\ref{table:pileup}.
\begin{center}
\begin{tabular}{cccc}
\hline
 & $pp$ & $^{14}$C (single) & $^{14}$C (double) {\rule{0pt}{2.9ex}} \\
\hline
Rates & 400 (cpd) & 687 Hz & 0.236 Hz \\
\hline
\end{tabular}
\captionof{table}{
Expected rates for $pp$ solar neutrino in counts per day (cpd), single 
and double
rates for $^{14}$C decays for a 2 kilo-tonne detector of 5-year 
running period.
The concentration of $^{14}\textrm{C}/^{12}\textrm{C}
= 2 \times 10^{-18}$ g/g is assumed.
}
\label{table:pileup}
\end{center}
\begin{center}
\includegraphics[width=\linewidth]{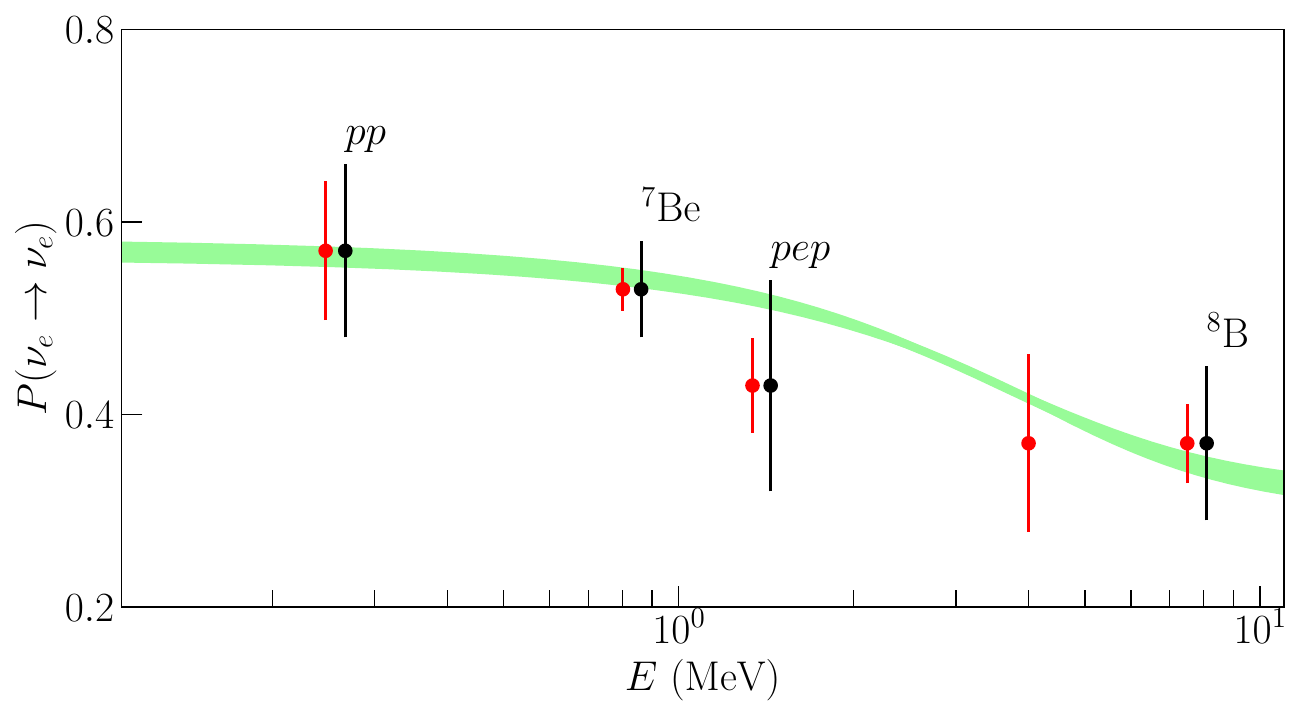}
\captionof{figure}
{
Black dots with error bars show a 
compilation of the $\nu_e$ survival
probability as a function of the neutrino energy measured
by Borexino. The red 
dots with error bars are expectations from the $\nEYE$ experiment,
shifted left a bit from Borexino central values,
assuming a 
similar running period.
The green curve represents the theoretical
prediction 
(\href{https://iopscience.iop.org/article/10.1088/1126-6708/2003/11/004}
{JHEP 11,004, 2003})
with \href{https://pdg.lbl.gov}{PDG} 
values of $\theta_{12}$ and $\Delta m^2_{21}$,  where its finite thickness
reflects their uncertainties. 
Two points next to $^8$B represents the $\nEYE$
experiment's expectation  in the [3,5] MeV
and [5,10] MeV bins.
}
\label{fig:MSW_LMA}
\end{center}

There is a statistical effect of this pile-up for the $pp$ solar neutrino
measurement. From fits using the $\nEYE$ detector simulation
that includes pile-up from $^{14}$C,  
we found that the statistical uncertainty
does not scale with $1/\sqrt{N}$ where $N$ is
the fit yield of the $pp$ neutrino due to the pile-up. This 
is reflected in
Fig.~\ref{fig:MSW_LMA}, which shows only a 20\% reduction
in statistical
uncertainty (instead of the expected 55\%). 
However, with longer data collection and 
pile-up removal using the 
detector directionality, we expect significant improvements
beyond this simulation stage.

\hfill \break
Two issues need to be addressed 
to achieve significantly more precise solar neutrino measurements. 
Since these measurements suffer predominantly from 
background neutrinos produced by radioactive isotope decays (See
Fig.~\ref{fig:solar_background}), an extremely
pure radio purification process must first be applied. 
Here we assume the concentration of $^{14}\textrm{C}/^{12}\textrm{C} =
2 \times 10^{-18}$ g/g
(\href{https://www.sciencedirect.com/science/article/abs/pii/S0370269397015657}
{PLB 422 349 1998}) and is most sensitive to $pp$ neutrino extraction.
The Borexino experiment has extensively  
studied this. The second issue is
distinguishing Cherenkov light from scintillation light to determine 
the direction of the neutrino and suppress the background. 
Note that while an event-by-event measurement is not yet feasible, a 
statistical measurement
of the direction of neutrinos has already been explored
in \href{https://journals.aps.org/prl/abstract/10.1103/PhysRevLett.128.091803}{PRL 128 091803, 2022}. 
Our proposal for  slow LS will be
discussed later for a possible event-by-event directionality.


In solar neutrino detection, the dominant background sources 
are internal radioactivity from the decay of radioactive 
isotopes within the scintillator and cosmogenic isotopes 
generated by  cosmic muon spallation. These backgrounds 
pose a significant challenge as they can mimic the 
signals from neutrino elastic scattering, making it challenging 
to isolate the true neutrino events. 
￼
The principal radioisotopes contributing to internal background 
include $^{40}$K, $^{85}$Kr, and members of the 
$^{232}$Th, $^{238}$U, and $^{210}$Pb decay chains. 
Mitigation of this internal background relies heavily on careful 
material selection and a multi-stage purification process 
applied to the liquid scintillator. This process typically 
involves distillation, liquid-liquid extraction (\textit{e.g.}, 
water extraction), and gas stripping. 
￼
Cosmogenic backgrounds are substantially suppressed by 
the detector's deep underground location, which significantly 
reduces the cosmic muon flux. Furthermore, residual 
muons traversing the detector can be efficiently detected 
and tracked, enabling the removal of most short-lived 
cosmogenic isotopes through a time veto applied around the 
muon track. The most persistent cosmogenic isotopes remaining after this 
veto are $^{10}$C and $^{11}$C, due to 
their relatively long half-lives. Further analysis 
techniques may be employed to discriminate against these 
longer-lived isotopes; for instance a Three-Fold-Coincidence (TFC) 
algorithm uses the spatial and time coincidence among those events
for enhanced discrimination. 
￼
The $\nEYE$ can achieve radio-purity levels of liquid 
scintillator comparable as those used in the Borexino experiment.
by adopting its purification techniques. Figure~\ref{fig:solar_background}
shows all known contributions to background alongside solar neutrinos.

\subsection{\fontsize{12}{15}\selectfont Reactor neutrino program}
 There are four nuclear power plants in South Korea and their distances
from Yemilab are summarized in Table~\ref{table:reactors}
with their thermal powers. 
As shown in Table~\ref{table:reactors}, the Hanul reactor complex,
the closest to Yemilab, is about 65 km away, which corresponds to
the first minimum of the survival probability of
$\bar{\nu}_e \rightarrow \bar{\nu}_e$. 
The geographical locations of reactors, and
survival probability as a function of the distance,
along with the distances four reactors from the Yemilab are 
shown in Fig.~\ref{fig:reactor_detection_yemilab}.
\begin{center}
\includegraphics[width=0.40\linewidth]{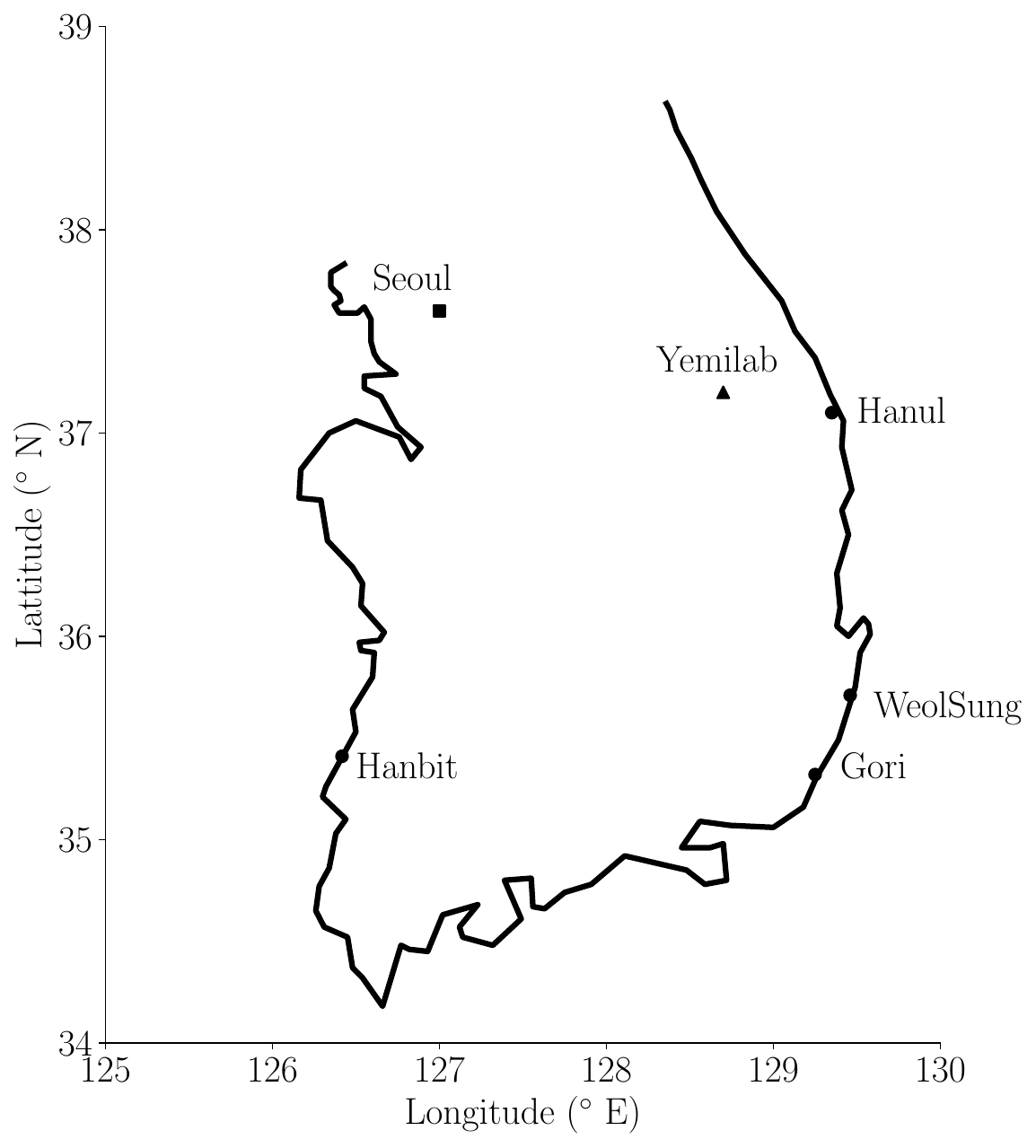}
\includegraphics[width=0.55\linewidth]{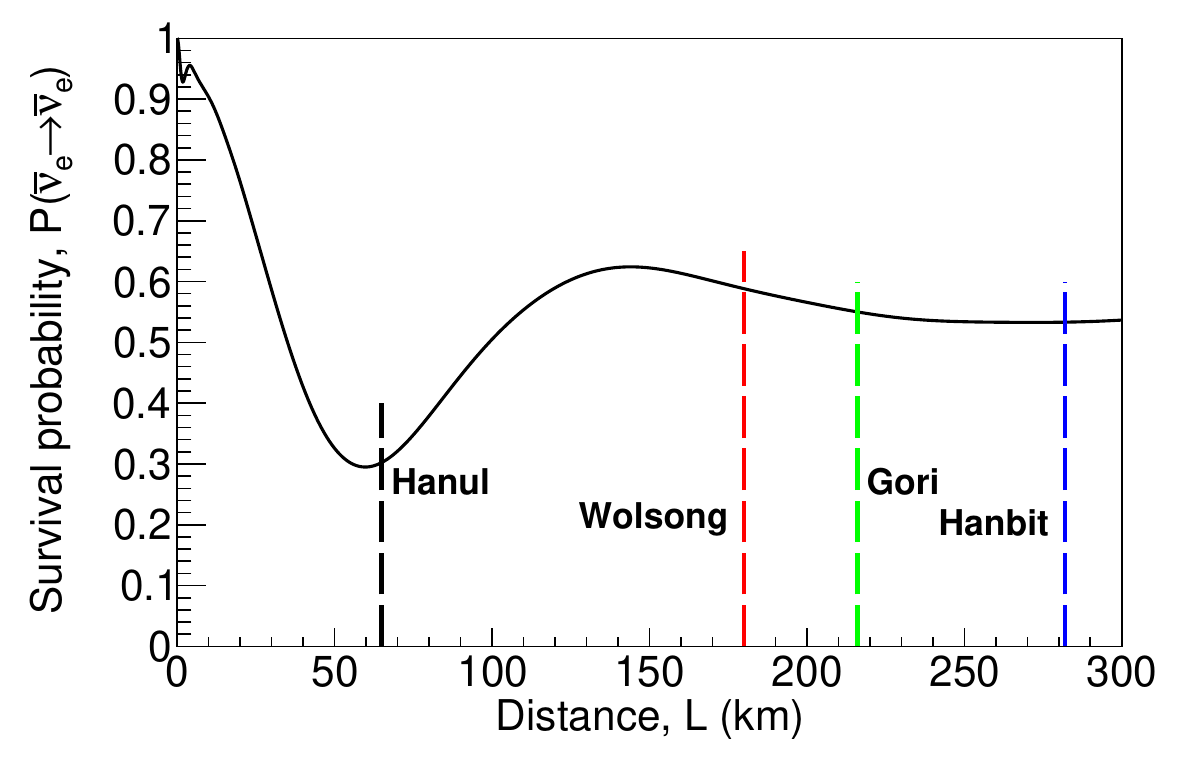}
\captionof{figure}{
Left: the geographical locations of four nuclear reactors, Yemilab,
and Seoul. Right:
our computation of the survival probability of 
anti-electron neutrino as a function
of the distance from nuclear reactors. The 
dashed lines indicate the distance between Yemilab and
the nuclear power plants: Hanul, Wolsong, Gori, and Hanbit.
}
\label{fig:reactor_detection_yemilab}
\end{center}
This positioning suggests that 
the $\nEYE$ detector would be one of the best locations to confirm neutrino
oscillations within the three-flavor paradigm.

\begin{center}
\begin{tabular}{lcccc}
\hline
    	& Hanul & Wolsong & Gori & Hanbit {\rule{0pt}{2.9ex}}\\
\hline
 Thermal & \multirow{2}{*}{20.8} & \multirow{2}{*}{11.8}
 & \multirow{2}{*}{21.3} & \multirow{2}{*}{16.9}\\ 
 Power (GW) & & & & \\ 
 Baseline (km) & 65 & 180 & 216 & 282 \\
\hline
\end{tabular}
\captionof{table}{
List of thermal power and baseline of reactor power plants.
}
\label{table:reactors}
\end{center}
The estimated event rate from the Hanul reactor is approximately
860 events per year. 
The expected sensitivities are $\Delta m^2_{21} = (7.51 \pm 0.06)
\times 10^{-5}~\textrm{eV}^2$ and $\sin^2 2\theta_{12} = 
0.848 \pm 0.010$ (statistical uncertainties only)
for a five-year run of $\nEYE$ experiment, as shown in Fig.~\ref{fig:fit_theta_12}.
Since the $\nEYE$ detector operates at a similar baseline to that 
of the JUNO detector, the $\nEYE$ experiment is uniquely positioned 
to validate results for neutrino oscillation
parameters from the JUNO experiment over the next decade.
\begin{center}
\includegraphics[width=0.8\linewidth]{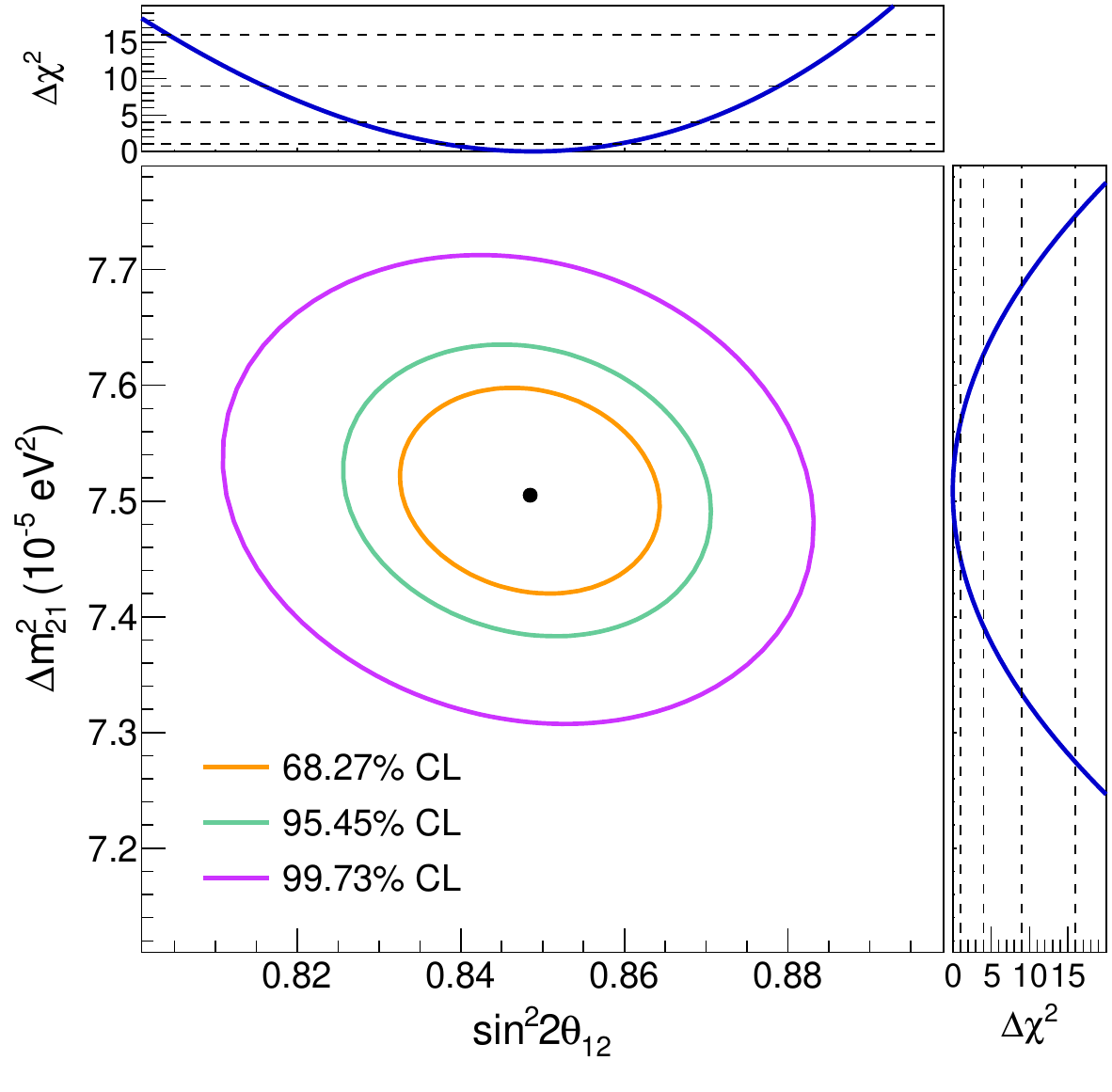}
\captionof{figure}{
A two-dimensional fit results on
$\Delta m^2_{21}$ vs. $\sin^2 \theta_{12}$ are shown.
Three contours represent different confidence levels. Also
projections of $\Delta \chi^2$ differences are shown on top and 
right.
}
\label{fig:fit_theta_12}
\end{center}
\subsection{\fontsize{12}{15}\selectfont Geoneutrinos and astronomical sources}
Understanding the inner structure and energy release mechanisms
of the Earth is one of the primary interests in geoscience.
The localized characteristics of the Earth's crust at various
locations can yield valuable insights into plate tectonics and geodynamics.
\hfill \break
Geo-neutrino signals originate 
from natural decays of $^{238}$U, $^{232}$Th, and
$^{40}$K, which 
Borexino 
(\href{https://journals.aps.org/prd/abstract/10.1103/PhysRevD.92.031101} 
{PRD 92 031101(R), 2015})
and 
KamLAND 
(\href{https://journals.aps.org/prd/abstract/10.1103/PhysRevD.88.033001}
{PRD 88 033001, 2013}) have detected from the first two isotopes.
The next generation underground neutrino experiments, including
$\nEYE$, should 
conduct further investigations of geo-neutrinos to answer such
questions. 
The most severe background, usually considered as an irreducible
background, is $\bar{\nu}_e$ from nuclear reactors.
Because of this, the expected signal-to-noise ratios for the
geo-neutrinos vary depending on the locations of
underground facilities. The best known site at present
is the Chinese 
\href{https://cjpl.tsinghua.edu.cn/column/home}
{Jinping undergound laboratory} (CJPL) in China, where one 
expects an almost background-free environment. On the
other hand, the Yemilab site is located approximately 65 km
from a nuclear power plant with 21 GW of thermal power. This
reduces the signal-to-background ratio to 1/10 initially, 
as indicated in Fig.~\ref{fig:geo_U_Th}.
\begin{center}
\includegraphics[width=0.8\linewidth]{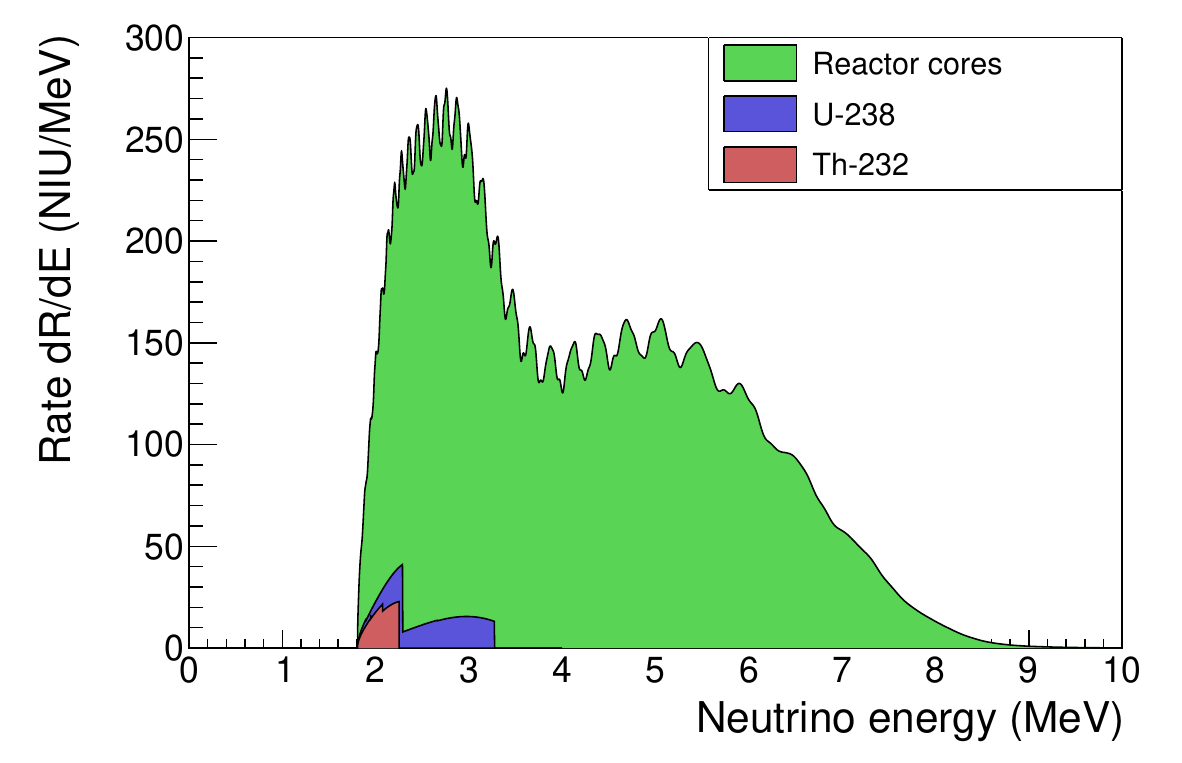}
\captionof{figure}
{Our estimate of geo and reactor neutrinos with the $\nEYE$ detector.
}
\label{fig:geo_U_Th}
\end{center}
We intend to implement the requirement for determining the direction 
of incoming $\bar{\nu}_e$ to 
significantly reduce the background further and extract
the geo-neutrino signal. 
Furthermore, data collected during periods when the
nuclear reactor is not operational--referred to as ``off'' data--will
enhance our understanding of background.
We expect to have higher statistics than CJPL but with
much larger backgrounds.
It is also essential to
gather local information about the geo-neutrino near 
Yemilab site to better understand local plate tectonics.
\hfill \break
At the core collapse of a supernova, an estimated of $10^{47}$ 
Joules of gravitational binding energy will be 
released via nuclear weak interactions in form of  MeV neutrinos.
Such neutrinos have been detected on one occasion by neutrino detectors
(\href{https://journals.aps.org/prl/abstract/10.1103/PhysRevLett.58.1490}
{PRL 58 1490, 1987}, 
\href{https://journals.aps.org/prl/abstract/10.1103/PhysRevLett.58.1494}
{PRL 58 1494, 1987}, 
\href{https://www.sciencedirect.com/science/article/abs/pii/0370269388916516}
{PLB 205 209, 1998}), albeit with only a few dozen events recorded. Further detection 
of supernova neutrinos is of great interest for the supernova physics  
and for probing novel particle physics properties, 
which are otherwise inaccessible in laboratories. The expected
detection rates for one supernova at a distance of 10 kpc is
$\mathcal{O}(100)$ events per kilo-tonne 
detector (\href{https://journals.aps.org/prd/abstract/10.1103/PhysRevD.66.033001}
{PRD 66, 033001, 2002}). 
We also plot
a possible coherent detection of neutrinos 
(\href{https://www.sciencedirect.com/science/article/abs/pii/S0927650512001132}
{Astroparticle Phys. 36 151, 2012},
\href{https://www.sciencedirect.com/science/article/pii/S0927650523000762}
{Astroparticle Phys. 153, 102890, 2023}) is shown 
in Fig.~\ref{fig:supernova_nEYE} along with the neutrino energy distribution.
\begin{center}
\includegraphics[width=0.49\linewidth]{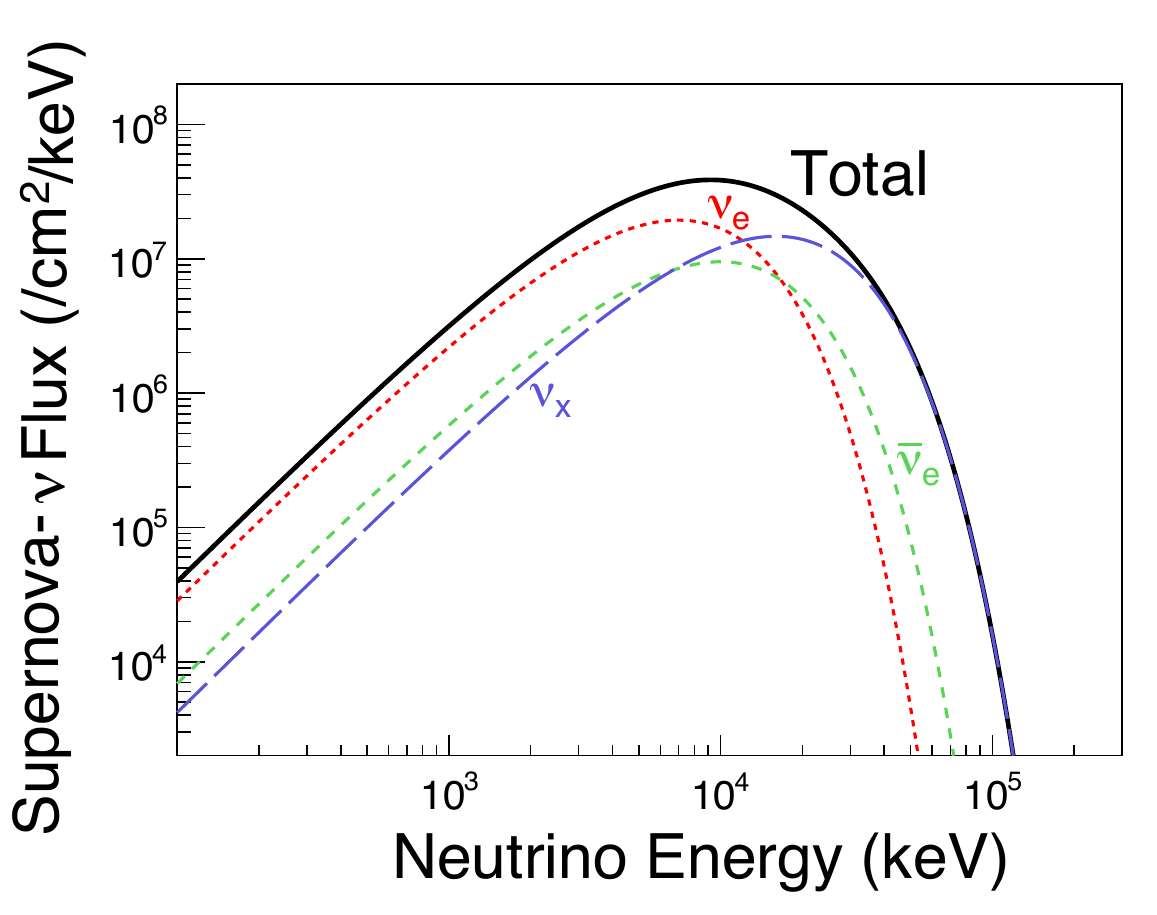}
\includegraphics[width=0.49\linewidth]{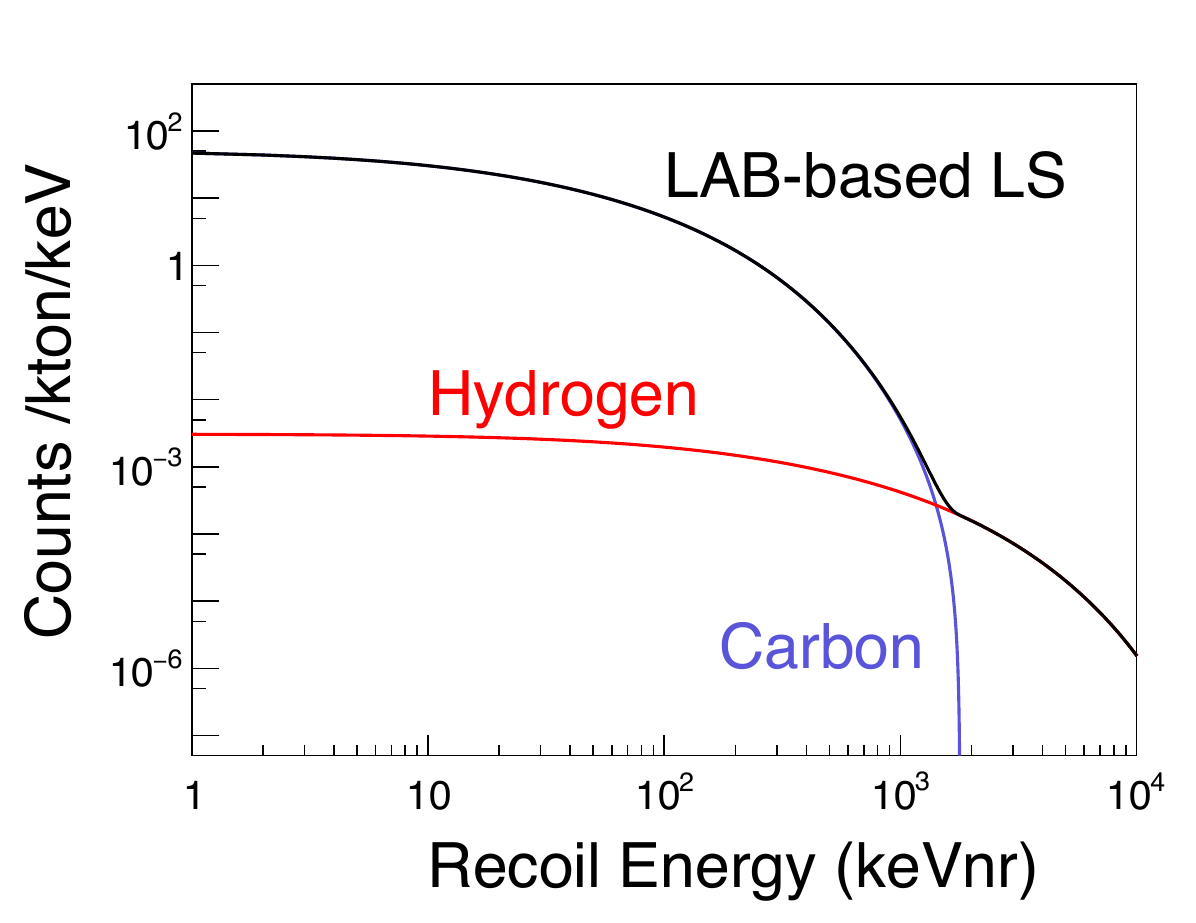}
\captionof{figure}
{
Energy spectra for the supernova neutrinos (left) and
the expected nuclear recoil energy spectra of coherent scattering
from supernova neutrinos.
}
\label{fig:supernova_nEYE}
\end{center}
Note that the $\nEYE$ will be one of the supernova burst detectors in the
world (less than 10) and first in South Korea.

\subsection{\fontsize{12}{15}\selectfont Neutrinoless double
beta decay}
The discovery of neutrinoless double beta decay process will
indicate that the neutrino is Majorana particle and will tell
us the mass scale of the neutrino through the effective
Majorana neutrino mass parameter. Therefore, this is considered
as one of the most important subjects in modern particle physics.
Experimental sensitivities for the search of neutrinoless double 
beta ($0\nu\beta\beta$) decay have begun to reach 
the inverted neutrino mass ordering; however, no evidence for 
$0\nu\beta\beta$ signal has yet been detected.
The current best upper limit on the effective 
Majorana neutrino mass, $m_{\beta\beta} < $ [28-122] meV at a 90\% confidence 
level (CL), was achieved by KamLAND-Zen 
(\href{https://journals.aps.org/prl/abstract/10.1103/jkf6-48j8}
{Phys. Rev. Lett. 135, 262501, 2025}).
It used 745 kg of enriched $^{136}$Xe loaded into the LS detector.


The $\nEYE$ detector can be turned into a 
decent $0\nu\beta\beta$ search detector that is competitive with 
existing and other planned experiments when it is loaded with 
an appropriate $\beta\beta$-decay candidate isotope.

One possible candidate isotope is 
Tin-124 ($^{124}$Sn) with $Q$=2.2 MeV.
It has abundance of 5.79\% 
but shows high solubility in aromatic solvent in 
the form of organo-tin such as tetramethyltin (TMSn) 
or tetrabutyltin (TBSn), and the loaded LS demonstrates relatively 
small light quenching according to a 
pioneering study 
(\href{https://www.sciencedirect.com/science/article/abs/pii/S0927650509000759}
{Astroparticle Phys. 6, 412, 2009}).

 Another possible choice is $^{130}$Te that has a $Q$ = 2.54 MeV.
It has relatively high natural abundance of 34.1\%. 
We plan to have a long-term R\&D for double beta candidates,
and loading methods into LS, in order to reach the half-life
time sensitivity of order $10^{28}$ years.

\subsection{\fontsize{12}{15}\selectfont Detector}
In this session, we discuss the $\nEYE$ detector design
R\&D and optimization status. Note that our default
detector is widely used LAB based LS. Below
we discuss R\&D beyond our default LS.
\subsubsection{The $\nEYE$ detector}
\tikz\draw[black,fill=black] (0,0) circle (.5ex); 
\colorbox{mycolor1!30}{\bf The detector geometry}
\hfill \break
%
Our proposed structure of the detector consists of
\begin{itemize}
\item Target : the central sensitive part of the detector 
with LS inside an acrylic tank.
\item Buffer : the buffer part that removes the background.
\item Veto : the veto of cosmic rays and external backgrounds. 
\end{itemize}

The currently considered geometries for the $\nEYE$ detector are 
illustrated in Fig.~\ref{fig:nEYE_all}.
The left panel shows a spherical frame for the LS,
while the right panel presents a cylindrical configuration.
In both designs,
the buffer consists of a stainless steel tank or structure filled with 
water or mineral oil, with a diameter of 17 m.
For the cylindrical option, the height is 17 m.  
Surrounding the buffer, a water tank serving as a  
veto system is positioned, measuring 20 m in height and
19.5 m in diameter. 
\hfill \break
Our preferred design is base on the spherical geometry, due to its
enhanced capability of the background reduction. 
Key design considerations include the number of PMTs, the volume
and mass of LS inside an acrylic tank,
and the effective target volume.
Table~\ref{table:two_options} summarizes the key parameters of 
the two target configurations.
\begin{center}
\includegraphics[width=\linewidth]{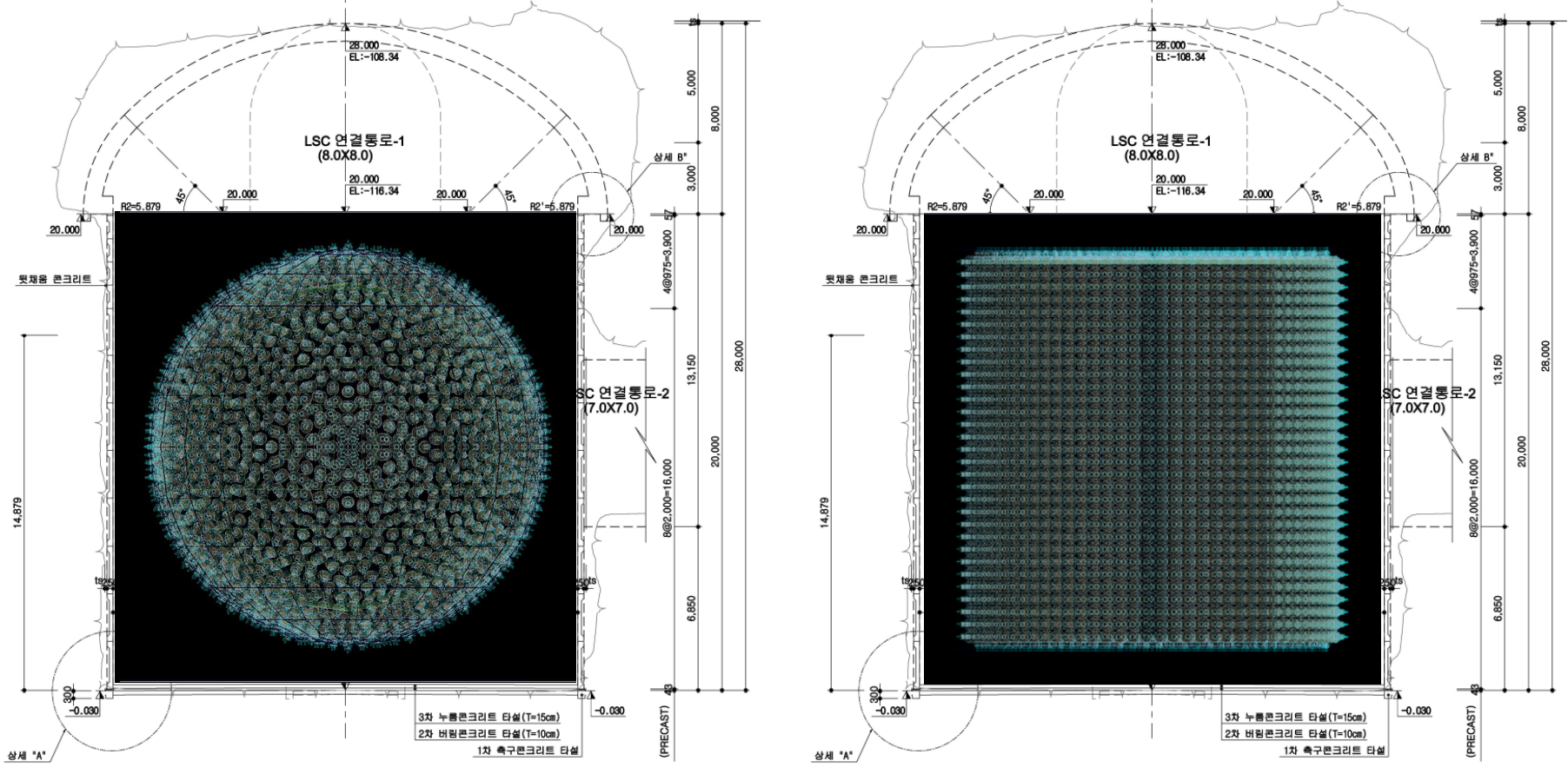}
\captionof{figure}
{
Left: the LS acrylic tank is placed in spherical frame, and
right: 
the LS is contained in a  cylindrical frame.
}
\label{fig:nEYE_all}
\end{center}
%
%
\begin{center}
\begin{tabular}{cll}
\hline
Shape & Cylindrical & Spherical {\rule{0pt}{2.9ex}} \\
\hline
Radius (m)     & 7     &  7.25  \\
Height (m)     & 14.5  &  -   \\
Volume (m$^3$) & 2232  & 1596 \\   
Mass (kg)      & 1920  & 1373 \\   
Number of PMTs & 3700  & 3000 \\
\hline
\end{tabular}
\captionof{table}{
A possible target (LS) geometries between
cylindrical and spherical shape.
}
\label{table:two_options}
\end{center}
\tikz\draw[black,fill=black] (0,0) circle (.5ex); 
\colorbox{mycolor1!30}{\bf Reconstruction of IBD
}
\hfill \break
 In order to develop the reconstruction algorithm for IBD events,
we first place the $^{144}$Ce radioactive source at the
coordinate (9.5,0,0) m in Geant4 where (0,0,0) m is the center
of the $\nEYE$ detector. The locations of the IBD events are
shown in Fig.~\ref{fig:IBD_xyz} where the event vertices are
more densely populated near the source.
\begin{center}
\includegraphics[width=\linewidth]{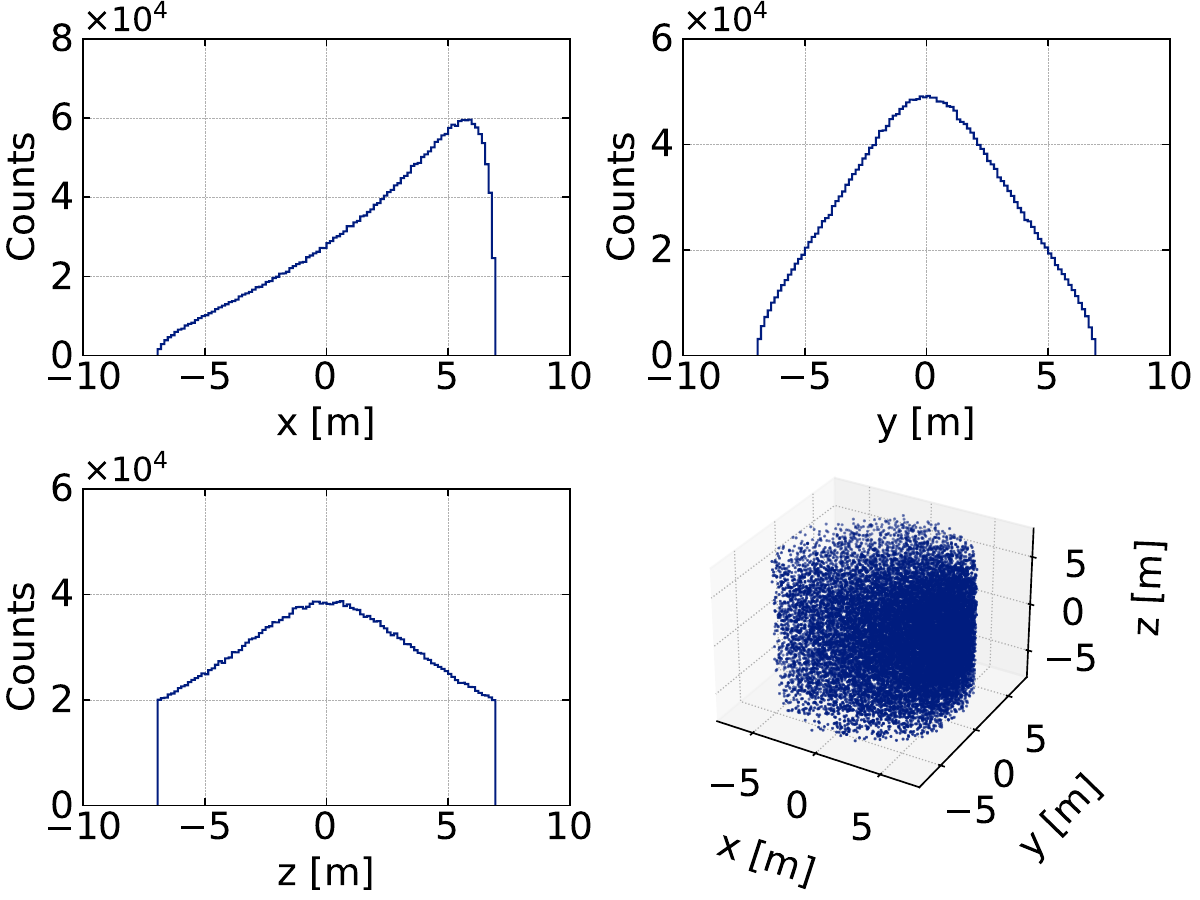}
\captionof{figure}
{From the left, $x$, $y$, and, $z$ position of IBD events
are shown. The rightmost scatter plot shows the 3-dimensional
plot of vertices.
}
\label{fig:IBD_xyz}
\end{center}
The detection principle of the IBD is the prompt detection
of light from $e^+$ annihilation, followed by the subsequent 
detection of light resulting from
delayed neutron capture.
The timing of the delayed capture is on the order of
$\mathcal{O}$(200) $\mu$s, with a
2.22 MeV gamma emission. Based on this detection principle, we implement the
following steps to trigger IBD events:
\begin{itemize}
\item Require hits in 200 ns.
\item If the number of photoelectrons (NPE) exceeds 
400, all hits in the 500 ns window
constitute the trigger cluster (S1).
\item The second cluster is required to have the same condition (S2).
\end{itemize}
Figure~\ref{fig:IBD_S1S2} shows the NPE distributions in 
250 $\mu$s window for one event (on the left). The
central (right) plot has a zoomed view of S1 (S2) NPEs. 
\begin{center}
\includegraphics[width=\linewidth]{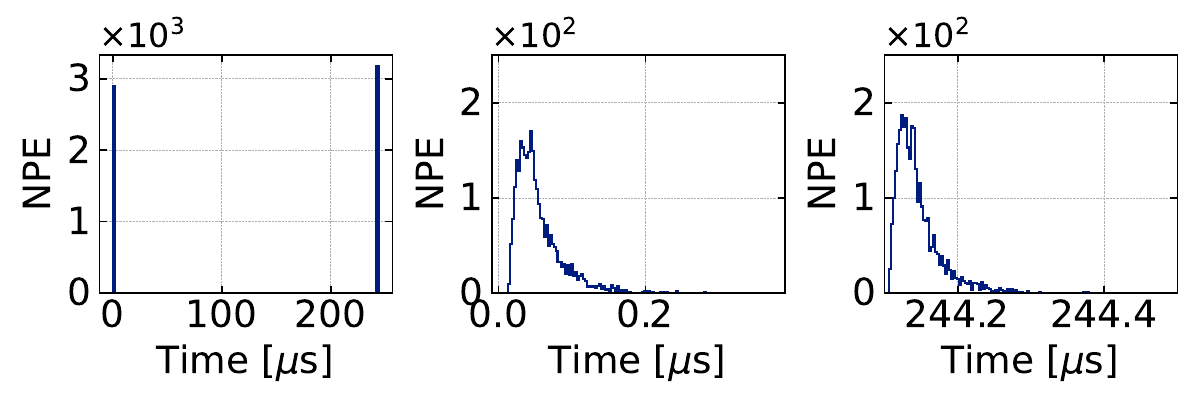}
\captionof{figure}
{
The NPE distributions in
250 $\mu$s window for one event (on the left). The
central (right) plot has a zoomed view of S1 (S2) NPEs.
}
\label{fig:IBD_S1S2}
\end{center}
Once S1 (prompt signal) and S2 (delayed signal)
are reconstructed, we analyze the time difference and deposited
energy distributions.
These are shown in Fig.~\ref{fig:IBD_tE}.
\begin{itemize}
\item The neutron capture time is found to be 214 $\mu$s on average.
\item The characteristic 2.22 MeV peak for S2 is clearly visible.
\item The overall efficiency of IBD reconstruction is approximately 94.6\%
\end{itemize}
This demonstrates the robustness of our detection 
and reconstruction approach for IBD events.
\begin{center}
\includegraphics[width=\linewidth]{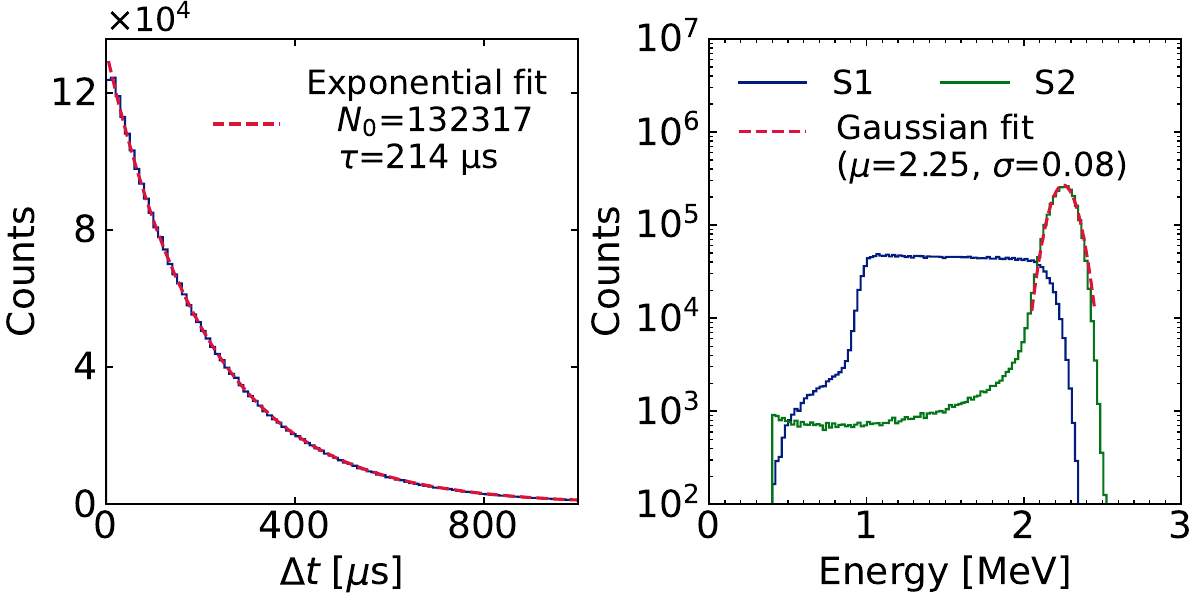}
\captionof{figure}
{
The time difference between S1 and S2 (left)
and their deposited energy distributions of them (right) are shown. 
}
\label{fig:IBD_tE}
\end{center}

%
%
%
\subsubsection{Slow LS}
In the order of $\mathcal{O}$(MeV) and
sub-MeV neutrino detection, a large part of the
background arises from radioactive decays in the
detector itself. Therefore, tremendous efforts have
been made to have a radio purification system
near the detector. The decay of radioactive material
is random in direction but the neutrinos of interest
are directional, such as neutrinos from a radioactive source or from 
the Sun. 
\hfill \break
The scattered electrons in the LS generate both scintillation and
Cherenkov lights, and neutrino's direction can be reconstructed
if the Cherenkov light is detected. However, the amount of 
Cherenkov light is far less than that of scintillation and usually
undetected. The separation of Cherenkov and scintillation
significantly improves background rejection and signal efficiency.
This hybrid detection can be achieved by following methods:
\begin{itemize}
\item Development of LS with slow scintillation time to separate Cherenkov light.
\item Fast photon detectors.
\item Spectral sorting.
\end{itemize}
\begin{center}
\includegraphics[width=\linewidth]{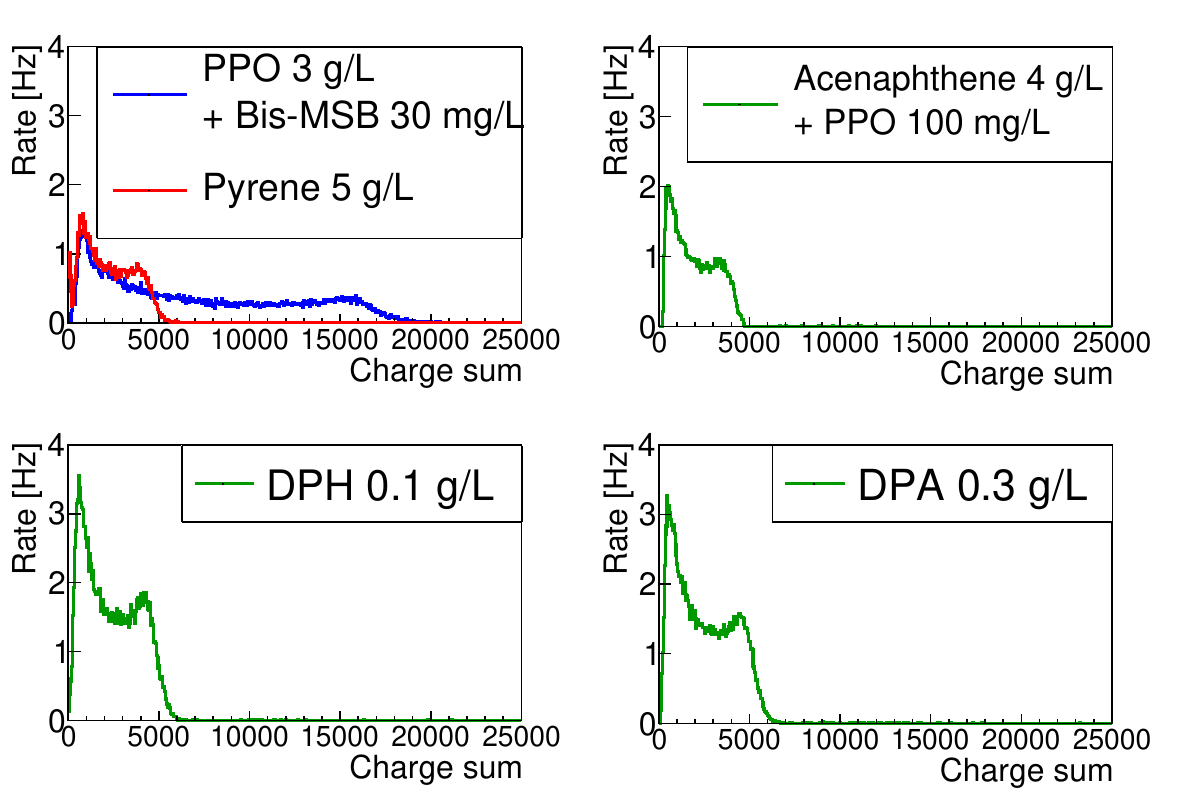}
\captionof{figure}
{
Relative light yields for PPO+Bis-MSB, pyrene, acenaphthene + PPO,
DPH, and DPA are shown. The combination of PPO+Bis-MSB shows
the highest light yield.
}
\label{fig:light_yield}
\end{center}
Our primary target R\&D is the first item, the slow LS. 
We will briefly mention our R\&D for the fast photon
detectors. The spectral sorting 
(\href{https://journals.aps.org/prd/abstract/10.1103/PhysRevD.101.072002}
{Phys. Rev. D 101 072002, 2020}) in principle works
but becomes very expensive for the $\nEYE$ experiment. We also
find the fast photon detector has a similar issue.
\begin{center}
\includegraphics[width=\linewidth]{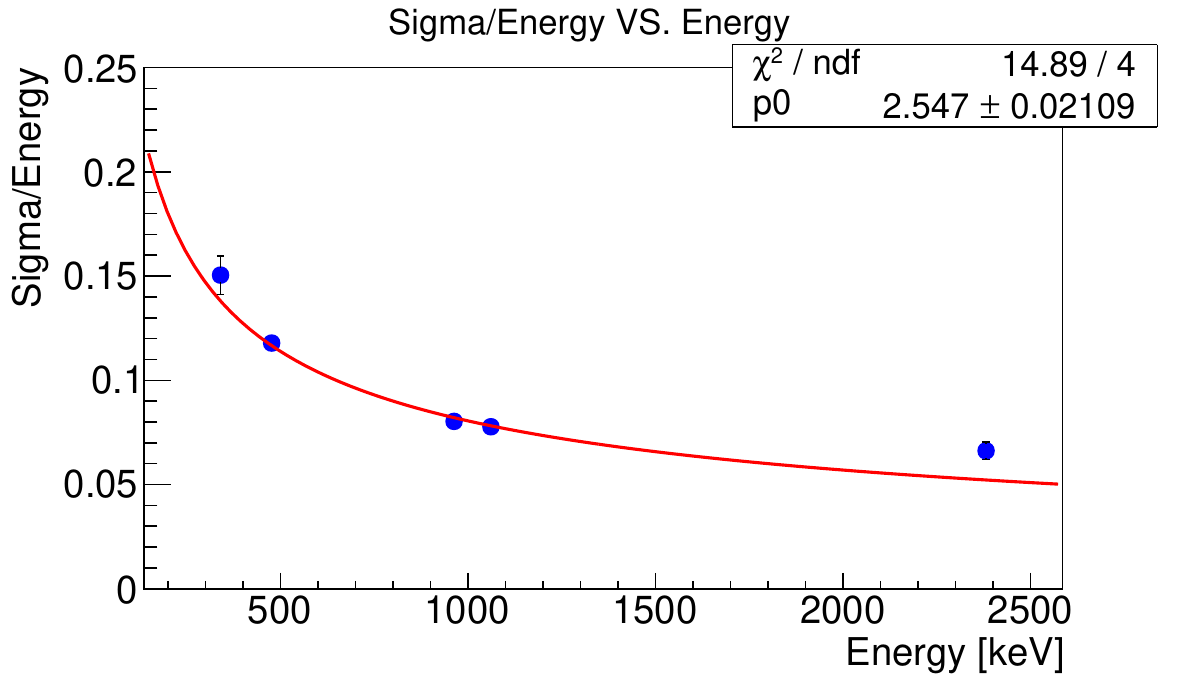}
\captionof{figure}
{
The relative energy resolution as a function of the source
energy is shown. Points are measurements with various radioactive
sources and the curve is the fit to points. 
}
\label{fig:LS_resolution}
\end{center}
For a slow LS R\&D, we start with mixtures of 
linear alkylbenzene (LAB) and fluors, motivated by 
\href{https://www.sciencedirect.com/science/article/pii/S0168900220305167?via%3Dihub}
{NIMA 972 164106, 2020}. For fluors, we use
PPO 3mg/L + Bis-MSB 30 mg/L, pyrene 5 g/L, acenaphthene 4 g/L + PPO 100 mg/L,
DPH 0.1 g/L or DPA 0.3 g/L, to measure relative light yields,
energy resolution, 
absorbance\footnote[1]{The logarithmic ratio of incoming and transmitted
radiant power.},
and emission spectra.
A small acrylic vessel with a diameter of 60 mm and a height of 50 mm is
prepared as a container for LS.
Relative light yields for PPO+Bis-MSB, pyrene, acenaphthene + PPO,
DPH, and DPA are shown in Fig.~\ref{fig:light_yield}. The combination 
of PPO+Bis-MSB shows
the highest light yield.
\begin{center}
\includegraphics[width=\linewidth]{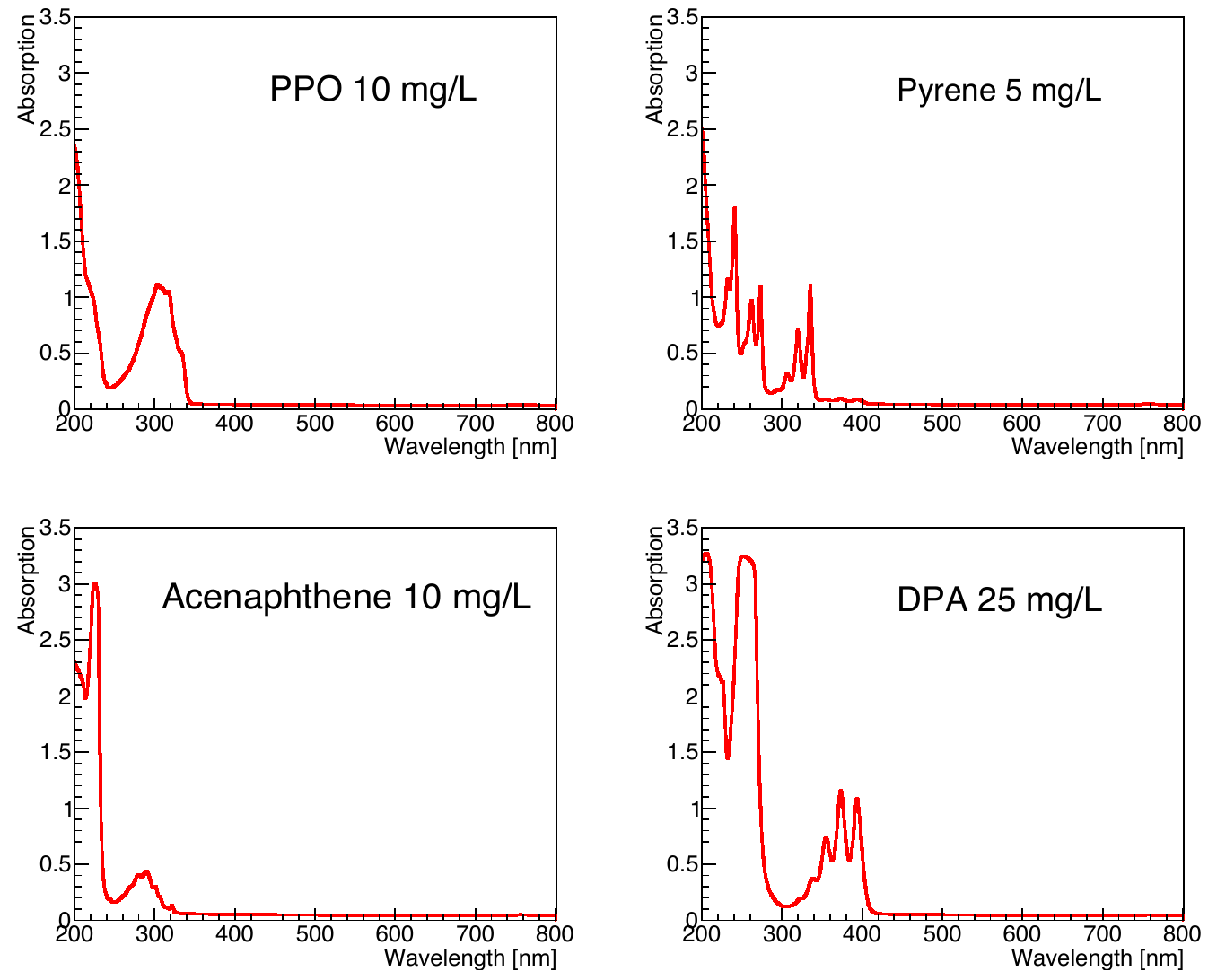}
\captionof{figure}
{
Absorbances for PPO+Bis-MSB, pyrene, acenaphthene + PPO,
and DPA as a function of the wavelength are shown. 
}
\label{fig:LS_absorbance}
\end{center}
The energy resolution is measured with radioactive sources
and the Geant4
simulation, and the result is shown in
Fig.~\ref{fig:LS_resolution}. 
The fit to the data points yields $\sigma/E = 2.55 \pm 0.02~(\sqrt{\textrm{keV}})
/\sqrt{E~\textrm{(keV)}}$.
Note that the container
is relatively small, so the resolution is
about 8\%  at 1 MeV.
\hfill \break
The absorbances
for all five LS targets are also measured
and shown in Fig.~\ref{fig:LS_absorbance}. Note that 
both the relative light yields and absorbances agree
with previous findings 
(\href{https://www.sciencedirect.com/science/article/pii/S0168900220305167?via%3Dihub}
{NIMA 972 164106, 2020}). 
\hfill \break 
We also developed a Geant4 simulation setup of a prototype
detector, for the simulation study of LS. The pyrene LS detector
is contained in an acrylic structure with 
a diameter of 1.2 m, a height of 1.2 m and a 
thickness of 1 cm. This corresponds
to 1.2 tonnes of LS. There is a buffer made of stainless steel
with a diameter of 2.2 m and a height
of 2.2 m. The buffer is currently filled
with water. There are 140 PMTs of 10 inches placed,
motivated by \href{https://hep.hamamatsu.com/eu/en/products/R7081.html}{Hamamatsu R7081 PMT}.
Note that the quantum efficiency is 25\% at the wavelength of 390 nm.
Fig.~\ref{fig:LS_G4_prototype} (left)
shows the geometry of the prototype LS detector. The PMT hit time
distribution is obtained from this geometry, shown in 
Fig.~\ref{fig:LS_G4_prototype} (right). The emission and re-emission spectra
are also checked.
We plan to construct this prototype detector in the 
early stages of the project to learn as many technical details
as possible about the slow LS properties.
\begin{center}
\includegraphics[angle=270,width=\linewidth]{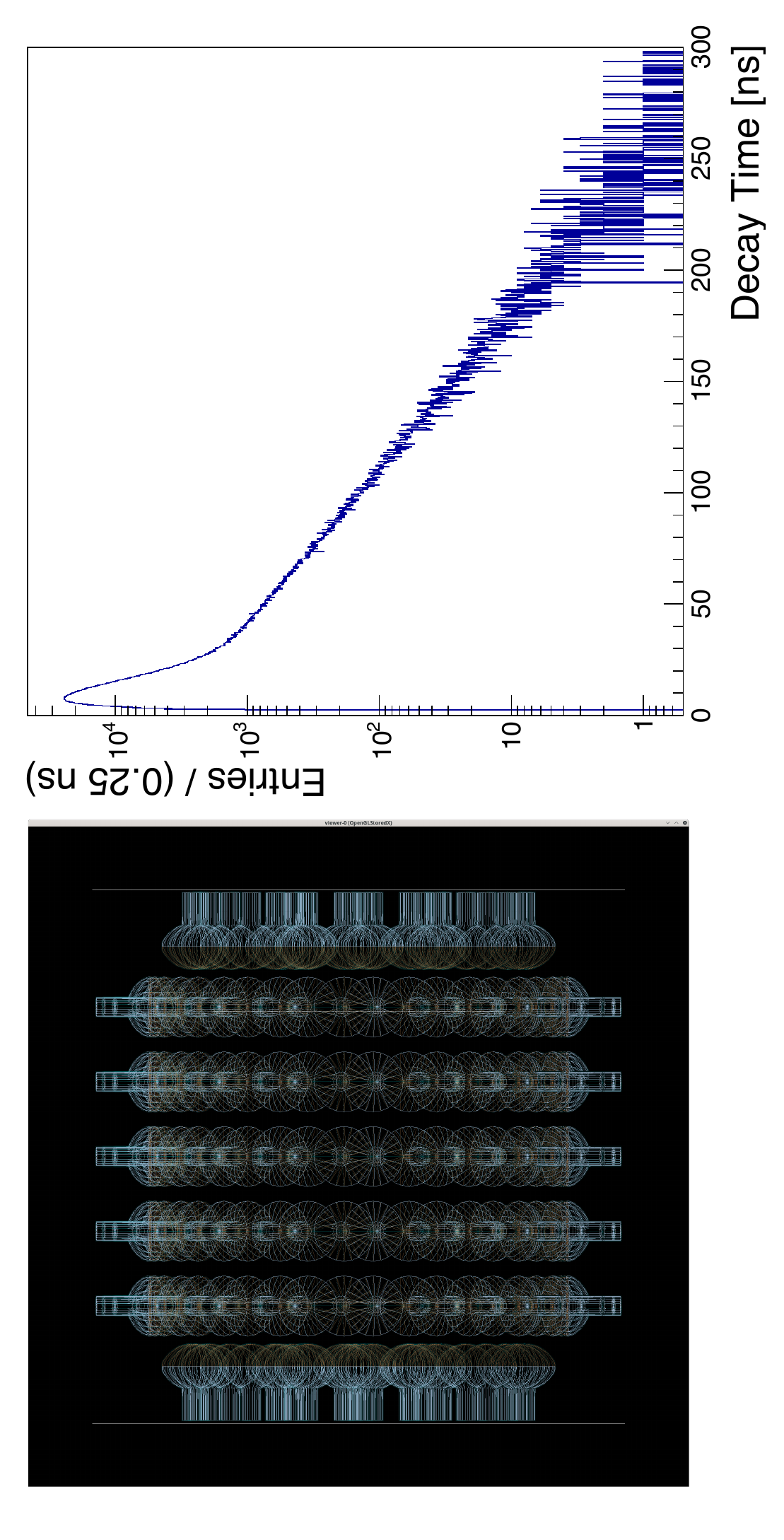}
\captionof{figure}
{
A prototype LS detector geometry is on the left. The LS scintillation
time seen by PMTs is on the right.
}
\label{fig:LS_G4_prototype}
\end{center}

\hfill \break
We also studied the Cherenkov light separation capability
in the Geant4 simulation. This was done with the
two-kilo-tonne $\nEYE$ detector geometry setup which will be described
in detail later. We introduce the PMT hit timing correction
with the formula
\begin{eqnarray}
t  =  t_\textrm{meas} - 
\frac{|\mathbf{r}_\textrm{rec} - \mathbf{r}_\textrm{PMT}|}
{c/n},
\end{eqnarray}
where $t_\textrm{meas}$ is the measured 
time. The $\mathbf{r}_\textrm{rec}$ and $\mathbf{r}_\textrm{PMT}$
are reconstructed vertex and each PMT location 3-vectors.
Here, $c$ and $n$ are the speed of light in vacuum and the index of refraction
of the LS.
Our Geant4 simulation of hit-time distributions with the $\nEYE$ 
detector setup, with (red) and without (blue) timing correction
are shown in Fig.~\ref{fig:LS_G4_timing}.
From the left, the pyrene concentrations are,
8, 4, and, 2 g/L, respectively.
Note that the separation depends on the concentration of pyrene in LS.
Here we assume 10 cm vertex reconstruction resolution and
the transient time spread of 1 ns. 
\begin{center}
\includegraphics[width=\linewidth]{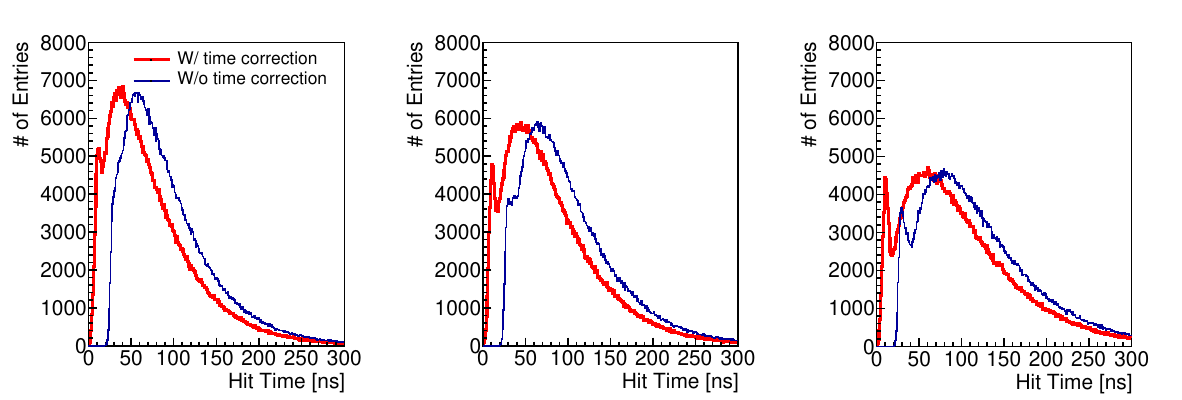}
\captionof{figure}
{
Our Geant4 simulation of hit-time distributions with the $\nEYE$ 
detector setup, with (red) and without (blue) timing correction
described in the text.
From the left, the pyrene concentrations are,
8, 4, and, 2 g/L, respectively.
}
\label{fig:LS_G4_timing}
\end{center}
We would like to emphasize that the increase in the separation
capability of Cherenkov light in the slow LS is obtained 
at the cost of the light yield in general. Our main R\&D 
is to see if there is an optimum balance between the two 
competing issues.

\subsubsection{Photodetector}

The photodetector is the most significant portion of the budget
in large-size neutrino telescope such as $\nEYE$.
Here we briefly mention two technologies.
\hfill \break
\hfill \break
\tikz\draw[black,fill=black] (0,0) circle (.5ex); 
\colorbox{mycolor1!30}{\bf PMT}
\hfill \break
In Table~\ref{table:PMT},
we list several candidate photodetectors and discuss
their properties. Hamamatsu PMTs have been well studied
already by Yemilab team and therefore not mentioned here.
We plan to study further 8 inch PMTs by NVT in near future 
since they offer competitive performance with reduced
production cost. This one will be used in 
\href{https://www.sciencedirect.com/science/article/abs/pii/S0168900224005527?via%3Dihub}
{Jinping Neutrino Experiment}.
\begin{center}
\begin{tabular}{lccc}
\hline
Model & R12860 & N6203 & N6082 {\rule{0pt}{2.9ex}} \\
\hline
Size (inch) & 20 & 20 & 8 \\
Peak wavelength (nm) & 420 & 380 & 380 \\
HV (V) & 2000 & 1900 & 1750 \\
Q.E. (\%) & 30 & 30 & 30 \\
TTS (ns) & 2.4 & 5 & 1.6 \\
Supplier & Hamamatsu & NVT & NVT \\
\hline
\end{tabular}
\captionof{table}{
Comparison between selected PMTs available
at present. 
Hamamatsu is a Japanese and NVT is a Chinese
company.
}
\label{table:PMT}
\end{center}
\hfill \break
\tikz\draw[black,fill=black] (0,0) circle (.5ex); 
\colorbox{mycolor1!30}{\bf LAPPD}
\hfill \break
The LAPPD (Large area picosecond photodetector, 
\href{https://www.sciencedirect.com/science/article/abs/pii/S0168900219312690}
{NIMA 958 162834, 2020}) is a planar geometry
photodetector based on the micro channel plate (MCP)
technology. The basic properties are summarized
in Table~\ref{table:LAPPD}.
\begin{center}
\begin{tabular}{lc|lc}
\hline
Area (mm$^2$) & $200\times 200$ & HV & $-2100$ V {\rule{0pt}{2.9ex}}\\
Gain & $10^7$                        & Q.E. & $>$ 30\%  \\
Timing & 55 ps & MCP & 2 layers \\
\hline
\end{tabular}
\captionof{table}{
A list of parameters for the LAPPD.
}
\label{table:LAPPD}
\end{center}
To understand the operational characteristics of LAPPD, we
rented a second-generation, capacitor-coupled 64-pad
LAPPD from \href{https://incomusa.com/lappd/}{Incom, Inc} in
the US. For triggers, a cubical bulk 
plastic scintillator is placed with top, bottom,
and side (right surface) PMTs. The LAPPD is
attached to the left surface of the scintillator as shown 
in Fig.~\ref{fig:LAPPD_bulk} left. In total 48 signals are
read out by CAEN 5742 digitizers. A cosmic muon signal is
shown in Fig.~\ref{fig:LAPPD_bulk} right. 
A couple of channels may have low gains.
\begin{center}
\includegraphics[width=0.48\linewidth]{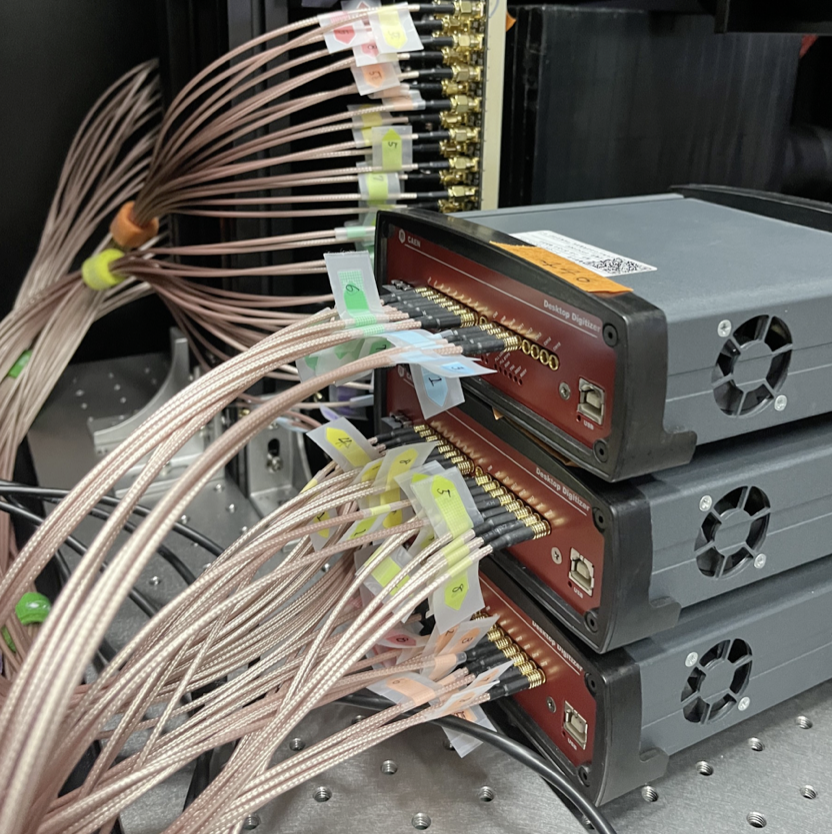}
\includegraphics[width=0.48\linewidth]{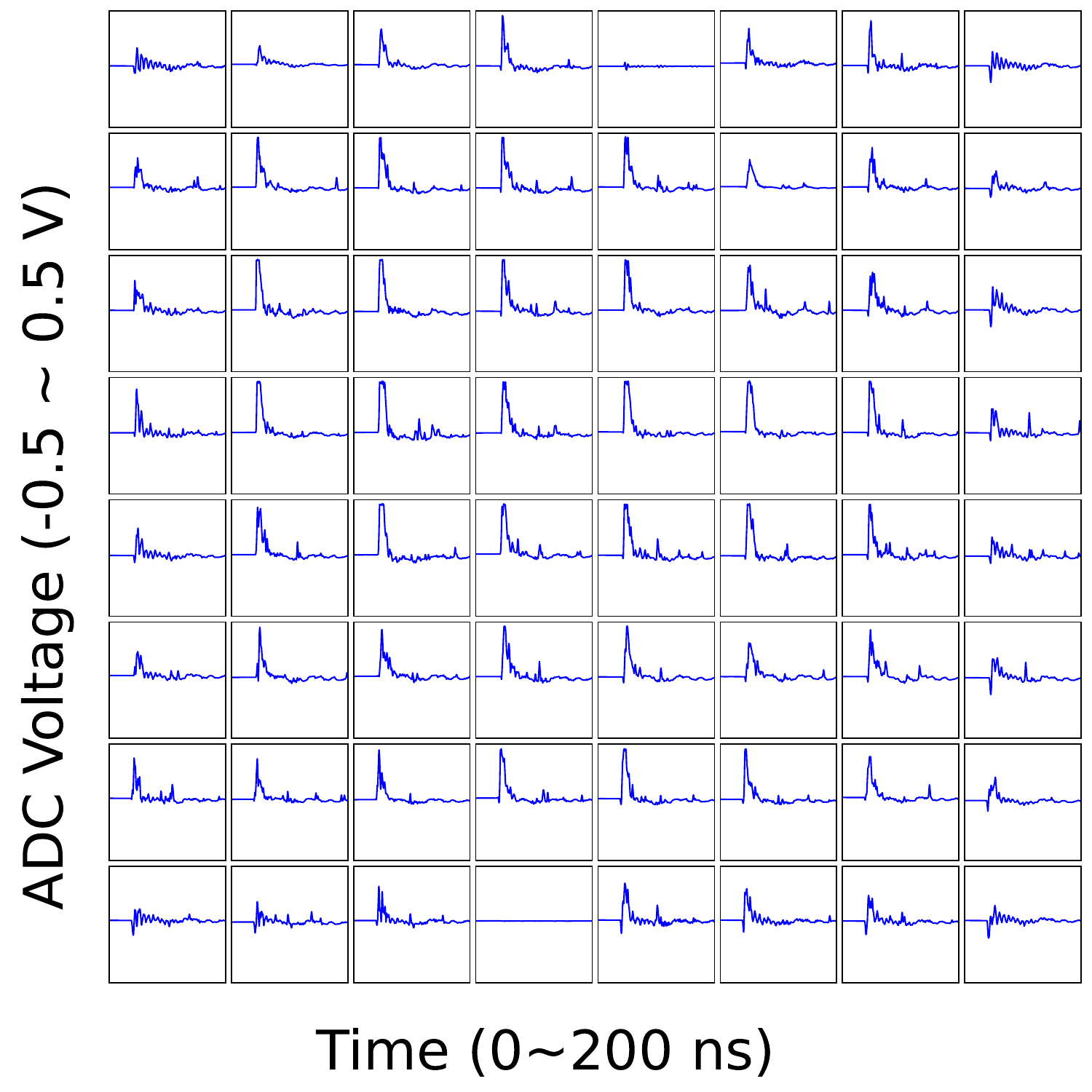}
\captionof{figure}
{
A cosmic muon-induced scintillation light detection setup (left). 
The right picture shows signals from 48 channels.
}
\label{fig:LAPPD_bulk}
\end{center}
We also prepared a motor-controlled $x-y$ stage with a mounted LED
for gain correction. The LAPPD is positioned face-up, while 
the $x-y$ stage moves the downward facing LED to illuminate 
specific pixels. The setup (left) and the calibrated
(set to 0) LED signals  are shown in Fig.~\ref{fig:LAPPD_LED}.
\begin{center}
\includegraphics[width=0.48\linewidth]{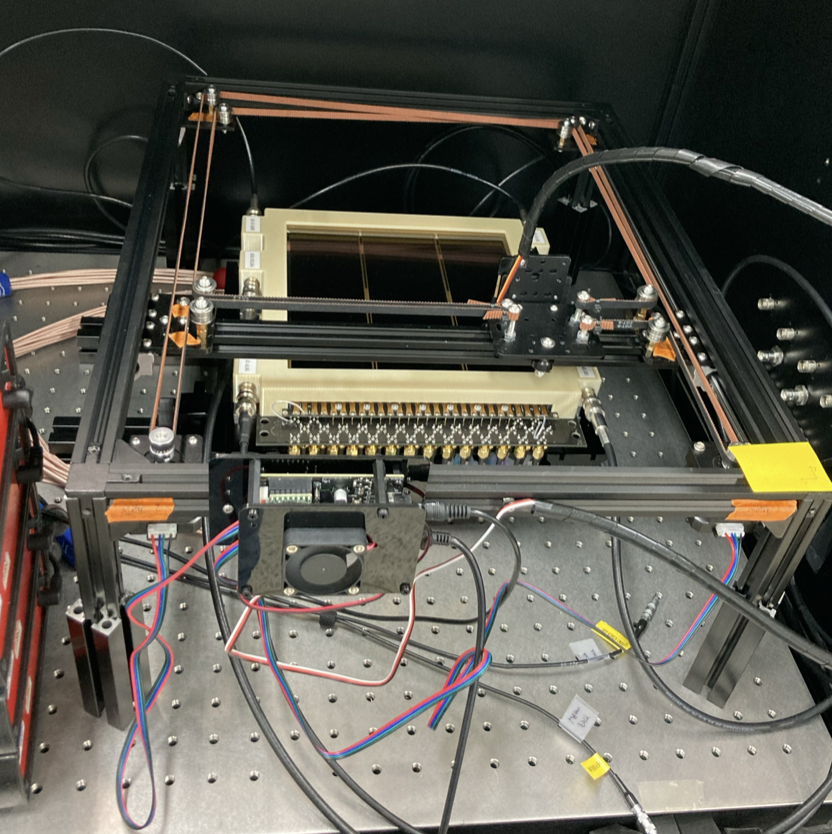}
\includegraphics[width=0.48\linewidth]{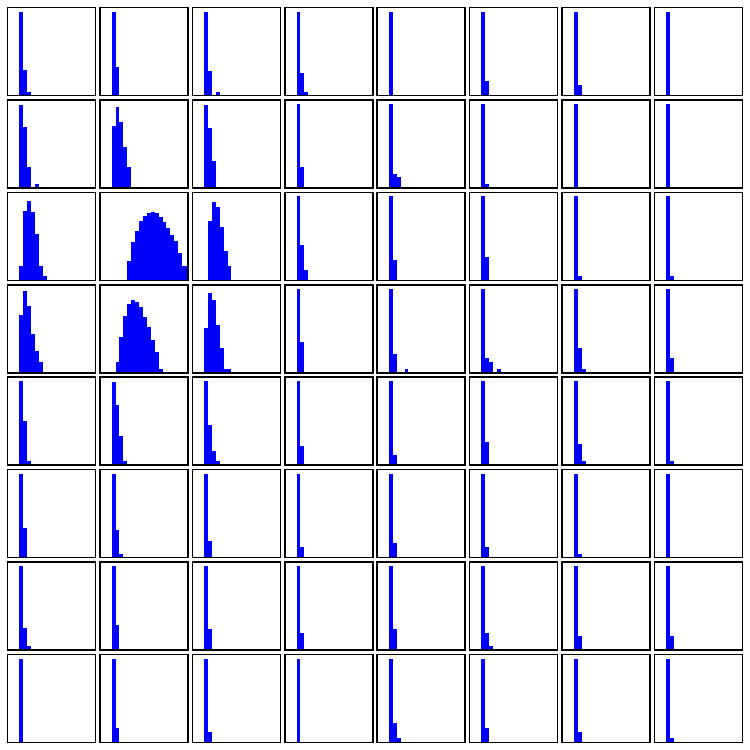}
\captionof{figure}
{
A picture of automatic $x-y$ stage (left) and the gain calibration result
with LED (right). The location of the LED is on the channel with the
largest signal (C2).
}
\label{fig:LAPPD_LED}
\end{center}
After four-month rental period, we tentatively 
concluded that LAPPD is an attractive option
for the $\nEYE$ experiment, assuming
its internal background is at least as
low as the case of the PMT. 
However, 
for LAPPD to be our primary choice
of the photodetector, a significant 
reduction in cost is necessary at this
stage. 
Additionally, a suitable ``base''
or voltage divider must be provided to supply different voltages to 
the photocathode, two MCPs (top and bottom), and 
the electric ground.
\subsubsection{Environmental Backgrounds}

 In the Yemilab $\nEYE$ experimental hall, samples
of shotcrete and rock are collected for analysis`, and the activities of
$^{238}$U,
$^{40}$K,
and
$^{232}$Th,
are measured. The results of these measurements 
are summarized in Table~\ref{table:environment},
where all units are in Bq/kg.
\begin{center}
\begin{tabular}{lccc}
\hline
  & $^{238}$U & $^{40}$K & $^{232}$Th {\rule{0pt}{2.9ex}}\\
\hline
Shotcrete & $16.7 \pm 0.6 $ & $447\pm 16$ & $25.3 \pm 0.6$ \\
Rock \# 1& $19 \pm 2  $ & $618 \pm 69$ & $22 \pm 2 $ \\
Rock \# 2& $18 \pm 2  $ & $872 \pm 98$ & $26 \pm 2 $ \\
Rock \# 3& $13 \pm 1  $ & $561 \pm 63$ & $15 \pm 1 $ \\
\hline
\end{tabular}
\captionof{table}{
Measured radio activities of $^{238}$U,
$^{40}$K,
and
$^{232}$Th in shotcrete and three rock samples. All numbers
are in units of Bq/kg.
}
\label{table:environment}
\end{center}
We also measured the background energy spectra at various locations in the
hall, including the floor, mid-tunnel, and top-tunnel areas. The results,
categorized into six
different energy bands, are presented in Fig.~\ref{fig:nEYE_HPGe}.
While different colors represent measurements taken at different locations, 
no significant variations in the spectra were observed across 
these areas.
\begin{center}
\includegraphics[width=\linewidth]{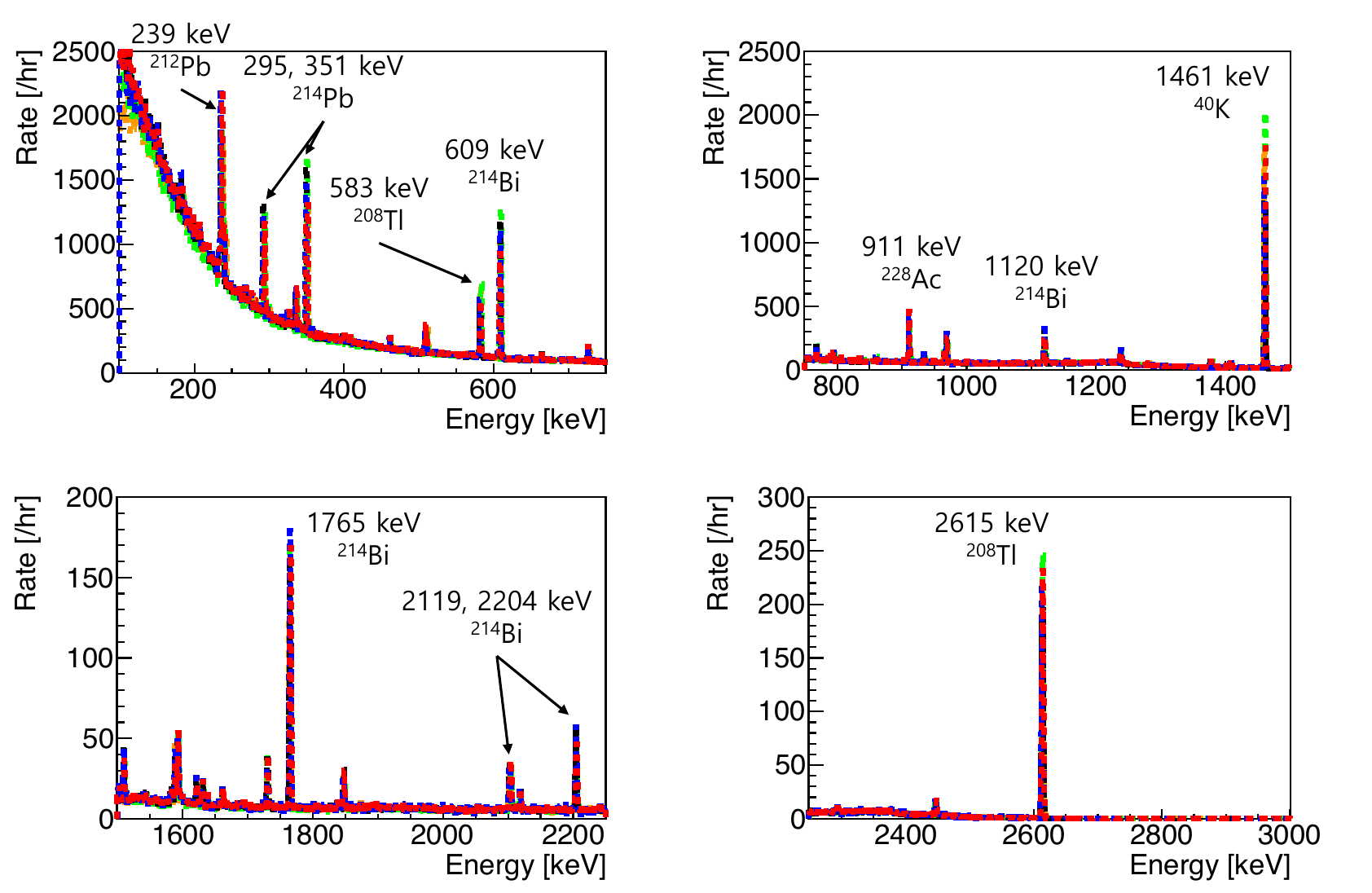}
\captionof{figure}
{
The HPGe measurements are shown. Histograms 
correspond to, from the top left to bottom right,
[100, 750] keV, 
[750, 1500] keV, 
[1.5, 2.3] MeV, [1.5, 2] MeV, and
[2.3, 3.0] MeV ranges. 
}
\label{fig:nEYE_HPGe}
\end{center}
We also measured the Radon activity in the hall and 
results are summarized in Table~\ref{table:radon}.
\begin{center}
\begin{tabular}{lc}
\hline
Location & Radon level (Bq/m$^3$) {\rule{0pt}{2.9ex}} \\
\hline
Up & $68\pm 14$ \\
Middle & $84 \pm 19 $ \\
Middle (opposite) & $118\pm 12 $ \\
Middle (low) & $125\pm 21 $ \\
Hall center & $148 \pm 17$ \\ 
\hline
\end{tabular}
\captionof{table}{
Measured Radon activity in several locations
in the $\nEYE$ hall.
}
\label{table:radon}
\end{center}
This measured environmental activity is being implemented
in our full Geant4 detector setup.
Note that we plan to install dedicated Radon reduction
systems in the experimental site to suppress the Radon level
sufficiently. 
\hfill \break

\subsection{\fontsize{12}{15}\selectfont R\&D on next generation 
neutrino detector}

The LS is our primary choice for the detector since it is
a well developed and understood technology, over the last
60 years. 
However, despite its advantage, the LS technology does not
allow the separation of Cherenkov from scintillation lights
easily and the particle identification is also difficult.
Due to 
Cherenkov radiation occurs at at faster rate than scintillation;
it is emitted instantaneously, whereas scintillation light results
from several molecular de-excitation processes that exhibit a
slower decay time constant with characteristic features,
new kinds of technologies have
been discussed in the neutrino community. Here we discuss
our next generation neutrino detector R\&D and approach.

\subsubsection{Water based LS}
Water-based Liquid Scintillator (WbLS) is a hybrid detection medium formulated by dispersing LS into pure water using surfactants. 
It has been proposed to combine the advantages of both water Cherenkov and LS detectors.
This novel medium is currently being implemented and demonstrated in several experimental 
efforts:
\href{https://doi.org/10.1103/PhysRevC.95.055801}
{Phys. Rev. C {\bf 95}, 055801 (2017)},
\href{https://doi.org/10.1103/PhysRevC.95.055801}
{Phys. Rev. C {\bf 95}, 055801 (2017)},
\href{https://doi.org/10.1140/epjc/s10052-020-7977-8}
{Eur. Phys. J. C {\bf 80}, 416 (2020)}, and
\href{https://doi.org/10.1088/1748-0221/18/02/P02009}
{JINST {\bf 18}, P02009 (2023)}.

Structurally, the organic scintillator molecules are encapsulated within micelles, 
which are suspended in the aqueous medium. 
This technique aims to take the advantages of both LS and water: the high light yield and low energy threshold of LS, and the high transparency, low cost, and directional capabilities of water Cherenkov detectors. 
WbLS is predominantly water, which allows us to optimize the scintillation light yield by adjusting its concentration depending on the physics goals of the experiment. 

The primary motivation for developing WbLS is to achieve ``hybrid detection'' with the 
discrimination of Cherenkov and scintillation signals.
In a conventional LS detector, the isotropic and intense scintillation light typically 
overwhelms the Cherenkov light. 
However, in WbLS, the two lights can be separated due to
the reduced scintillation light yield and the distinct features of the two lights. 
Cherenkov radiation occurs at a faster rate than scintillation;
it is emitted instantaneously, whereas scintillation light results from
several molecular de-excitation processes that exhibit a slower decay time
constant with characteristic features.
Utilizing these distinct time profiles, the prompt Cherenkov photons can be distinguished from the delayed scintillation photons.
In addition, the characteristic shape of Cherenkov cone and an isotropic shape of scintillation, and photon wavelength profile also can be used to separate the Cherenkov and scintillation signals.
This discrimination capability enhances the signal-to-background ratio. 
It allows us to reconstruct the direction of incident particle with better energy resolution than pure water Cherenkov detectors due to the additional scintillation photons.

While traditional organic LS requires complex chemical formulations to dissolve metallic isotopes, the water phase of WbLS allows for the direct dissolution of hydrophilic metallic salts. 
This feature is particularly important for loading neutron-capturing agents such as Gadolinium (Gd), which significantly enhances the neutron capture cross-section.
The neutron-capture signal provides a robust delayed coincidence tag for identifying Inverse Beta Decay (IBD).
From an engineering and safety perspective, WbLS offers distinct advantages over pure LS. 
Since WbLS typically comprises of 90\% water, it is non-flammable 
and environmentally safer than oil-based scintillators. 
Additionally, the cost of water is negligible compared to organic solvents, 
rendering WbLS a cost-effective solution for massive detectors.
Furthermore, WbLS offers additional advantages in terms of maintaining the medium by circulation and purification. 
The aqueous and organic components of WbLS can be separated 
using nanofiltration system, enabling the removal of the radioactive impurities and purify each phase independently before re-mixing, ensuring long-term stability.

\subsubsection{Opaque LS}

A new detection technique for neutrinos with
an opaque scintillator is proposed recently
(\href{https://www.nature.com/articles/s42005-021-00763-5}
{Nature, Commun. Phys. 4, 273 2021}). This technique,
known as LiquidO in the neutrino community,
can provide efficient identification of particles
in event-by-event basis through a high-resolution
imaging based on optical fibers. A 10 litter scale
prototype was fabricated and characterized in detail
(\href{
https://doi.org/10.48550/arXiv.2503.02541}
{\tt arxiv:2503.02541}).

 In order to explore this new technology and 
 study a possibility to use it to a 2 killo tonne scale
 neutrino detector, we started R\&D on the LiquidO 
 technology. 
The R\&D efforts encompass the fabrication of a small-scale 
prototype as well as considerations for scaling up to larger
detector systems.

\subsubsection{Segmented Detector Option}

We also look at the original idea from Raghavan's 
$^{115}$In loading 
with a segmented detector~\href{https://arxiv.org/abs/0705.2769}
{\tt arXiv:0705.2769}.  The main idea is to look at the
triple-coincidence signals from $\nu_e + ^{115}$In 
$\rightarrow ^{115}
\textrm{Sn}^* + e^-$ and the delayed (4.76 $\mu$s) reaction of
$^{115}\textrm{Sn}^* \rightarrow ^{115}\textrm{Sn} + 2\gamma$.
In order to realize the delayed triple-coincidence of two $\gamma$s and $e^-$,
the detector must be segmented or to be ``liquidO'' like, in order to
reconstruct the event topology. The segmentation with total
internal reflection has been explored~\href{https://arxiv.org/abs/2507.07397}
{\tt arXiv:2507.07397} and
\href{https://arxiv.org/abs/1501.06935}{\tt arXiv:1501.06935}.
The main issue in this case is the huge $\beta$ decay of $^{115}$In 
background, and the finer segmentation has stronger background rejection
in general.
We start to look at this approach to see if this approach with 
the $^{115}$In loading is feasible or not, for the $pp$ solar neutrino
detection in particular.


Solar pp neutrinos provide a direct probe of the primary fusion process in the
Sun. Since their energies lie in the sub-MeV region, their detection requires a
target with a sufficiently low energy threshold and a strong rejection
capability against radioactive backgrounds. An indium-based detector has long
been considered an attractive option for this purpose, because
electron-neutrino capture on $^{115}\mathrm{In}$ can proceed with a low
threshold of about 114 keV
(\href{https://journals.aps.org/prl/abstract/10.1103/PhysRevLett.37.259}
{PRL 37, 259 (1976)}):
\begin{equation}
\nu_e + {}^{115}\mathrm{In}
\rightarrow
{}^{115}\mathrm{Sn}^{**} + e^- .
\label{eq:in115_capture}
\end{equation}
This low threshold makes the reaction sensitive to solar pp neutrinos.
The excited $^{115}\mathrm{Sn}^{**}$ state de-excites through a cascade:
\begin{align}
{}^{115}\mathrm{Sn}^{**}
&\rightarrow
{}^{115}\mathrm{Sn}^{*} + \gamma~(115~\mathrm{keV}),
\label{eq:sn115_gamma_115}
\\
{}^{115}\mathrm{Sn}^{*}
&\rightarrow
{}^{115}\mathrm{Sn} + \gamma~(497~\mathrm{keV}) .
\label{eq:sn115_gamma_497}
\end{align}
The lifetime of the excited $^{115}\mathrm{Sn}^{**}$ state provides the delayed
timing tag
(\href{https://journals.aps.org/prl/abstract/10.1103/PhysRevLett.37.259}
{PRL 37, 259 (1976)}). 
The first transition has a half-life of about
$3.3~\mu\mathrm{s}$, corresponding to a mean lifetime of about
$4.8~\mu\mathrm{s}$. Therefore, the delayed emission time is not fixed
event by event, but follows an exponential decay-time distribution. The second
transition occurs on a much shorter time scale of about $100~\mathrm{ps}$.
Therefore,
on the detector time scale considered in this study, the signal can be regarded
as a prompt electron followed by delayed de-excitation products. The delayed
component is identified through the cascade products rather than by the
115 keV gamma ray alone.

The nuclear-level scheme relevant to the $^{115}\mathrm{In}$ capture signal is
shown in Fig.~\ref{fig:in115_nuclearlevel}. Electron-neutrino capture on
$^{115}\mathrm{In}$ populates an excited state of $^{115}\mathrm{Sn}$, whose
delayed de-excitation products provide a coincidence signature. Thus, the
indium capture signal is not only an energy signature but also a space-time
coincidence signature.

\begin{figure}[tbp]
  \centering
  \includegraphics[width=0.75\textwidth]{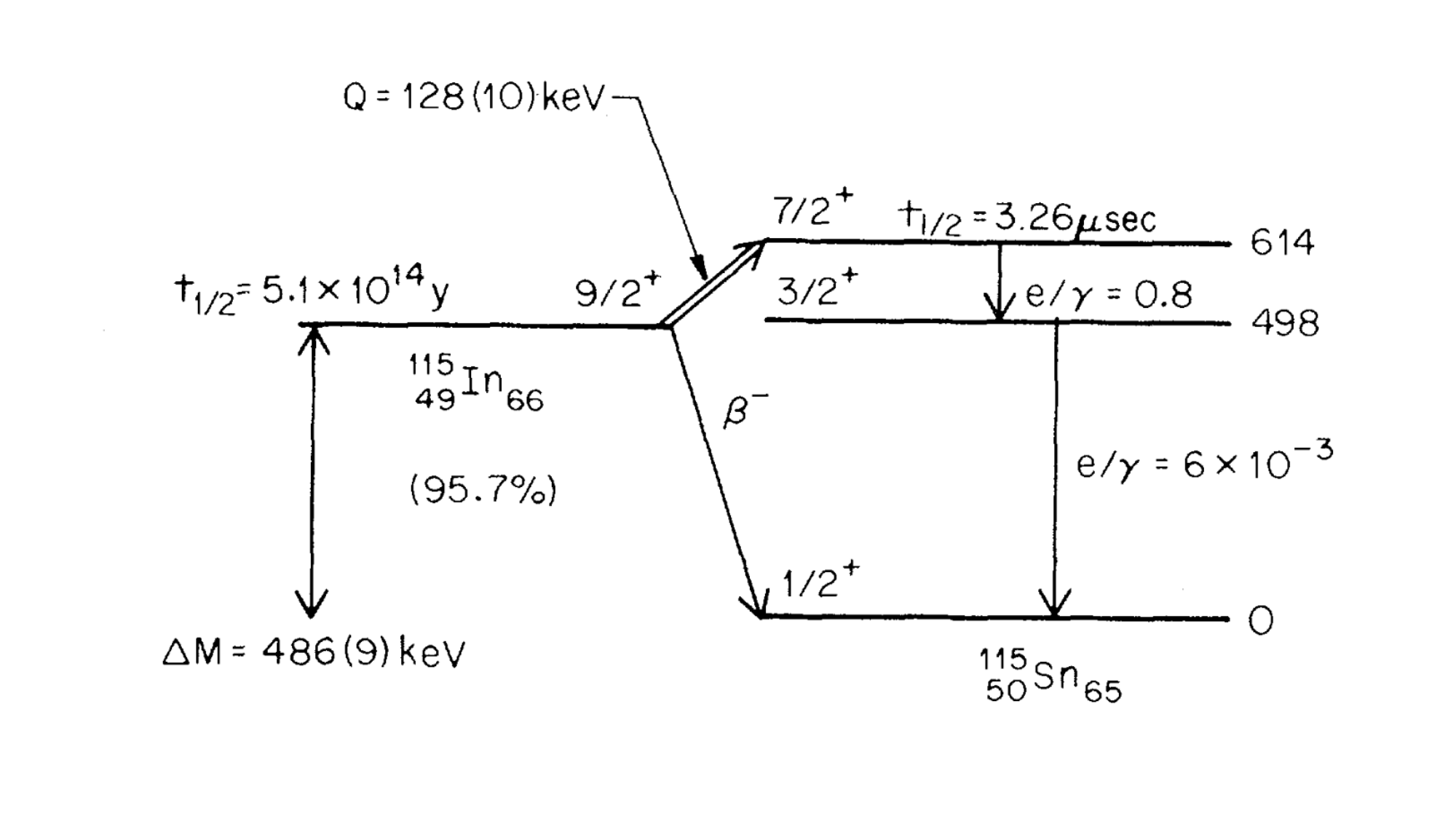}
  \caption{
  Nuclear-level scheme of the $^{115}\mathrm{In}$ neutrino-capture signal and
  the intrinsic $^{115}\mathrm{In}$ beta-decay background. Electron-neutrino
  capture on $^{115}\mathrm{In}$ populates an excited state of
  $^{115}\mathrm{Sn}$, whose delayed de-excitation products provide a
  coincidence signature for background rejection. The scheme is adapted from
  Fig.~4 of Ref.
(\href{https://journals.aps.org/prl/abstract/10.1103/PhysRevLett.37.259}
{PRL 37, 259 (1976)}).
  }
  \label{fig:in115_nuclearlevel}
\end{figure}

A practical realization of this concept is challenging. The expected solar
pp-neutrino capture rate is very small, whereas natural $^{115}\mathrm{In}$ is
intrinsically radioactive. The beta decay of $^{115}\mathrm{In}$,
\begin{equation}
{}^{115}\mathrm{In}
\rightarrow
{}^{115}\mathrm{Sn} + e^- + \bar{\nu}_e ,
\label{eq:in115_beta_decay}
\end{equation}
forms an unavoidable internal background. Although the half-life of
$^{115}\mathrm{In}$ is extremely long,
\begin{equation}
T_{1/2}=4.41\times10^{14}~\mathrm{years},
\end{equation}
its activity becomes significant for detector-scale indium masses. The
half-life and natural abundance of $^{115}\mathrm{In}$ are taken from nuclear
data
(\href{https://atom.kaeri.re.kr/cgi-bin/nuclide?nuc=In115}
{KAERI Nuclear Data Center}). Therefore, the feasibility of an indium-based
solar-neutrino detector depends critically on whether the delayed coincidence
signature can be identified with sufficient efficiency and background rejection.

Segmented scintillator detectors provide a possible way to preserve partial
information about the energy-deposition topology. In NuLat-like detectors,
scintillator cells are optically separated so that scintillation photons can be
confined within a cell or along a limited set of neighboring cells. Such an
optical-lattice structure can guide scintillation light through total internal
reflection and provide spatial information about the energy-deposition
position
(\href{https://arxiv.org/abs/1501.06935}
{arXiv:1501.06935}). This motivates a Geant4 study of optical photon
propagation in a segmented liquid-scintillator detector for
indium-capture-like events.

The goal of this work is to study the optical response of a NuLat-like
segmented liquid-scintillator detector using
Geant4
(\href{https://doi.org/10.1016/S0168-9002(03)01368-8}
{NIM A 506, 250 (2003)};
\href{https://doi.org/10.1016/j.nima.2016.06.125}
{NIM A 835, 186 (2016)}). In particular, we focus
on the propagation of scintillation photons, optical confinement by
fluorinated ethylene propylene (FEP) film and air gaps, and the behavior of
optical photons at liquid-scintillator--FEP--air boundaries. Since the
neutrino-capture interaction on $^{115}\mathrm{In}$ is not directly simulated
by Geant4 in this study, capture-like final states are generated at the
primary-generator level. The subsequent particle transport, scintillation
production, and optical-photon propagation are handled by Geant4.
The present work should therefore be interpreted as an optical-response and
boundary-validation study for future indium-capture event studies. The main
questions addressed here are:
\begin{enumerate}
  \item whether the segmented LS/FEP/air geometry provides optical confinement;
  \item whether Geant4 reproduces the expected Fresnel reflection, refraction,
        and total internal reflection at the optical boundaries;
  \item how optical photons generated by signal-like events propagate to the PMT
        arrays;
  \item what signal-rate and intrinsic-background conditions motivate future
        coincidence, reconstruction, and background-rejection studies.
\end{enumerate}



A NuLat-like segmented liquid-scintillator detector was modeled in this study.
The detector geometry follows the basic idea of an optical-lattice detector, in
which scintillator cells are optically separated to preserve local information
about scintillation-light production
(\href{https://arxiv.org/abs/1501.06935}
{arXiv:1501.06935}). In the present work,
the detector is used to study optical photon propagation in a segmented
scintillator volume. The segmentation is intended to provide optical confinement
through refractive-index contrast between the scintillator, FEP film, and air
gaps.

The detector envelope is modeled as an approximately cubic volume with a
nominal side length of 1 m. It is divided into cubic scintillator cells with a
side length of 6 cm. Each detector side contains 16 cells, giving
\begin{equation}
16 \times 16 \times 16 = 4096
\end{equation}
scintillator cells in total. Neighboring cells are separated by FEP film and an
air gap. The FEP film thickness is $50~\mu\mathrm{m}$, and the air gap thickness
is $300~\mu\mathrm{m}$. The baseline detector geometry is summarized in
Table~\ref{tab:detector_geometry}.

\begin{table}[tbp]
  \centering
  \caption{
  Baseline geometry parameters of the segmented liquid-scintillator detector.
  }
  \label{tab:detector_geometry}
  \begin{tabular}{lc}
    \hline
    Parameter & Value \\
    \hline
    Detector envelope &
    nominally $1~\mathrm{m} \times 1~\mathrm{m} \times 1~\mathrm{m}$ \\
    Cell size & $6~\mathrm{cm} \times 6~\mathrm{cm} \times 6~\mathrm{cm}$ \\
    Number of cells per side & 16 \\
    Total number of cells & 4096 \\
    FEP film thickness & $50~\mu\mathrm{m}$ \\
    Air gap thickness & $300~\mu\mathrm{m}$ \\
    \hline
  \end{tabular}
\end{table}

Each cell is implemented as a nested optical structure consisting of a
scintillator volume surrounded by FEP film and an air layer. This structure
defines the LS--FEP--air optical boundaries between neighboring cells. The
purpose of the geometry is not to model mechanical support structures in detail,
but to reproduce the basic refractive-index boundary configuration relevant for
optical confinement and total internal reflection.

The detector geometry is shown in Fig.~\ref{fig:detector_geometry}. PMT arrays
are placed on the outer surfaces of the detector and are used to record the
optical response of the segmented scintillator volume.

\begin{figure}[tbp]
  \centering
  \includegraphics[width=0.60\textwidth]{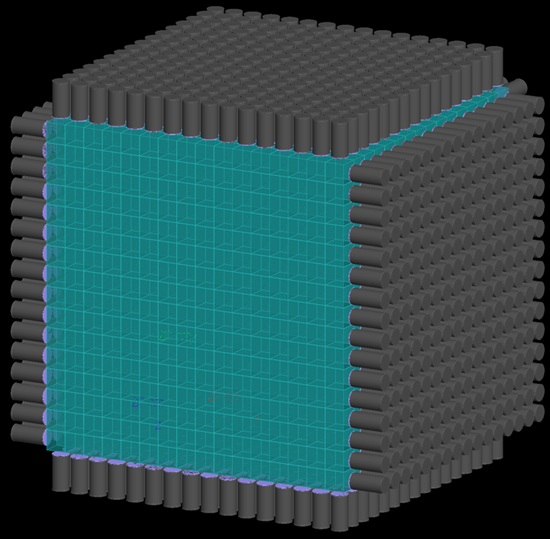}
  \caption{
  Geant4 geometry of the segmented liquid-scintillator detector. The detector
  consists of 4096 scintillator cells with a side length of 6 cm. PMT arrays are
  placed on the six detector faces. The front PMTs are omitted in this display
  to show the internal segmented structure.
  }
  \label{fig:detector_geometry}
\end{figure}

The active scintillator is modeled as a LAB-based liquid scintillator containing
PPO and Bis-MSB as wavelength shifters. In the present detector geometry,
indium is not loaded into the scintillator material. The detector is therefore
used as an optical-response model for indium-capture-like events rather than as
a full material model of an indium-loaded scintillator. The optical effects of
realistic indium loading, such as changes in attenuation length, light yield,
emission spectrum, scintillation time profile, and possible absorption features,
are not included in the present simulation.

The optical properties of the detector materials are assigned through their
refractive indices and scintillation properties in the simulation. The
refractive indices used in the optical model are summarized in
Table~\ref{tab:refractive_indices}. These values define the optical boundary
behavior of the segmented detector. The resulting Fresnel reflection,
refraction, and total internal reflection are discussed 
later.

\begin{table}[tbp]
  \centering
  \caption{
  Refractive indices used in the optical simulation.
  }
  \label{tab:refractive_indices}
  \begin{tabular}{lc}
    \hline
    Material & Refractive index \\
    \hline
    LAB-based liquid scintillator & 1.49 \\
    FEP film & 1.34 \\
    Air & 1.00 \\
    PMT window glass & 1.47 \\
    \hline
  \end{tabular}
\end{table}

Photon detection is modeled using PMT arrays placed on the six detector faces.
Each PMT has a diameter of 2 inches, comparable to the 6 cm cell size. A
$16 \times 16$ PMT array is placed on each detector face, giving
\begin{equation}
16 \times 16 \times 6 = 1536
\end{equation}
PMTs in total. The PMT hit pattern is used as the optical readout observable in
this study.

In this simulation, a PMT hit represents an optical photon reaching the PMT
sensitive surface. It should not be interpreted as a fully modeled
photoelectron signal. A detailed photoelectron conversion model, including
wavelength-dependent quantum efficiency, collection efficiency, transit-time
spread, dark noise, afterpulsing, electronics thresholds, and trigger response,
is not included.

The present detector model is intentionally simplified. Mechanical support
structures, detailed optical coupling layers, electronics response, PMT dark
noise, afterpulsing, and realistic indium-loading effects are not modeled. These
effects must be included in future studies before making a quantitative detector
performance or background-rejection estimate.



The neutrino-capture interaction on $^{115}\mathrm{In}$ is not generated as a
standard Geant4 physics process in this study. Instead, the final-state
particles of the capture reaction are generated at the primary-generator level.
This approach is used because the present study focuses on detector response and
optical photon propagation rather than the absolute neutrino interaction rate.

For event samples involving radioactive nuclei, the implementation was
developed with reference to the Geant4 radioactive-decay example,
\texttt{rdecay01}
(\href{https://doi.org/10.1016/S0168-9002(03)01368-8}
{NIM A 506, 250 (2003)};
\href{https://doi.org/10.1016/j.nima.2016.06.125}
{NIM A 835, 186 (2016)}). In
particular, this structure is used for the \texttt{nuCapture},
\texttt{in115\_beta}, and \texttt{radioactive\_decay} event modes, where
radioactive ions are generated and their subsequent decay or de-excitation
products are handled by the Geant4 radioactive-decay process.

The signal-like event sample, denoted as \texttt{nuCapture}, is generated by
placing a prompt electron and an excited $^{115}\mathrm{Sn}$ nucleus at the same
vertex. The vertex position is randomly sampled inside the active detector
volume. The excited nucleus is then allowed to de-excite through the cascade
described later 
and the subsequent de-excitation
products and particle transport are handled by Geant4 radioactive-decay and
electromagnetic physics processes.

The incident solar neutrino energy is sampled from the solar pp-neutrino
spectrum
(\href{https://doi.org/10.3847/1538-4357/835/2/202}
{ApJ 835, 202 (2017)}). Events with neutrino energy below the
capture threshold are rejected. For accepted events, the kinetic energy of the
prompt electron is calculated as
\begin{equation}
T_e = E_\nu - Q_{\mathrm{thr}},
\label{eq:prompt_electron_energy}
\end{equation}
where $Q_{\mathrm{thr}} = 114~\mathrm{keV}$
(\href{https://journals.aps.org/prl/abstract/10.1103/PhysRevLett.37.259}
{PRL 37, 259 (1976)}).
The electron direction is generated isotropically in the present implementation.
The weak-interaction angular dependence of the capture reaction is not included.
This approximation is sufficient for the present optical-response study, but a
more physical event generator would be required for a detailed signal
kinematics study.

The \texttt{in115\_beta} mode generates $^{115}\mathrm{In}$ ions at rest in the
detector volume. Their beta decay is then handled by the Geant4
radioactive-decay process. This sample is used to study the optical response to
the intrinsic beta-decay background of $^{115}\mathrm{In}$. The
\texttt{radioactive\_decay} mode provides an auxiliary radioactive-decay sample,
where the ion species and excitation energy can be specified by the user.

Several additional event modes are implemented for comparison and optical
validation. The \texttt{external\_gamma} mode generates gamma rays from outside
the detector volume to study the optical response to gamma-induced
electromagnetic interactions. The \texttt{optical\_photon} mode generates
optical photons for dedicated optical-boundary validation. In this mode, the
photon position and direction can be either randomly generated or specified by
user macro commands, allowing both general propagation tests and fixed-incident-
angle boundary tests.
The event samples and their purposes 
are listed in Table~\ref{tab:event_samples}.
\begin{table}[tbp]
  \centering
  \caption{
  Event samples implemented in the primary generator.
  }
  \label{tab:event_samples}
  \begin{tabular}{lll}
    \hline
    Event type & Primary generation & Purpose \\
    \hline
    \texttt{nuCapture} &
    $e^- + {}^{115}\mathrm{Sn}^{**}$ &
    Signal-like optical response \\
    \texttt{in115\_beta} &
    $^{115}\mathrm{In}$ at rest &
    Intrinsic beta-decay background \\
    \texttt{radioactive\_decay} &
    user-defined ion &
    Auxiliary radioactive-decay sample \\
    \texttt{external\_gamma} &
    external $\gamma$ ray &
    External gamma response \\
    \texttt{optical\_photon} &
    optical photon &
    Optical-boundary validation \\
    \hline
  \end{tabular}
\end{table}
The simulation records the generated vertex, the sampled neutrino energy, the
transport of optical photons, the optical-boundary status at material
interfaces, and the PMT hit positions and times. Gamma and electron information
is also stored when needed to identify the source of scintillation photons.
Detailed topology reconstruction or topology-based background rejection is not
performed in this study.



The segmented detector relies on optical confinement produced by the
refractive-index contrast between the LAB-based liquid scintillator, FEP film,
and air gaps. Therefore, before studying event-level optical response, it is
necessary to validate that the Geant4 optical-boundary process correctly
reproduces reflection and refraction at the LS--FEP--air interfaces.

The basic optical-boundary configuration is shown in
Fig.~\ref{fig:ls_fep_air_boundary}. A scintillator cell is surrounded by FEP
film and an air gap. Optical photons generated in the liquid scintillator first
encounter the LS--FEP boundary. Depending on the incident angle, photons can be
transmitted into the FEP film, reflected at the boundary, or totally internally
reflected. Photons transmitted into the FEP film may subsequently encounter the
FEP--air boundary, where additional refraction or total internal reflection can
occur.
\begin{figure}[tbp]
  \centering
  \includegraphics[width=0.40\textwidth]{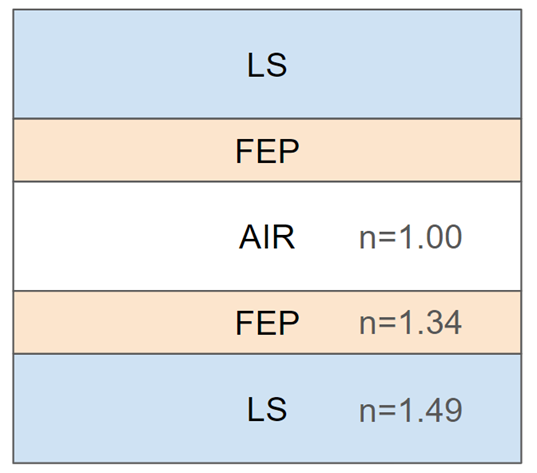}
  \caption{
  Schematic view of the LS--FEP--air optical-boundary structure used in the
  segmented detector. The refractive indices of the LAB-based liquid
  scintillator, FEP film, and air gap are approximately 1.49, 1.34, and 1.00,
  respectively. Optical confinement arises from Fresnel reflection and total
  internal reflection at these dielectric boundaries.
  }
  \label{fig:ls_fep_air_boundary}
\end{figure}
In the baseline optical model, the material boundaries are defined by the
refractive indices listed in Table~\ref{tab:refractive_indices}. No additional
surface reflectivity is imposed for the boundary-validation sample. Thus,
reflection and refraction at dielectric--dielectric interfaces are expected to
be governed by the Fresnel equations and by total internal reflection when the
incident angle exceeds the critical angle.

For a photon incident from a material with refractive index $n_1$ to a material
with refractive index $n_2$, the critical angle is given by
\begin{equation}
\theta_c
=
\sin^{-1}
\left(
\frac{n_2}{n_1}
\right),
\qquad
(n_1 > n_2).
\label{eq:critical_angle}
\end{equation}
Using the refractive indices in the simulation, the critical angle for the
LS--FEP boundary is
\begin{equation}
\theta_c^{\mathrm{LS\rightarrow FEP}}
=
\sin^{-1}
\left(
\frac{1.34}{1.49}
\right)
\simeq
64^\circ .
\end{equation}
Similarly, the critical angle for the FEP--air boundary is approximately
$48^\circ$, and that for a direct LS--air boundary is approximately $42^\circ$.
These values define the expected angular behavior of photon confinement in the
segmented detector.

Single optical photons were generated to visualize photon propagation through
the LS/FEP/air optical lattice. The photon energy was set to 3 eV,
corresponding to a wavelength of approximately 413 nm. This energy is
representative of optical photons in the scintillation wavelength region. The
photon position, direction, and polarization can be controlled by macro
commands, allowing both general propagation tests and fixed-incident-angle
boundary tests.
Figure~\ref{fig:single_photon_tracks} shows representative single-photon tracks
in the segmented geometry. These tracks illustrate the basic optical behavior
expected from the refractive-index structure: photons may be guided along the
cell direction by repeated reflections, transmitted through the FEP layer, or
leak into neighboring cells depending on their incident angles at the
LS--FEP--air boundaries. The purpose of this test is to verify the qualitative
photon-transport behavior before performing a quantitative boundary comparison.

\begin{figure}[tbp]
  \centering
  \includegraphics[width=0.80\textwidth]{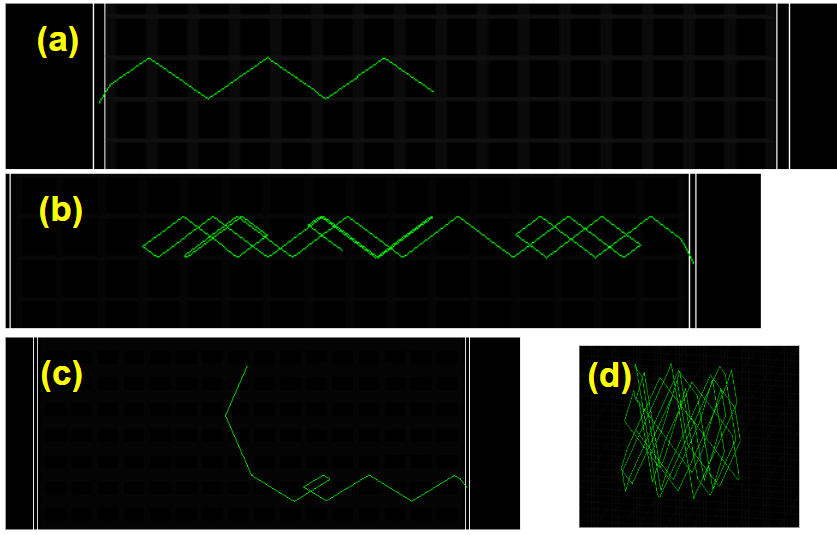}
  \caption{
    Representative single optical-photon tracks in the segmented LS/FEP/air
    geometry. (a) A photon propagates from the detector center toward a PMT on the
    left detector face while remaining confined along a single cell column.
    (b) A photon initially propagating to the left is reflected at an optical
    boundary and redirected to the right. (c) A photon initially propagating
    downward is reflected at an optical boundary and redirected to the right.
    (d) A photon is trapped within a cell by repeated reflections and is expected
    to be absorbed eventually in the scintillator. These examples illustrate
    reflection, refraction, leakage, and trapping at the optical boundaries defined
    by the refractive-index contrast between the liquid scintillator, FEP film, and
    air gaps.
  }
  \label{fig:single_photon_tracks}
\end{figure}

The Fresnel reflection coefficients for $s$- and $p$-polarized light are
\begin{equation}
R_s
=
\left|
\frac{
n_1 \cos\theta_i - n_2 \cos\theta_t
}{
n_1 \cos\theta_i + n_2 \cos\theta_t
}
\right|^2 ,
\label{eq:fresnel_rs}
\end{equation}
and
\begin{equation}
R_p
=
\left|
\frac{
n_2 \cos\theta_i - n_1 \cos\theta_t
}{
n_2 \cos\theta_i + n_1 \cos\theta_t
}
\right|^2 ,
\label{eq:fresnel_rp}
\end{equation}
where $\theta_i$ is the incident angle and $\theta_t$ is the transmitted angle
obtained from Snell's law,
\begin{equation}
n_1\sin\theta_i
=
n_2\sin\theta_t .
\label{eq:snell_law}
\end{equation}
For unpolarized light, the expected reflected fraction is
\begin{equation}
R
=
\frac{1}{2}
\left(
R_s + R_p
\right).
\label{eq:fresnel_unpolarized}
\end{equation}
To validate the optical-boundary behavior quantitatively, single optical
photons were generated at fixed incident angles toward an LS--FEP boundary. The
reflected and refracted fractions were measured for photons incident near the
critical angle. The reflected fractions obtained from the Geant4 simulation
were 9.18\%, 29.5\%, and 71.9\% for incident angles of $60^\circ$,
$63^\circ$, and $64^\circ$, respectively. The corresponding refracted fractions
were 90.7\%, 70.3\%, and 27.8\%.
The small difference between the sum of the reflected and refracted fractions
and unity is due to photon absorption in the liquid scintillator before reaching
the boundary. In this validation sample, the absorption length of the liquid
scintillator was set to 10 m. Therefore, the missing fraction is not interpreted
as a boundary-process loss, but as ordinary absorption during photon transport
inside the scintillator.
The comparison between the Geant4 result and the Fresnel expectation is shown in
Fig.~\ref{fig:fresnel_validation}. As the incident angle approaches the critical
angle, the reflected fraction increases while the refracted fraction decreases.
This behavior confirms that the optical-boundary process correctly reproduces
the angular dependence expected from Fresnel reflection and total internal
reflection at the LS--FEP interface.
\begin{figure}[tbp]
  \centering
  \includegraphics[width=0.75\textwidth]{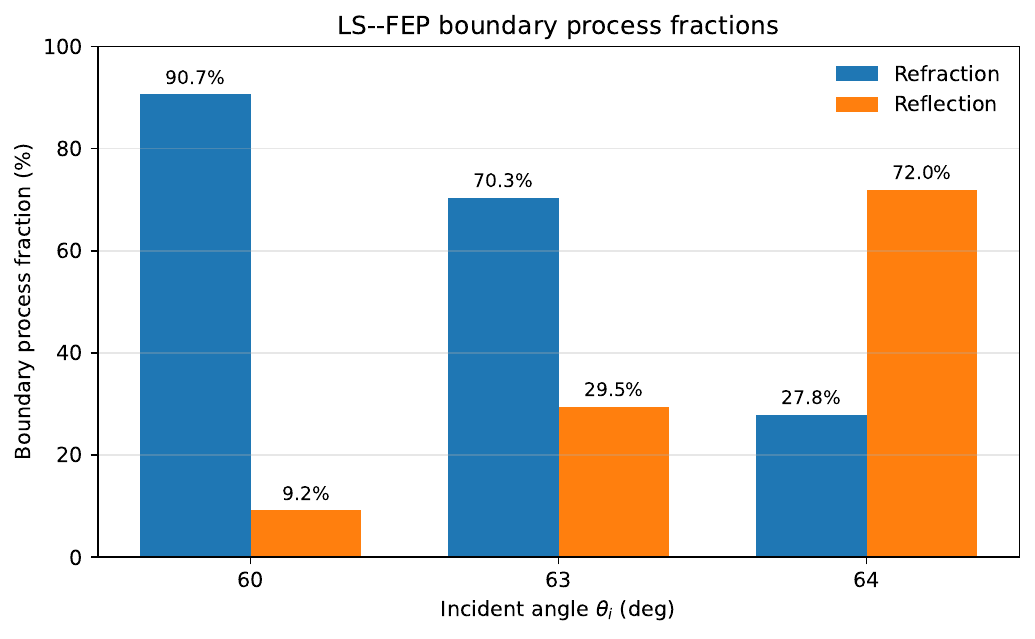}
  \caption{
  Reflected and refracted fractions of single optical photons incident on the
  LS--FEP boundary for fixed incident angles near the critical angle. As the
  incident angle increases from $60^\circ$ to $64^\circ$, the reflected fraction
  increases while the refracted fraction decreases, consistent with the expected
  Fresnel behavior. The small missing fraction is due to photon absorption in
  the liquid scintillator before reaching the boundary.
  }
  \label{fig:fresnel_validation}
\end{figure}
After validating the single-boundary behavior, optical photons produced by
signal-like scintillation events were propagated through the full segmented
detector geometry. In this geometry, photons undergo repeated interactions with
LS--FEP--air boundaries. Photons emitted within a cell can therefore be guided
preferentially along the segmented optical channels, while photons with angles
outside the trapped phase space can leak into neighboring cells or escape toward
the detector boundary.

Figure~\ref{fig:signal_like_optical_event} shows a representative signal-like
event in the segmented detector. The event display illustrates the generation
vertex and the optical photons produced by the prompt electron and delayed
de-excitation products. The labeled 115 keV and 497 keV components indicate the
delayed cascade used in the signal-like event sample. The scintillation photons
are partially confined and transported along the LS/FEP/air optical lattice.
This figure is used only to visualize photon propagation in the segmented
geometry. No event reconstruction or topology-based background rejection is
performed in this study.

\begin{figure}[tbp]
  \centering
  \includegraphics[width=0.62\textwidth]{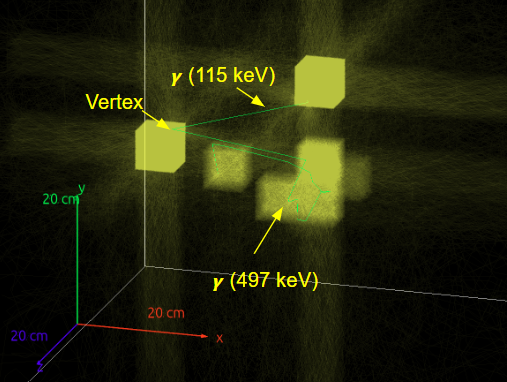}
  \caption{
  Representative signal-like event in the segmented detector. The event display
  shows the generation vertex, the prompt-electron component, and delayed
  de-excitation components labeled by their characteristic transition energies.
  The yellow tracks represent optical photons propagating through the segmented
  LS/FEP/air optical lattice. The figure is used as an optical-propagation
  visualization and not as a quantitative topology-reconstruction result.
  }
  \label{fig:signal_like_optical_event}
\end{figure}

The optical response can also be represented by the PMT hit pattern on the
detector faces. In this study, a PMT hit corresponds to an optical photon
reaching the PMT sensitive surface in the simulation
as is described later.
Therefore, the PMT hit pattern should be
interpreted as a geometrical optical-response observable, not as a fully modeled
photoelectron distribution.
Figure~\ref{fig:pmt_hit_pattern} shows the PMT hit pattern for a representative
signal-like event. The upper panels show early PMT hits with hit times below
200 ns, dominated by scintillation photons produced by the prompt electron. The
lower panels show delayed PMT hits with hit times above 200 ns, corresponding
to photons produced by the delayed de-excitation cascade. The hit counts are
shown on a logarithmic scale for each PMT plane. The distribution illustrates
how the prompt and delayed components produce different optical hit patterns on
the six detector faces. This observable is used here only to visualize the
optical response and is not used to perform event classification or background
rejection.

\begin{figure}[tbp]
  \centering
  \includegraphics[width=0.75\textwidth]{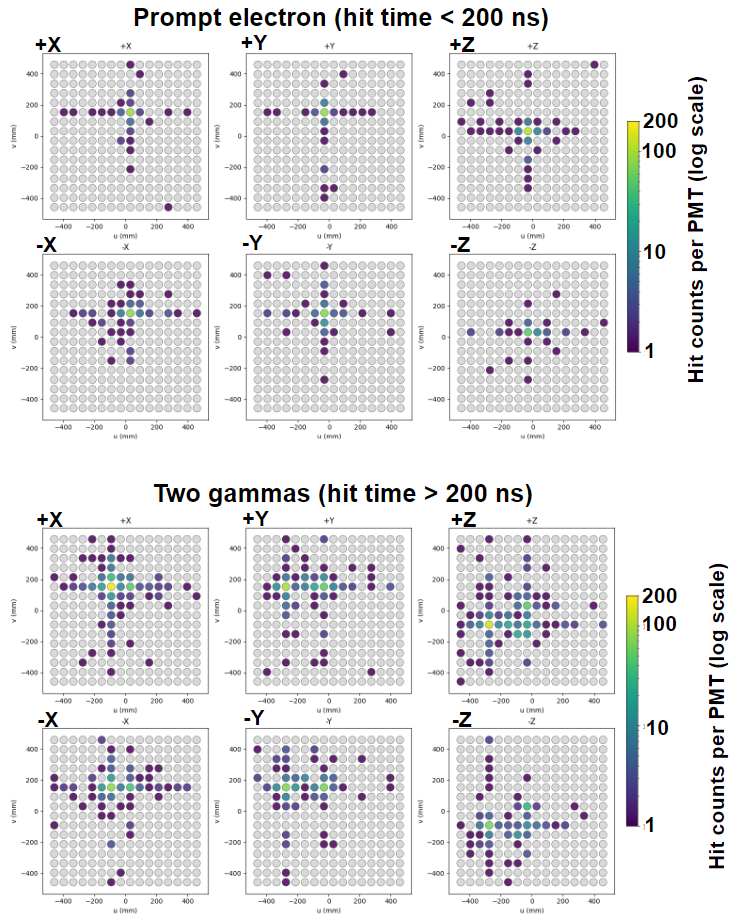}
  \caption{
  PMT hit patterns for a representative signal-like event. The upper panels
  show early hits with hit times below 200 ns, dominated by the prompt-electron
  component. The lower panels show delayed hits with hit times above 200 ns,
  corresponding to the delayed de-excitation cascade. Each panel shows one
  detector face, and the color scale indicates the number of optical photons
  reaching each PMT on a logarithmic scale. The figure is used as an
  optical-response visualization and not as a quantitative event-reconstruction
  result.
  }
  \label{fig:pmt_hit_pattern}
\end{figure}

These validation results show that the segmented LS/FEP/air geometry is modeled
as a refractive-index-based optical lattice in Geant4. The observed photon
channeling and confinement are consistent with Fresnel reflection and total
internal reflection at the material boundaries. The signal-like event display
and PMT hit pattern further illustrate how this optical lattice maps local
scintillation light production onto the detector faces. This validated optical
model is therefore used as the baseline configuration for the event-level
optical response studies in the present work.



Although the present study focuses on optical photon propagation and boundary
validation, the expected signal rate and intrinsic background rate are important
for understanding the motivation of the detector concept. An indium-based
solar-neutrino detector must identify a very small solar pp-neutrino capture
signal in the presence of frequent beta decays from the target isotope itself.

In this section, simple rate estimates are made for indium-loaded liquid
scintillator configurations. If the total liquid-scintillator mass is
$M_{\mathrm{LS}}$ and the indium loading fraction by mass is $w_{\mathrm{In}}$,
the corresponding loaded indium mass is
\begin{equation}
M_{\mathrm{In}}
=
w_{\mathrm{In}} M_{\mathrm{LS}} .
\label{eq:indium_loading_mass}
\end{equation}
The number of $^{115}\mathrm{In}$ nuclei is then
\begin{equation}
N_{^{115}\mathrm{In}}
=
\frac{
M_{\mathrm{In}} f_{115}
}{
M_{\mathrm{In,mol}}
}
N_A ,
\label{eq:number_in115_loaded}
\end{equation}
where $f_{115}=0.957$ is the natural abundance of $^{115}\mathrm{In}$,
$M_{\mathrm{In,mol}}\simeq115~\mathrm{g/mol}$ is the molar mass of indium, and
$N_A$ is Avogadro's number
(\href{https://atom.kaeri.re.kr/cgi-bin/nuclide?nuc=In115}
{KAERI Nuclear Data Center}). Unless otherwise stated, the
loaded indium masses quoted below assume natural isotopic abundance.
For the baseline configuration considered in this study, we assume a
1-ton liquid-scintillator detector with 10 wt.\% indium loading. This
corresponds to
\begin{equation}
M_{\mathrm{In}}
=
0.10 \times 1~\mathrm{ton}
=
100~\mathrm{kg}
\end{equation}
of loaded indium. This configuration is used only for rate estimates. As
discussed later,
the optical simulation geometry
itself does not include the optical-property changes expected from realistic
indium loading.

The solar pp-neutrino capture rate on $^{115}\mathrm{In}$ can be estimated from
the capture rate expressed in solar neutrino units. One solar neutrino unit is
defined as
\begin{equation}
1~\mathrm{SNU}
=
10^{-36}~\mathrm{s^{-1}}
\quad
\mathrm{per~target~atom}.
\label{eq:snu_definition}
\end{equation}
Using a capture rate of 690 SNU
(\href{https://journals.aps.org/prl/abstract/10.1103/PhysRevLett.37.259}
{PRL 37, 259 (1976)}), the capture probability
per $^{115}\mathrm{In}$ nucleus is
\begin{equation}
R_{\nu}
=
690 \times 10^{-36}~\mathrm{s^{-1}}
=
6.9 \times 10^{-34}~\mathrm{s^{-1}} .
\label{eq:snu_rate}
\end{equation}
For 100 kg of loaded indium, assuming natural isotopic abundance, the
corresponding number of $^{115}\mathrm{In}$ nuclei is approximately
\begin{equation}
N_{^{115}\mathrm{In}}
\simeq
5.0 \times 10^{26}.
\end{equation}
The expected solar pp-neutrino capture rate is therefore
\begin{equation}
R_{\mathrm{sig}}
=
R_{\nu}N_{^{115}\mathrm{In}}
\simeq
3.5 \times 10^{-7}~\mathrm{s^{-1}},
\end{equation}
or approximately
\begin{equation}
R_{\mathrm{sig}}
\simeq
10~\mathrm{yr^{-1}} .
\label{eq:signal_rate_100kg}
\end{equation}
For comparison, a 100 ton loaded-indium configuration would increase this raw
capture rate by a factor of $10^3$, giving
\begin{equation}
R_{\mathrm{sig}}
\simeq
10^4~\mathrm{yr^{-1}} .
\label{eq:signal_rate_100ton}
\end{equation}
The dominant intrinsic background is the beta decay of $^{115}\mathrm{In}$.
Using the half-life
\begin{equation}
T_{1/2}=4.41\times10^{14}~\mathrm{years},
\end{equation}
the decay constant is
\begin{equation}
\lambda =
\frac{\ln 2}{T_{1/2}} .
\label{eq:in115_decay_constant}
\end{equation}
The intrinsic beta-decay activity is
\begin{equation}
A =
\lambda N_{^{115}\mathrm{In}} .
\label{eq:in115_activity}
\end{equation}
For 100 kg of loaded indium, assuming natural isotopic abundance, this gives
\begin{equation}
A
\simeq
2.5\times10^4~\mathrm{s^{-1}},
\end{equation}
corresponding to a mean interval between beta decays of
\begin{equation}
\tau_\beta =
\frac{1}{A}
\simeq
40~\mu\mathrm{s}.
\end{equation}
For 100 ton of loaded indium, the beta activity scales linearly to
\begin{equation}
A
\simeq
2.5\times10^7~\mathrm{s^{-1}} .
\end{equation}
A coincidence time window must be chosen to contain the delayed de-excitation
products from the excited $^{115}\mathrm{Sn}^{**}$ state. In this rate
estimate, we use
\begin{equation}
t_{\min}
=
0.2~\mu\mathrm{s},
\qquad
t_{\max}
=
20~\mu\mathrm{s}.
\label{eq:coincidence_time_window}
\end{equation}
The lower bound is used to separate the delayed-cascade search from the prompt
optical response, while the upper bound contains most of the delayed decays.
The delayed de-excitation time follows an exponential distribution with mean
lifetime
\begin{equation}
\tau_{\mathrm{Sn}}
=
\frac{3.3~\mu\mathrm{s}}{\ln 2}
\simeq
4.8~\mu\mathrm{s}.
\end{equation}
The delayed-cascade acceptance within the selected coincidence time window is
therefore
\begin{equation}
\epsilon_t
=
\exp\left(
-\frac{t_{\min}}{\tau_{\mathrm{Sn}}}
\right)
-
\exp\left(
-\frac{t_{\max}}{\tau_{\mathrm{Sn}}}
\right),
\label{eq:coincidence_time_window_acceptance}
\end{equation}
which gives
\begin{equation}
\epsilon_t
\simeq
0.94 .
\end{equation}
Thus, the expected accepted solar-neutrino capture rates are approximately
\begin{equation}
R_{\mathrm{sig}}^{\mathrm{acc}}
\simeq
9.4~\mathrm{yr^{-1}}
\end{equation}
for 100 kg of loaded indium, and
\begin{equation}
R_{\mathrm{sig}}^{\mathrm{acc}}
\simeq
9.4\times10^3~\mathrm{yr^{-1}}
\end{equation}
for 100 ton of loaded indium.
The width of the coincidence time window is
\begin{equation}
\Delta t
=
t_{\max}-t_{\min}
=
19.8~\mu\mathrm{s}.
\label{eq:coincidence_time_window_width}
\end{equation}
The expected number of intrinsic beta decays in this window is
\begin{equation}
\mu_\beta
=
A\Delta t .
\label{eq:beta_window_mean}
\end{equation}
For the 100 kg loaded-indium configuration,
\begin{equation}
\mu_\beta
=
(2.5\times10^4~\mathrm{s^{-1}})
(19.8\times10^{-6}~\mathrm{s})
=
0.495 .
\end{equation}
For the 100 ton configuration,
\begin{equation}
\mu_\beta
=
(2.5\times10^7~\mathrm{s^{-1}})
(19.8\times10^{-6}~\mathrm{s})
=
495 .
\end{equation}
The probability of observing $k$ unrelated beta decays in the coincidence time
window is given by the Poisson distribution,
\begin{equation}
P(k)
=
\frac{
\mu_\beta^k e^{-\mu_\beta}
}{
k!
}.
\label{eq:beta_pileup_poisson}
\end{equation}
The probability of observing more than one unrelated beta decay is
\begin{equation}
P(k>1)
=
1
-
e^{-\mu_\beta}
\left(
1+\mu_\beta
\right).
\label{eq:beta_pileup_kgt1}
\end{equation}
For the 100 kg configuration, this gives
\begin{equation}
P(k>1)
\simeq
8.9\%.
\end{equation}
For the 100 ton configuration, $\mu_\beta=495$, and therefore
\begin{equation}
P(k>1)
\simeq
1 .
\end{equation}

A window-based pileup rate can be estimated as
\begin{equation}
R_{\beta,\mathrm{pileup}}
=
\frac{P(k>1)}{\Delta t}.
\label{eq:window_based_beta_pileup_rate}
\end{equation}
This definition avoids multiple counting of overlapping beta decays in the
high-occupancy regime. It should therefore be interpreted as a raw
pileup-occupancy rate, not as a reconstructed background rate.

For the 100 kg loaded-indium configuration, the expected beta-pileup window
rate is
\begin{equation}
R_{\beta,\mathrm{pileup}}
\simeq
\frac{0.089}{19.8\times10^{-6}~\mathrm{s}}
\simeq
4.5\times10^3~\mathrm{s^{-1}},
\end{equation}
or
\begin{equation}
R_{\beta,\mathrm{pileup}}
\simeq
1.4\times10^{11}~\mathrm{yr^{-1}} .
\end{equation}
For the 100 ton configuration, $P(k>1)\simeq1$, giving
\begin{equation}
R_{\beta,\mathrm{pileup}}
\simeq
\frac{1}{19.8\times10^{-6}~\mathrm{s}}
\simeq
5.1\times10^4~\mathrm{s^{-1}},
\end{equation}
or
\begin{equation}
R_{\beta,\mathrm{pileup}}
\simeq
1.6\times10^{12}~\mathrm{yr^{-1}} .
\end{equation}
Figure~\ref{fig:capture_vs_beta_pileup_rate} compares the expected accepted
solar-neutrino capture rate with the expected beta-pileup window rate for
loaded indium masses of 100 kg and 100 ton, assuming natural isotopic
abundance. The comparison shows that, even with this window-based definition,
the expected intrinsic beta-pileup occupancy rate is many orders of magnitude
larger than the accepted neutrino-capture rate.
\begin{figure}[tbp]
  \centering
  \includegraphics[width=0.80\textwidth]{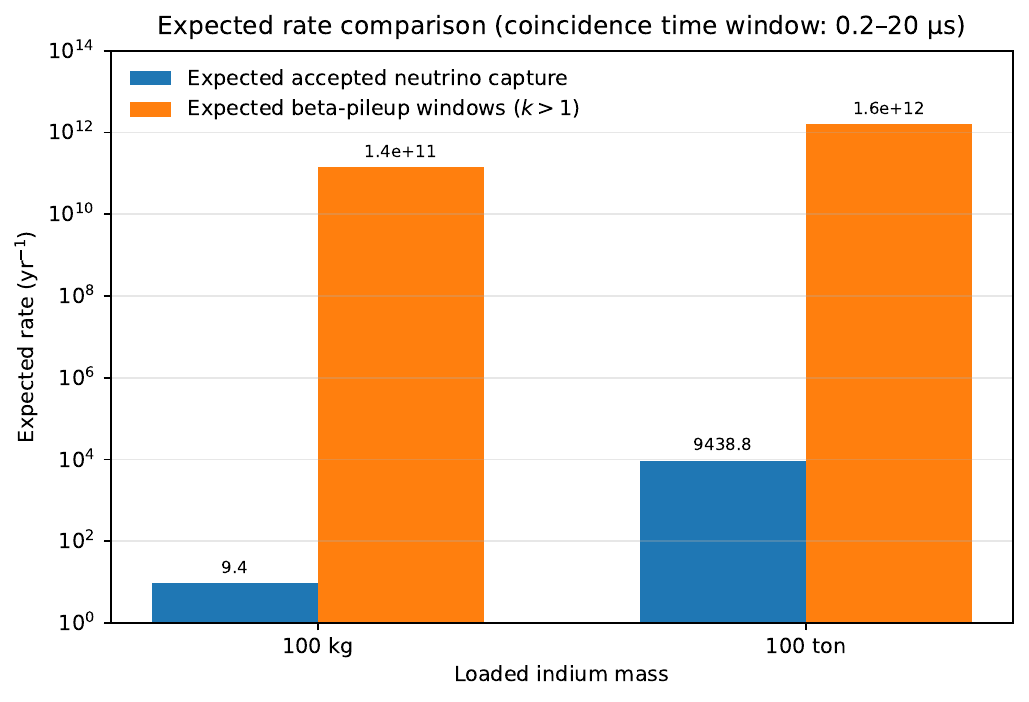}
  \caption{
  Expected rate comparison between the accepted solar-neutrino capture signal
  and the intrinsic $^{115}\mathrm{In}$ beta-pileup window rate for loaded
  indium masses of 100 kg and 100 ton, assuming natural isotopic abundance. A
  coincidence time window of $0.2$--$20~\mu\mathrm{s}$ is assumed. The
  neutrino-capture rate includes the delayed-cascade time-window acceptance.
  The beta-pileup rate is calculated as $P(k>1)/\Delta t$, where $P(k>1)$ is
  the Poisson probability for more than one beta decay in a coincidence time
  window of width $\Delta t=19.8~\mu\mathrm{s}$. This window-based rate avoids
  multiple counting of overlapping beta decays and should be interpreted as a
  raw pileup-occupancy estimate before applying energy, spatial-correlation, or
  topology selections.
  }
  \label{fig:capture_vs_beta_pileup_rate}
\end{figure}
Figure~\ref{fig:beta_pileup_poisson_comparison} shows the corresponding Poisson
distributions for the number of intrinsic beta decays in the coincidence time
window. For the 100 kg loaded-indium configuration, the mean number of beta
decays in the window is $\mu_\beta=0.495$, so zero or one beta decay is still
likely. For the 100 ton configuration, the mean increases to $\mu_\beta=495$,
and the distribution is centered around hundreds of beta decays per window.

\begin{figure}[tbp]
  \centering
  \includegraphics[width=0.97\textwidth]{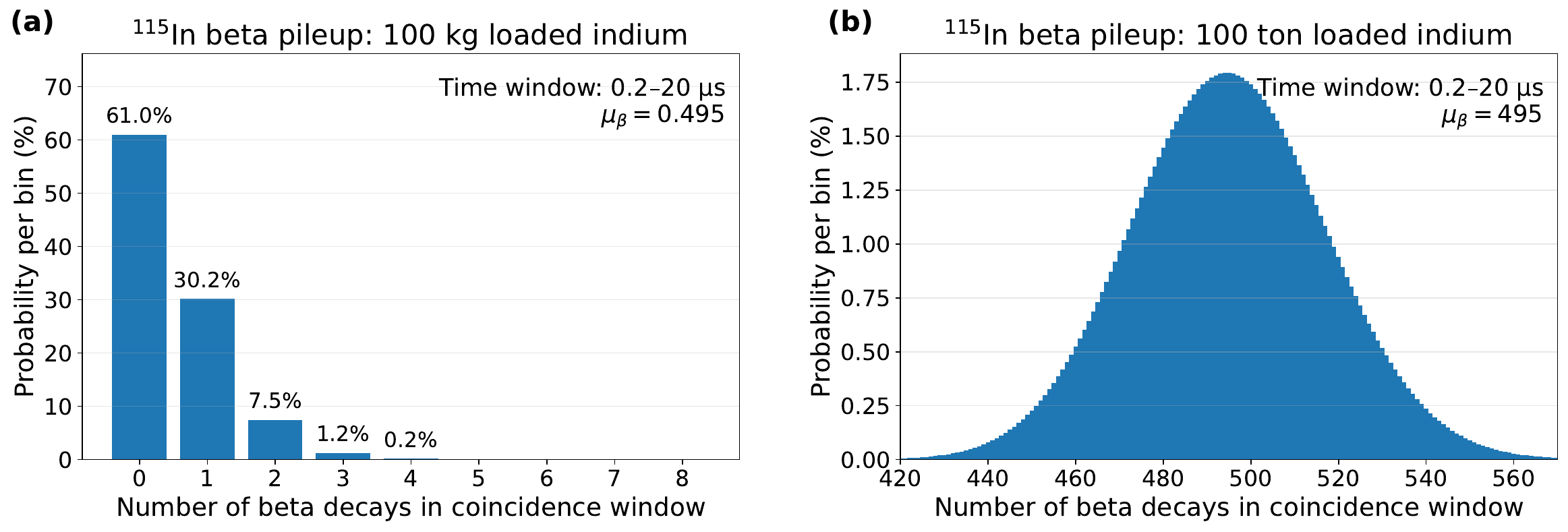}
  \caption{
  Poisson distributions for intrinsic $^{115}\mathrm{In}$ beta-decay pileup in
  the coincidence time window. A window of $0.2$--$20~\mu\mathrm{s}$ is
  assumed. Panel (a) shows the 100 kg loaded-indium configuration, for which
  the mean number of beta decays in the window is $\mu_\beta=0.495$. Panel (b)
  shows the 100 ton loaded-indium configuration, for which the mean increases
  to $\mu_\beta=495$. The comparison illustrates the drastic increase of
  intrinsic beta occupancy with loaded indium mass.
  }
  \label{fig:beta_pileup_poisson_comparison}
\end{figure}

The rate imbalance is severe. For the 100 kg loaded-indium configuration, the
expected accepted solar pp-neutrino capture rate is of order
$10~\mathrm{yr^{-1}}$, whereas the raw beta-pileup window rate is of order
$10^{11}~\mathrm{yr^{-1}}$ for the chosen coincidence time window. For the
100 ton configuration, the accepted signal rate increases to about
$10^4~\mathrm{yr^{-1}}$, but the beta occupancy becomes essentially continuous
on the microsecond time scale. This large intrinsic activity motivates
coincidence-based and spatial-correlation methods for future signal selection
studies. In the present work, however, no quantitative reconstruction or
background-rejection analysis is performed.


The present study was performed as an optical-response and boundary-validation
study for a segmented liquid-scintillator detector motivated by
$^{115}\mathrm{In}$ solar-neutrino detection. The main result is that optical
photon propagation in the LS/FEP/air segmented geometry is governed by the
expected refractive-index structure. Single-photon validation shows that the
Geant4 optical-boundary process reproduces the expected Fresnel reflection,
refraction, and total internal reflection behavior. Therefore, the photon
confinement observed in the segmented detector can be understood as a physical
consequence of the LS/FEP/air optical boundaries rather than as an imposed
surface-reflectivity effect.

This optical validation is important because the indium-capture concept relies
on spatial information. The signal consists of a prompt electron followed by
delayed de-excitation products from the excited $^{115}\mathrm{Sn}^{**}$ state.
In a segmented detector, the prompt electron is expected to produce a relatively
localized energy deposition, while the delayed de-excitation cascade can produce
a spatially extended pattern through gamma transport and Compton scattering.
The segmented optical lattice is intended to preserve enough spatial
information to support such a delayed-coincidence tag.
However, the present work should not be interpreted as a demonstration of
signal-background discrimination. The PMT hit patterns and optical photon
distributions shown in this study are used only as optical-response
observables. No reconstruction-level separation of prompt-electron,
delayed-cascade, and intrinsic-beta event classes is performed. In particular,
no quantitative triple-coincidence selection, likelihood-based event
association, or topology-based background rejection is evaluated in this work.

The simple rate estimates illustrate
why such reconstruction-level studies are essential. For a 100 kg natural
indium configuration, the accepted solar pp-neutrino capture rate is expected
to be of order $10~\mathrm{yr^{-1}}$, while the intrinsic beta-decay activity
is about $2.5\times10^4~\mathrm{s^{-1}}$. In a delayed window of
$0.2$--$20~\mu\mathrm{s}$, the mean number of unrelated beta decays is
$\mu_\beta=0.495$. Thus, even in this relatively small configuration, accidental
beta occupancy in the delayed window is not negligible.
The situation becomes more severe for a 100 ton natural indium configuration.
Although the expected neutrino-capture rate increases linearly with indium mass
to about $10^4~\mathrm{yr^{-1}}$, the intrinsic beta occupancy also increases.
For the same delayed window, the mean number of beta decays becomes
$\mu_\beta=495$. In this regime, essentially every microsecond-scale delayed
window contains many unrelated beta decays. Therefore, the problem is no longer
simply whether an unrelated beta decay occurs in the coincidence window, but how
to associate a prompt candidate and delayed-cascade candidate in a continuously
occupied beta environment.
This point is particularly important because the delayed de-excitation is not
emitted at a fixed time. The delayed component follows the lifetime distribution
of the excited nuclear state, with a mean lifetime of about
$4.8~\mu\mathrm{s}$. Therefore, a practical triple-coincidence search cannot
use a single fixed delay time to associate the prompt electron with the delayed
cascade. Timing information must be combined with deposited energy, spatial
correlation, and cascade topology. The timing window provides only a statistical
selection, not a unique event association.

Previous LENS studies addressed this problem using the triple-coincidence
signature of $^{115}\mathrm{In}$ capture and developed background-rejection
strategies based on timing, energy, spatial coincidence, and gamma-shower
topology
(Rountree, Ph.D. thesis, Virginia Tech, (2009)). Those studies indicate that segmentation
and topology information are essential ingredients for an indium-based
solar-neutrino detector. The present work is more limited in scope. It validates
the optical behavior of a NuLat-like segmented LS/FEP/air geometry, but it does
not reproduce the full LENS-style background-rejection analysis.

The present simulation also contains several simplifying assumptions. First,
the neutrino-capture reaction itself is not simulated as a Geant4
weak-interaction process. Instead, capture-like final-state particles are
generated at the primary-generator level. The prompt-electron direction is
generated isotropically, and the weak-interaction angular dependence is not
included. Second, the active detector material in the current geometry is
LAB-based liquid scintillator without explicit indium loading. The optical
effects of realistic indium loading, including changes in attenuation length,
light yield, refractive index, emission spectrum, scintillation time profile,
and possible absorption features, are not included. Third, detector effects such
as PMT quantum efficiency, collection efficiency, transit-time spread, dark
noise, afterpulsing, electronics thresholds, and trigger logic are not modeled
in detail.

These limitations define the next steps. First, realistic optical properties of
indium-loaded liquid scintillator should be implemented, preferably based on
measured attenuation length, light yield, refractive index, emission spectrum,
and scintillation time constants. Second, the PMT and electronics response
should be modeled in sufficient detail to convert optical photon arrival
patterns into realistic photoelectron and trigger observables. Third,
signal-like events, intrinsic $^{115}\mathrm{In}$ beta decays, and external
gamma backgrounds should be processed through the same reconstruction chain.
Finally, the delayed triple-coincidence tag should be evaluated quantitatively
using timing, energy, spatial correlation, and topology information.

In summary, this work validates the optical-boundary behavior of the segmented
LS/FEP/air detector model and confirms that photon confinement arises naturally
from the refractive-index configuration. The result supports the use of this
geometry as a platform for future indium-capture event studies. At the same
time, the expected event-rate and pileup estimates show that intrinsic
$^{115}\mathrm{In}$ beta activity is a severe challenge. Therefore, the present
study should be regarded as a prerequisite optical simulation, not as a
feasibility demonstration of an indium-loaded solar-neutrino detector.


\subsubsection{Others}

\begin{itemize}
\item $^7$Li loading
\end{itemize}

 There has been attempt to load $^7$Li to water based LS since it allows
the charged current interaction with about 60 times larger cross section
than that from nutrino-electron elastic 
scattering,
as in ~\href{https://link.springer.com/article/10.1140/epjc/s10052-023-11950-9}{Eur. Phys. J. C (2023) 83:799}. 
It is relatively easy to make LiCl water solution but it is uncertain
how one loads $^7$Li in pure LS. Our chemistry team starts to look
at possibility to load $^7$Li to LS to increase the solar neutrino
sensitivity.

\begin{itemize}
\item A new Radon reduction system R\&D 
\end{itemize}

 All low-background underground experiments require 
significant Radon reduction system and since the Radon
concentration in ambient air of Yemilab 
ranges from 100 to 2000 Bq/m$^3$, similar Radon reduction system
is required. We study a new method based on silver ion-exchange
zeolites,
as in~\href{https://academic.oup.com/ptep/article/2024/2/023C01/7504754}
{Prog. Theor. Exp. Phys, {\bf 2024} 023C01},
and
\href{https://academic.oup.com/ptep/article/2025/1/013H01/7911846}
{Prog. Theor. Exp. Phys, {\bf 2025} 013H01}.
To study this new approach, a pilot setup with silver ion-exchanged
zeolites has been constructed as shown in Fig.~\ref{fig:zeolite_Sejong}, 
at the Sejong campus of Korea University.

\begin{center}
\includegraphics[width=0.8\linewidth]{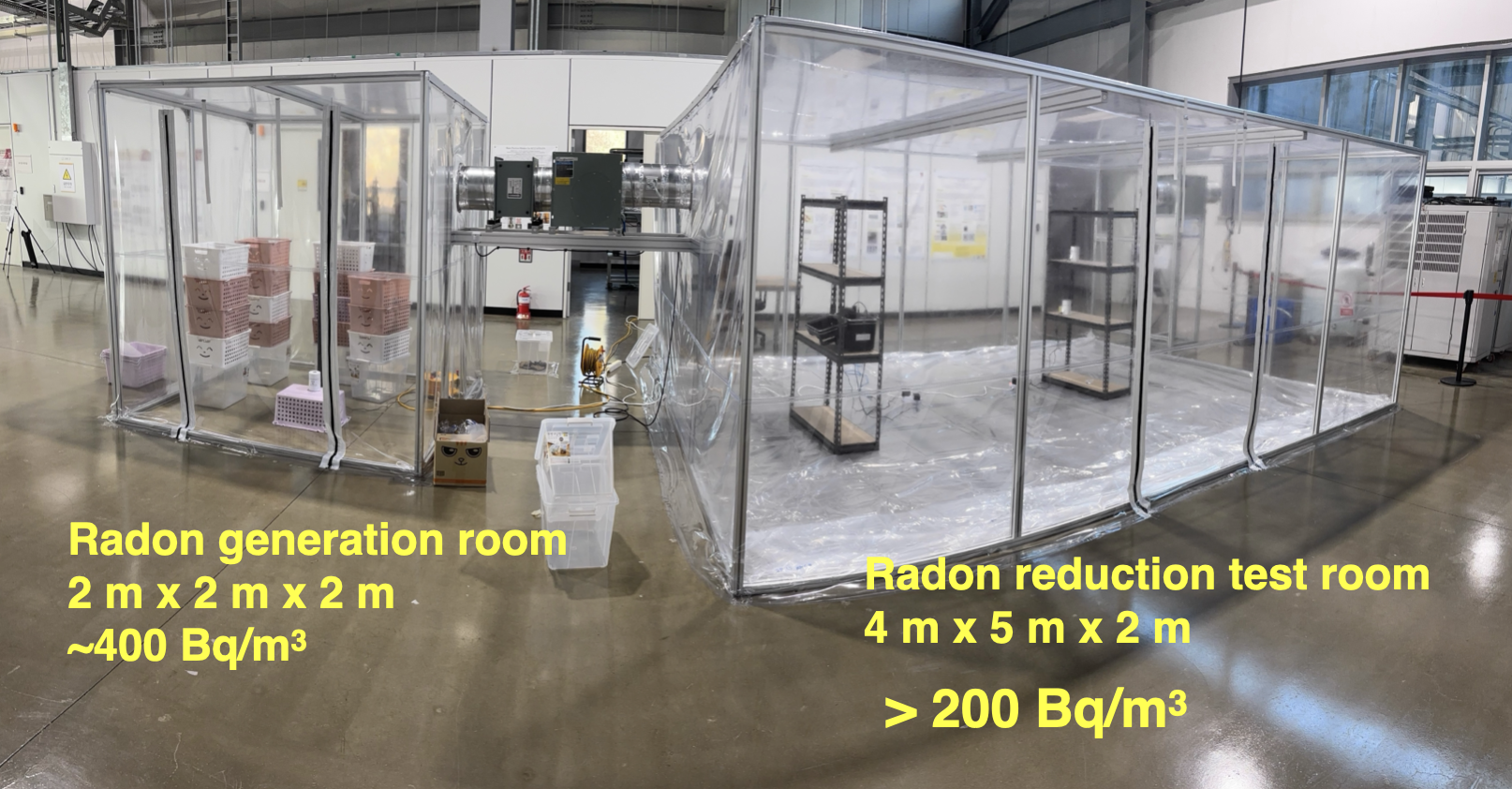}
\captionof{figure}
{A picture of 
a pilot setup with silver ion-exchanged
zeolites is shown. It consists of a Radon generation room 
of $2\times 2\times 2$ m$^3$ and a larger reduction test
room of $4\times 5\times 2$ m$^3$.
}
\label{fig:zeolite_Sejong}
\end{center}
R\&D activities 
aimed at implementing this method are currently in progress.
Compared with an activated charcoal filter, a silver 
ion-exchange zeolite filter can reduce the required filter 
volume in a radon mitigation system by approximately an 
order of magnitude, resulting in lower operating costs 
and a more compact system design.

\newpage

\subsubsection{One-Tonne Prototype Detector}

The one-tonne prototype detector is being commissioned in the LSC hall. The
prototype is not intended to be a final physics detector by itself. Its purpose
is to test, in an integrated underground system, the liquid handling, shielding,
photosensor operation, trigger, waveform readout, data transfer, and background
analysis chain needed for the larger $\nEYE$ program. The limited target mass is
therefore part of the design: the detector is large enough to operate as a real
underground scintillator detector, while remaining small enough for
construction, modification, and repeated commissioning. The data will be used
to measure backgrounds in the prototype geometry and to build operating
experience before scaling to the full detector.
\hfill \break

First of all a polyvinyl structure is constructed at the $\nEYE$ experimental
site. The area is 12 m $\times$ 6 m with the baseline height of 4 m. The top
area is extended as an asymmetric triangular type, allowing maximum total
height of 5.5 m. A bird's-eye view of this tent structure is shown in
Fig.~\ref{fig:one_tonne_tent}. This allows us to have a temperature-controlled
space to operate the one-tonne detector as well as to isolate it from the
outside to a certain extent.
\begin{center}
\includegraphics[width=\linewidth]{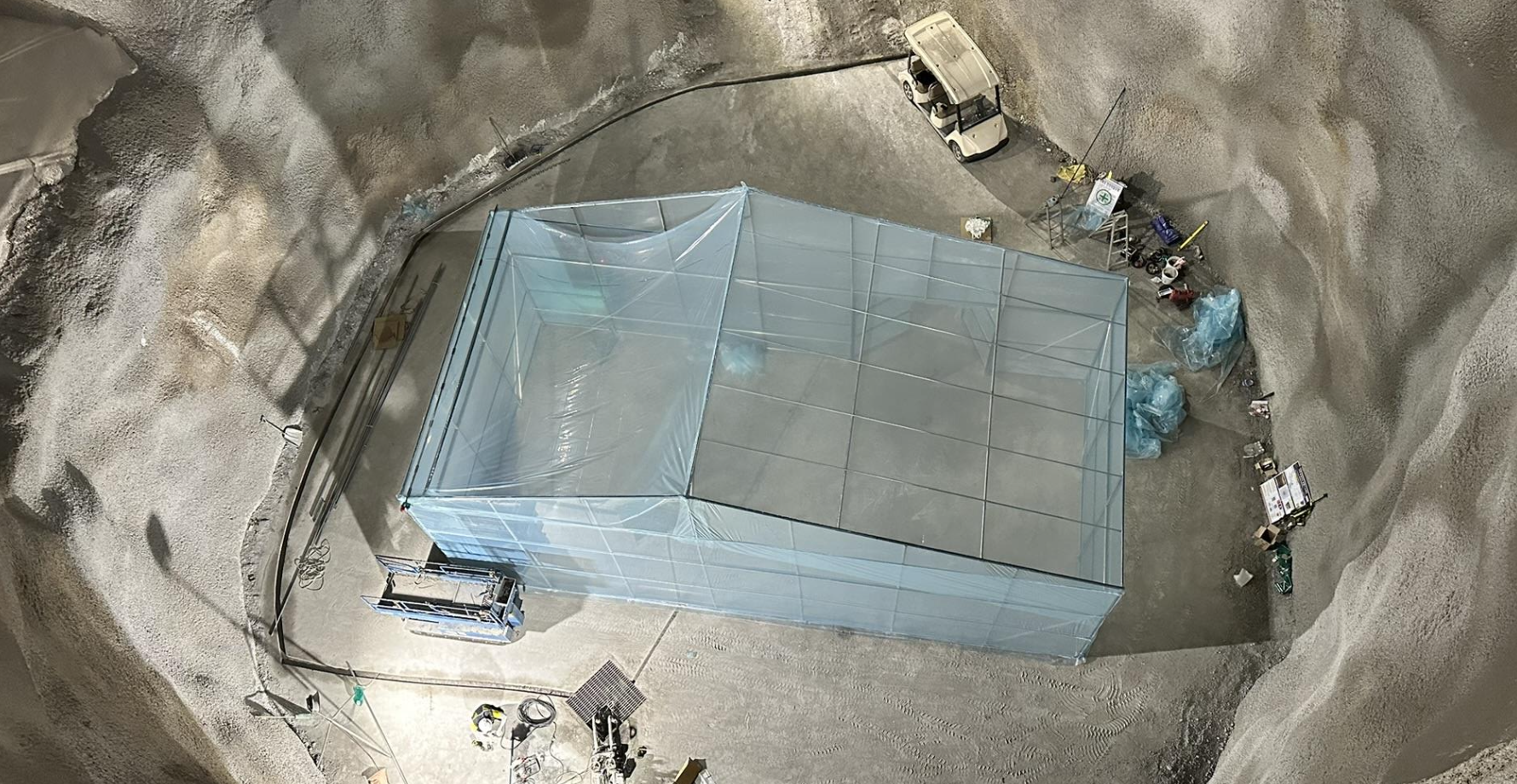}
\captionof{figure}
{
A picture of the housing structure for the one-tonne prototype detector is
shown. The area is 12 m $\times$ 6 m with the baseline height of 4 m. The top
size is extended as an asymmetric triangular type, allowing maximum total height
of 5.5 m.
}
\label{fig:one_tonne_tent}
\end{center}

The detector uses a nested liquid target and buffer geometry. The inner target
vessel is made of PMMA (acrylic), so that scintillation light from the target
can reach the surrounding PMTs while the liquid scintillator remains separated
from the water buffer. The PMMA vessel has a $1200~\mathrm{mm}$ outer diameter
and a $1200~\mathrm{mm}$ cylindrical body height. A central chimney provides
the filling and access path, with an $80~\mathrm{mm}$ outer diameter and a
$60~\mathrm{mm}$ inner diameter. The outer cylindrical stainless-steel tank has
an inner diameter of $2189~\mathrm{mm}$, an outer diameter of
$2201~\mathrm{mm}$, and a height of $2180~\mathrm{mm}$. A
$5~\mathrm{cm}$-thick layer of lead blocks is installed below the tank bottom
to reduce gamma-ray backgrounds from the floor side.
\hfill \break

A Geant4 implementation of the outer stainless-steel tank with the target
acrylic tank and PMTs is shown in the left of
Fig.~\ref{fig:one_tonne_container}, and a photograph of the outer container is
shown in the right. The drawing dimensions define a target volume of about
$1.31~\mathrm{m^3}$ and a water-buffer volume of about
$6.84~\mathrm{m^3}$ before subtracting small displaced volumes from PMT glass,
PMT bases, support structures, cabling, and nozzles.
\begin{center}
\includegraphics[width=0.52\linewidth]{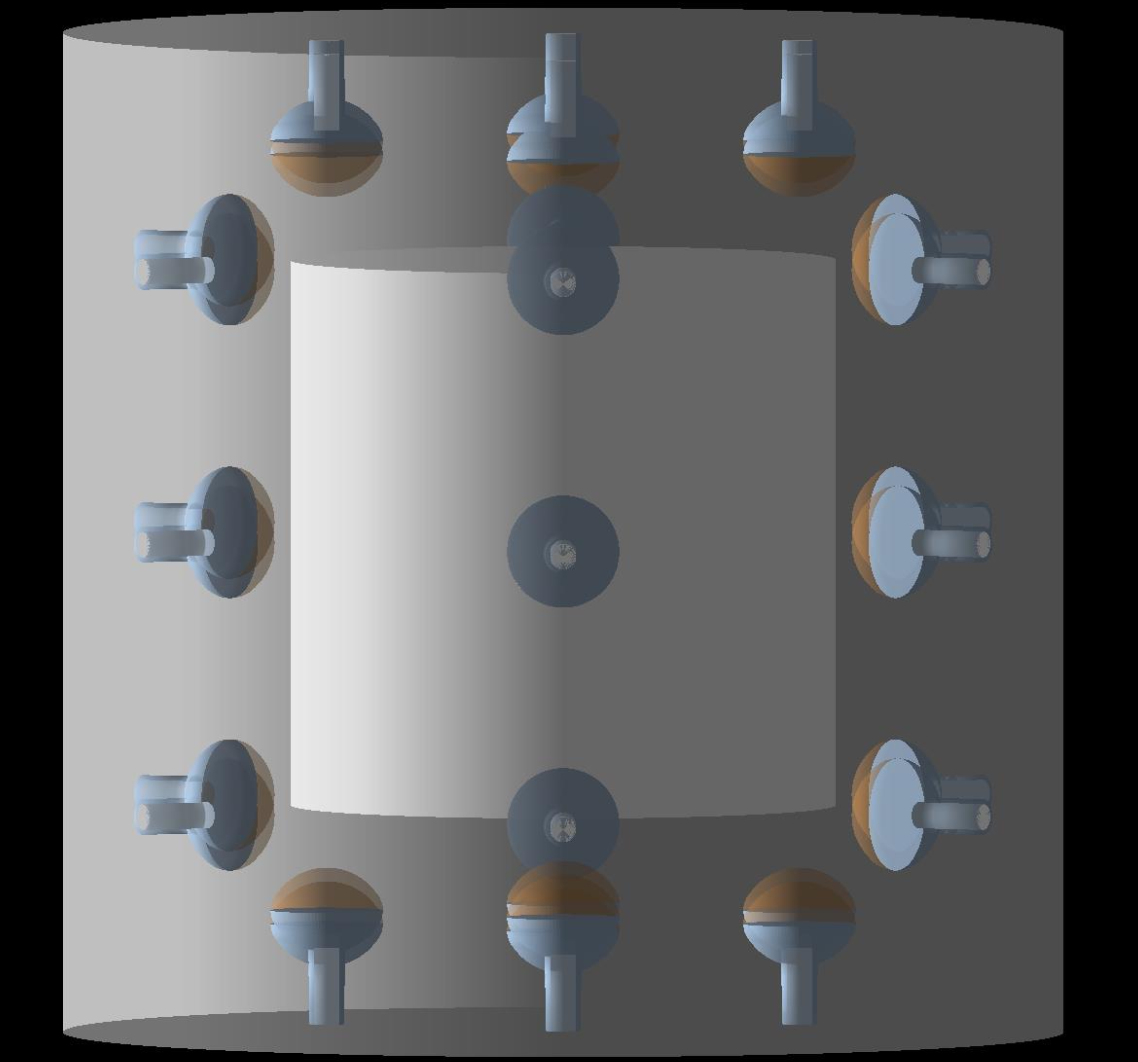}
\includegraphics[width=0.44\linewidth]{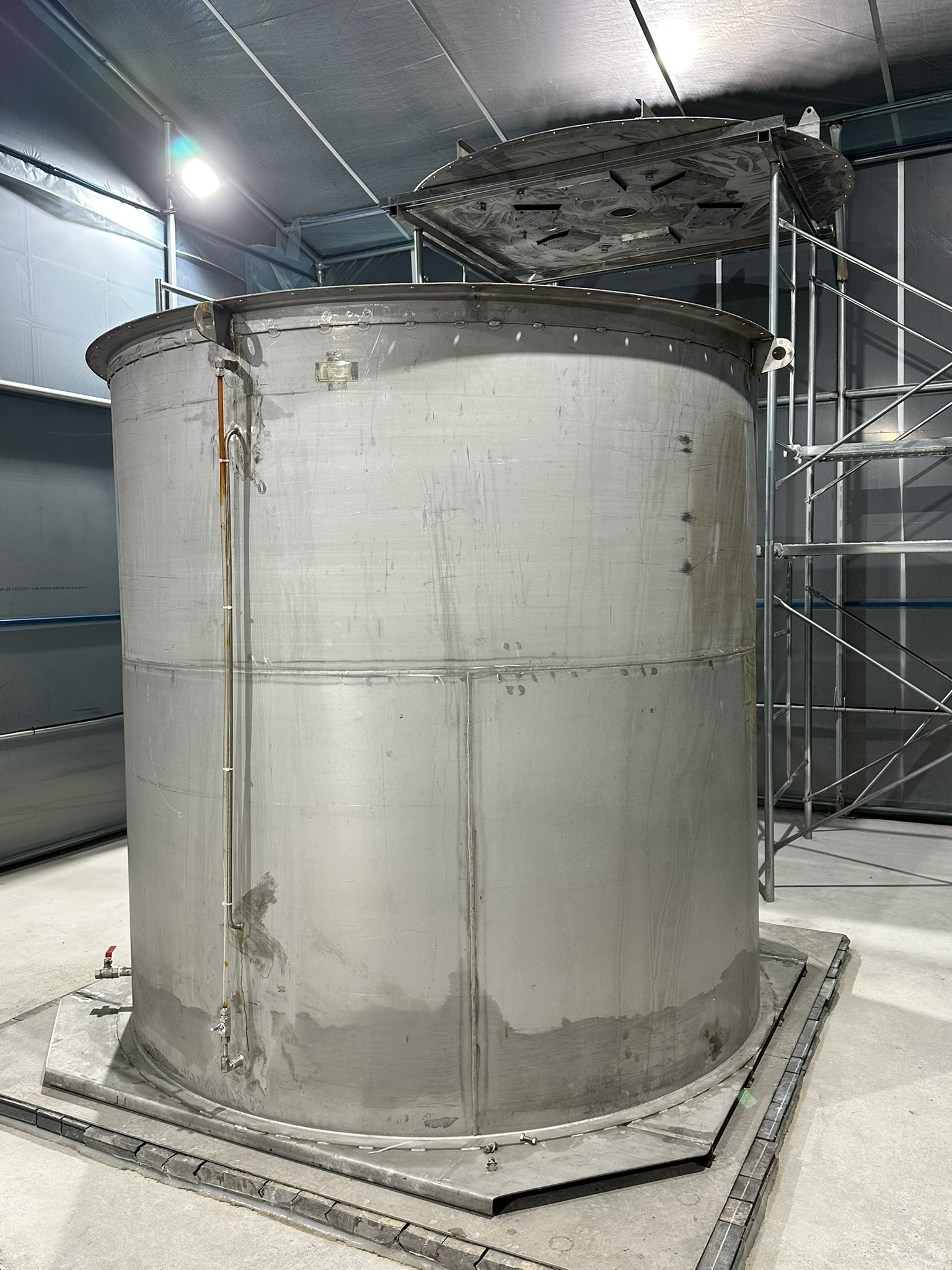}
\captionof{figure}
{
A Geant4 implementation of the outer stainless steel with the target acrylic
tank and PMTs is shown in the left. A photograph of the outer container is
shown in the right.
}
\label{fig:one_tonne_container}
\end{center}

In order to place PMTs accurately, we designed a two-dimensional CAD drawing
of the inner acrylic container, PMTs, and outer tank, as shown in
Fig.~\ref{fig:one_tonne_cad}. The photosensor system consists of 31 Hamamatsu
R7081 10-inch PMTs arranged in top, barrel, and bottom regions. This geometry
is a commissioning layout rather than a final optimized physics array. It
provides three-dimensional optical sampling of the target and buffer volumes,
checks channel-to-channel variations, and supports comparisons with detector
simulation.
\begin{center}
\includegraphics[width=\linewidth]{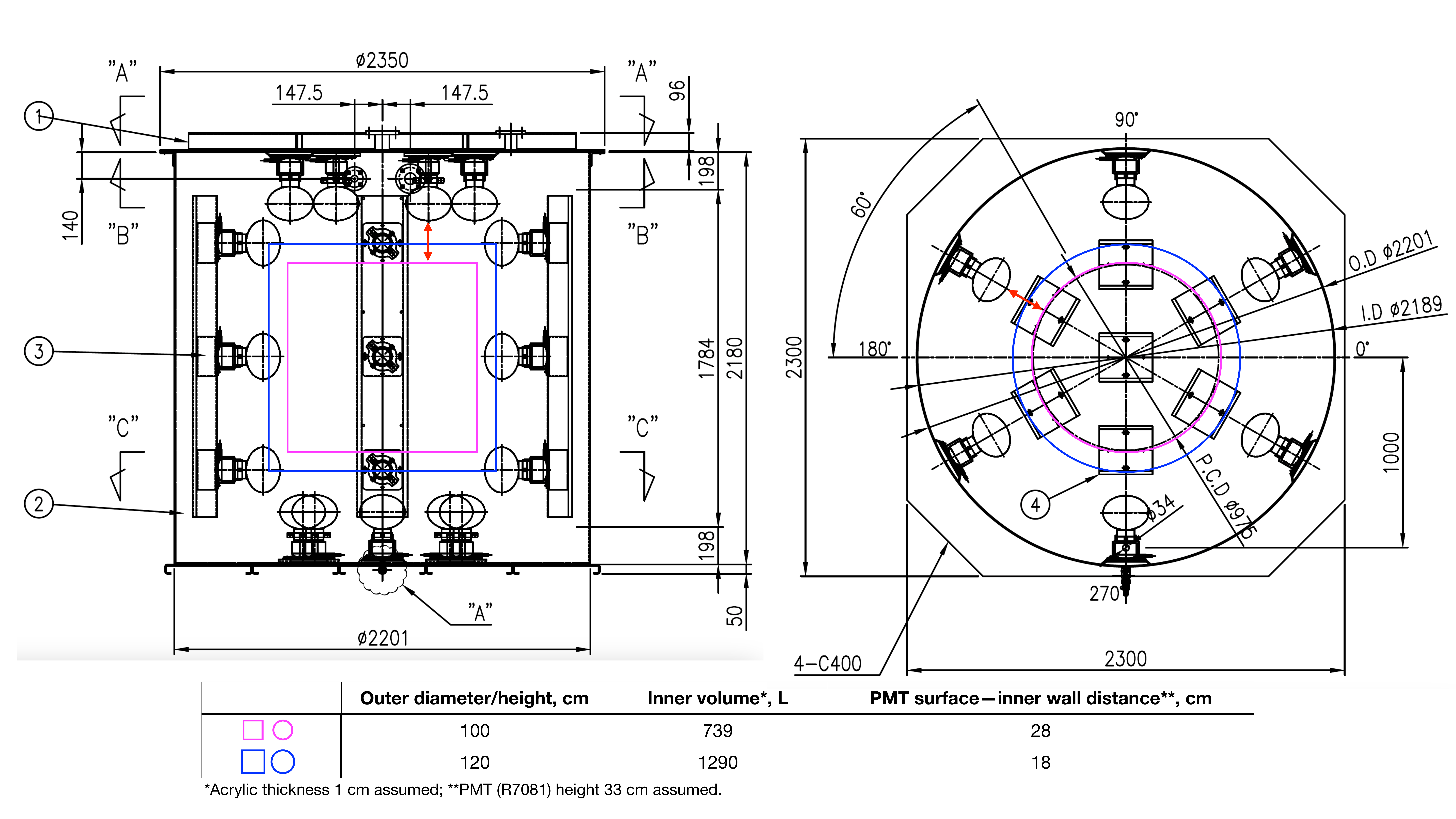}
\captionof{figure}
{
CAD drawings of outer and target containers with PMTs are shown. The left
figure is a side view and the right one is the top view.
}
\label{fig:one_tonne_cad}
\end{center}

Figure~\ref{fig:one_tonne_assembled_open_top} shows the fabricated detector
during assembly before the top lid was closed. The central transparent PMMA
target tank and chimney, the inner stainless-steel wall, the PMT support
structures, the reflective PMT housings, and the cable routing are visible in
the photograph.
\begin{center}
\includegraphics[width=0.8\linewidth]{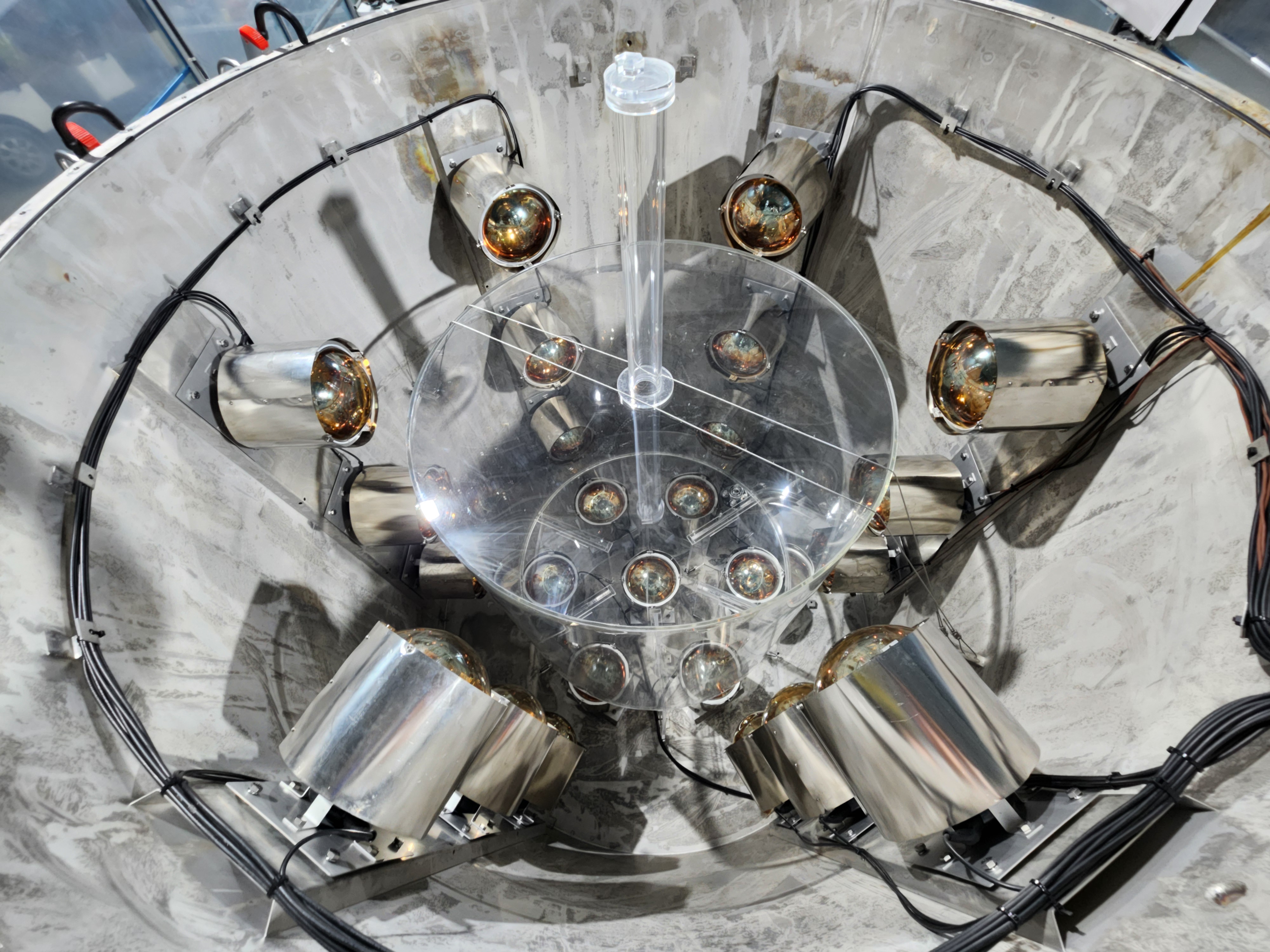}
\captionof{figure}
{
Photograph of the fabricated one-tonne prototype detector before closure of
the top lid. The central PMMA target tank and chimney, PMT supports,
reflective PMT housings, and cable routing are visible.
}
\label{fig:one_tonne_assembled_open_top}
\end{center}

We re-use 10-inch PMTs from the RENO experiment (Hamamatsu R7081). To do that,
the PMTs from the RENO detector were dismounted from the support structure,
transported to the Yemilab site, and cleaned. In total, we collected 67
water-proof PMTs when we need to select 31 PMTs for the prototype detector. We
also collected 354 oil-proof PMTs for future usage.
\begin{center}
\includegraphics[width=\linewidth]{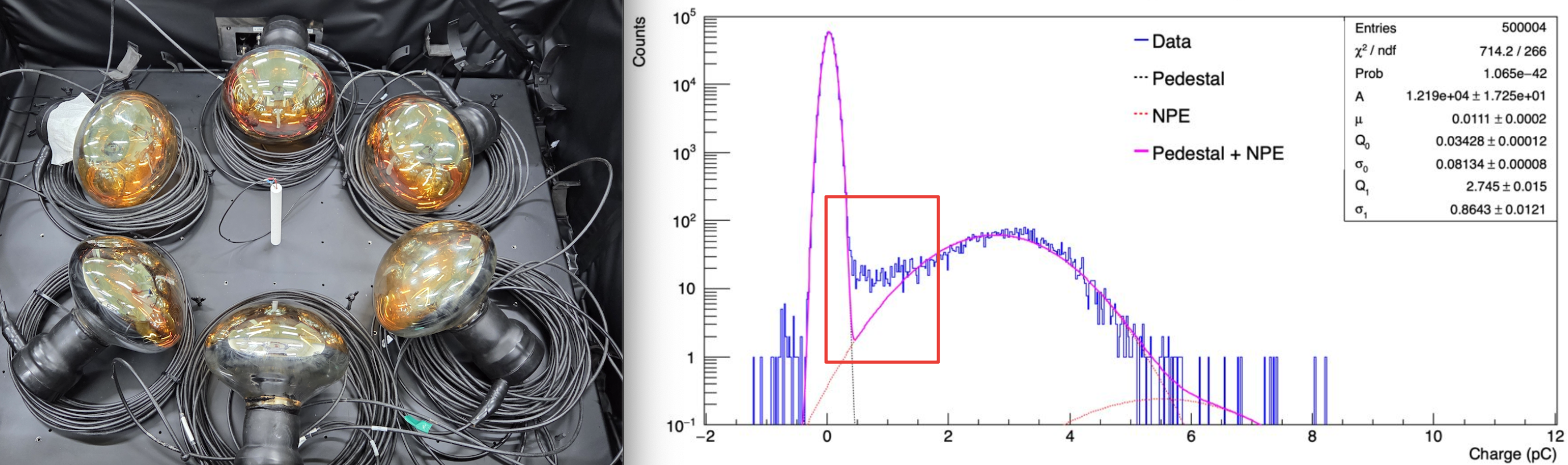}
\captionof{figure}
{
A picture of six PMTs in the dark box for the calibration with LED is shown.
An example of the single-photoelectron signal with the pedestal distribution
is shown with the fit in the right histogram.
}
\label{fig:one_tonne_PMT_fit}
\end{center}

The PMT commissioning program checks gain, dark rate, channel mapping,
threshold settings, and waveform quality. During the dry run, all 31 installed
PMT channels were operational. The high voltages were adjusted so that each
PMT operated near a gain of $10^7$, and the dark-rate threshold was set to the
ADC value corresponding to 1~PE. The gain was obtained from a fit to the 1~PE
region of the dark-pulse spectrum. Gain maps were also prepared over the range
from $10^6$ to $5\times10^7$ for later liquid-scintillator operation, where
larger light signals are expected. These dark-rate and gain measurements will
be repeated after major configuration changes and during long-term operation
as stability checks.
\begin{center}
\includegraphics[width=\linewidth]{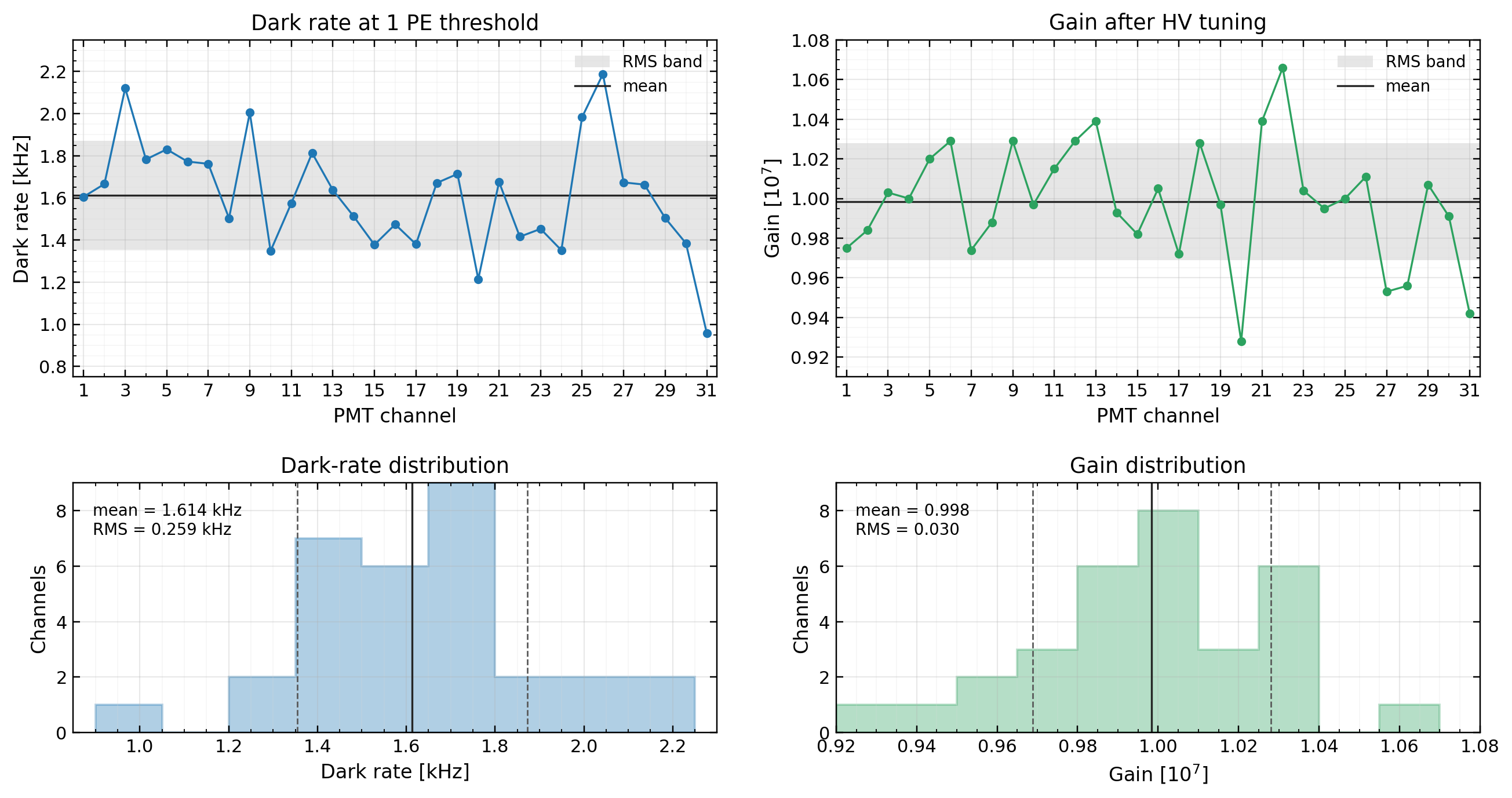}
\captionof{figure}
{
Dry-run PMT commissioning summary. The upper panels show the channel-by-channel
dark rate at the 1~PE threshold and the PMT gain after HV tuning. The lower
panels show the corresponding channel distributions. Solid lines indicate the
detector averages, and the shaded or dashed bands indicate the RMS spreads
across the PMT array.
}
\label{fig:one_tonne_dryrun_dark_gain}
\end{center}

The DAQ system supplies PMT high voltage, forms triggers, records FADC
waveforms, and transfers data for storage and analysis. The HV system uses
CAEN modules with 24-channel and 12-channel outputs, providing 36 available
HV channels for the 31-PMT detector. The signal readout uses four NOTICE
NKFADC500 modules, each with eight channels, giving 32 available waveform
channels. A NOTICE trigger control board distributes trigger and timing
signals to the FADC modules. For the water-filled commissioning runs, the event
trigger required at least four PMT channels with signals at or above 1~PE
within a $64~\mathrm{ns}$ coincidence window. A $1~\mu\mathrm{s}$ event window
was recorded for each triggered event. The data are first stored on a local
server at the experimental site and then backed up to a cloud server for
long-term storage and remote analysis.

\begin{center}
\begingroup
\small
\setlength{\tabcolsep}{3pt}
\renewcommand{\arraystretch}{1.15}
\begin{tabular}{@{}p{0.24\linewidth}p{0.40\linewidth}p{0.17\linewidth}p{0.13\linewidth}@{}}
\hline
Subsystem & Installed modules & Available channels & Used channels \\
\hline
HV & CAEN 24-channel + 12-channel modules & 36 & 31 \\
Waveform readout & Four NOTICE NKFADC500 modules & 32 & 31 \\
\hline
\end{tabular}
\endgroup
\captionof{table}
{
Channel allocation of the one-tonne prototype HV and waveform-readout systems.
}
\label{tab:one_tonne_daq_channels}
\end{center}

In parallel with the construction of the one-tonne prototype detector, an LS
purification system for one-tonne LS is developed and installed in the housing.
The liquid system is separated into the water-handling line and the
liquid-scintillator preparation line. The water line includes a deionized-water
production facility with a production rate of $24~\mathrm{L/min}$. The
baseline scintillator plan uses LAB with PPO and bis-MSB. The working recipe
is PPO at $3~\mathrm{g/L}$ and bis-MSB at $30~\mathrm{mg/L}$, to be confirmed
by the LS preparation test. The purification skid includes stainless-steel
preparation vessels, Al$_2$O$_3$ purification columns, a downstream
particulate-filter stage, valves, drains, transfer lines, and N$_2$ gas lines
for purging and pressure control. The baseline LAB circulation flow is
$2~\mathrm{L/min}$, and the downstream cartridge filter is
$1.0~\mu\mathrm{m}$.
\begin{center}
\includegraphics[width=0.48\linewidth]{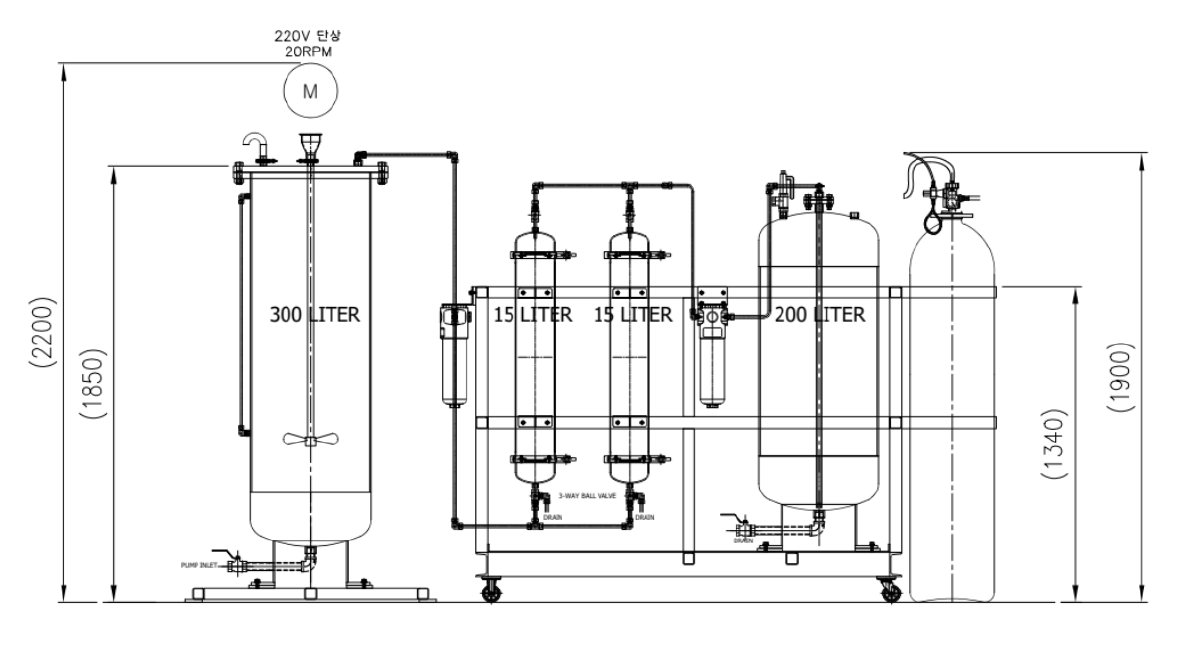}
\includegraphics[width=0.48\linewidth]{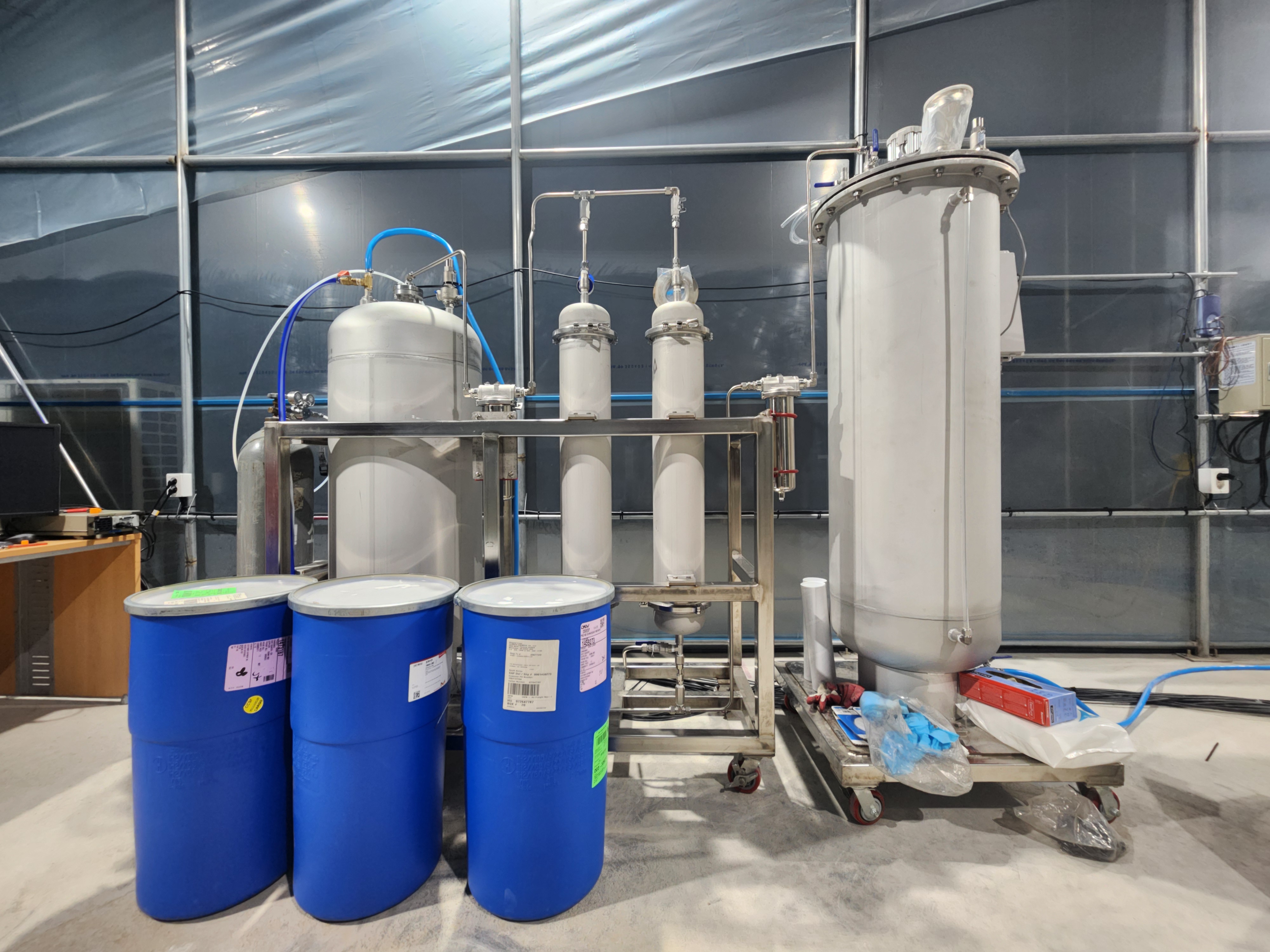}
\captionof{figure}
{
Liquid-scintillator purification equipment for the one-tonne prototype. The
left panel shows the design drawing of the purification and liquid-handling
system, and the right panel shows the fabricated purification equipment.
}
\label{fig:one_tonne_purification}
\end{center}

The commissioning program proceeds in stages. The first stage was the dry run
described above, in which PMT dark rates, gain maps, channel mapping, HV
settings, trigger thresholds, and FADC waveform quality were checked before
liquid filling. The second stage is water-filled operation. In this stage the
full detector is filled with water and operated with the four-channel, 1~PE,
$64~\mathrm{ns}$ coincidence trigger. The water data measure the detector-wide
trigger rate, PMT coincidence response, timing alignment, DAQ continuity, and
the background baseline in the LSC hall. The data set contains cosmic-muon
candidates together with low-energy backgrounds such as rock-gamma events,
radon-related activity, and radioactivity from detector materials.
\begin{center}
\includegraphics[width=0.8\linewidth]{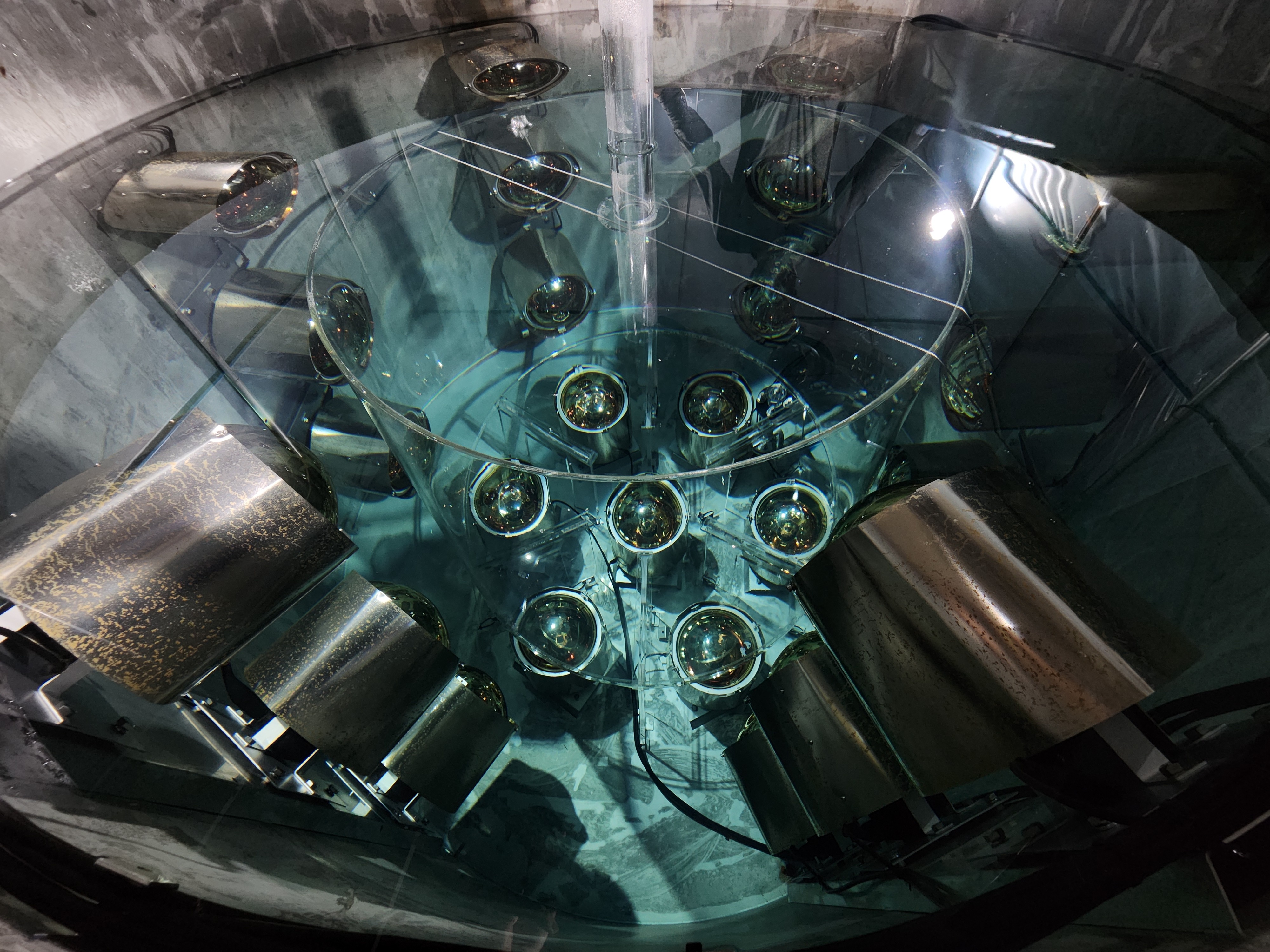}
\captionof{figure}
{
Photograph of the one-tonne prototype during the water-filled commissioning
phase. The stainless-steel tank is filled with DI water, and the PMMA target
tank, chimney, PMT housings, and water-buffer region are visible.
}
\label{fig:one_tonne_water_filled_photo}
\end{center}
\hfill \break

The first water-filled data sample corresponds to about $48~\mathrm{h}$ of
system operation. This time includes DAQ blind time and is not yet the
live-time-corrected exposure. The event trigger rate was approximately
$350~\mathrm{Hz}$. The trigger-rate history was averaged in 1-hour intervals
for the stability check. During this period the detector cooled after DI-water
filling toward the temperature maintained by the constant-temperature system,
approximately $23^\circ\mathrm{C}$. A linear temperature correction reduced
the RMS spread of the 1-hour averaged trigger rate from $7.0~\mathrm{Hz}$ to
$4.2~\mathrm{Hz}$. A residual hour-of-day dependence remains under study and
is treated as an operational diagnostic rather than as a final background-rate
correction.
\begin{center}
\includegraphics[width=\linewidth]{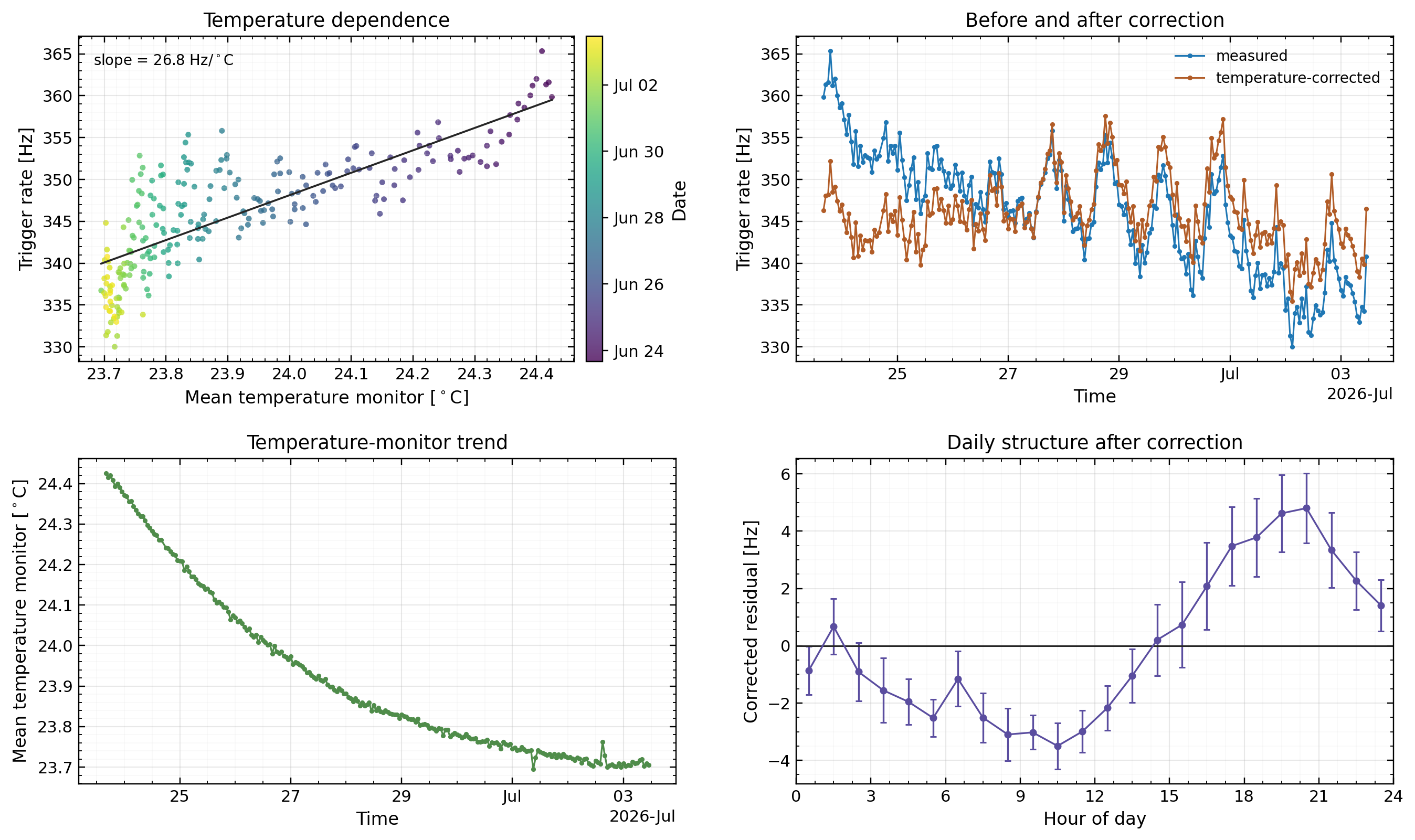}
\captionof{figure}
{
Temperature dependence of the water-filled trigger rate. The analysis uses
1-hour averaged trigger-rate data and the mean of three temperature monitors.
The figure shows the linear temperature dependence, the measured and
temperature-corrected trigger rates, the temperature-monitor trend, and the
residual hour-of-day structure after the linear temperature correction.
}
\label{fig:one_tonne_trigger_temperature}
\end{center}

Figure~\ref{fig:one_tonne_water_spectra} shows initial total-PE and hit
multiplicity spectra for selected water-filled events. The logarithmic
event-count axes show both the dominant low-charge, low-multiplicity
population and the smaller high-charge or high-multiplicity tail. The shaded
regions mark commissioning selections for muon-like samples. Since the data
were recorded with a trigger requiring at least four PMT channels above 1~PE,
the selected spectra have no entries below $N_\mathrm{hit}=4$; the total-charge
axis is drawn from $1~\mathrm{PE}$ only to show the trigger threshold region.
\begin{center}
\includegraphics[width=\linewidth]{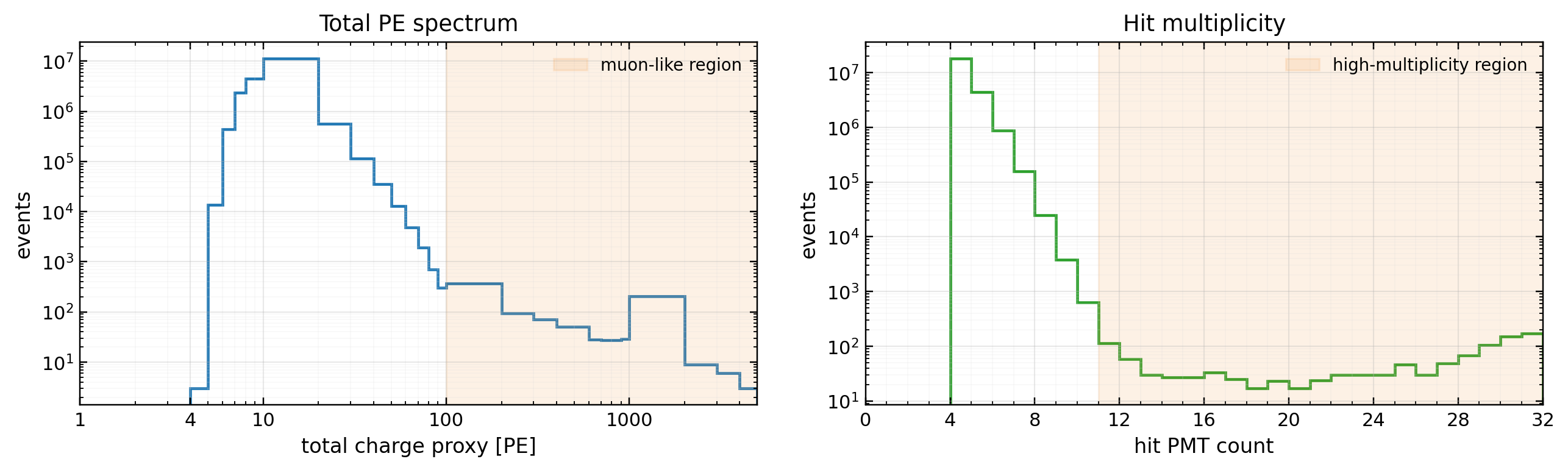}
\captionof{figure}
{
Initial water-filled commissioning spectra. The left panel shows the
event-by-event total charge proxy in PE units, and the right panel shows the
number of hit PMT channels. The shaded bands indicate high-charge and
high-multiplicity regions used to select muon-like candidates for
commissioning studies.
}
\label{fig:one_tonne_water_spectra}
\end{center}

The water-filled data are also compared with Geant4 detector simulations to
separate event classes and to validate the background model.
Figure~\ref{fig:one_tonne_water_mc} compares the measured spectra with the
Geant4 source templates. The measured data are shown together with
\texttt{sim\_dark}, \texttt{sim\_rock\_gamma}, and \texttt{sim\_muon}
components. The rock-gamma component mainly populates the low-charge and
low-multiplicity region, while the muon component extends to higher PE and
larger hit multiplicity. The live-time-corrected rates, systematic
uncertainties, and data--simulation agreement metric will be evaluated after
the complete analysis.
Even at the one-tonne scale, the same data set can provide useful information
on high-energy underground muons and related air-shower questions, including
the so-called muon puzzle discussed in
\href{https://arxiv.org/abs/2510.16341}{\tt arXiv:2510.16341}.
\begin{center}
\includegraphics[width=\linewidth]{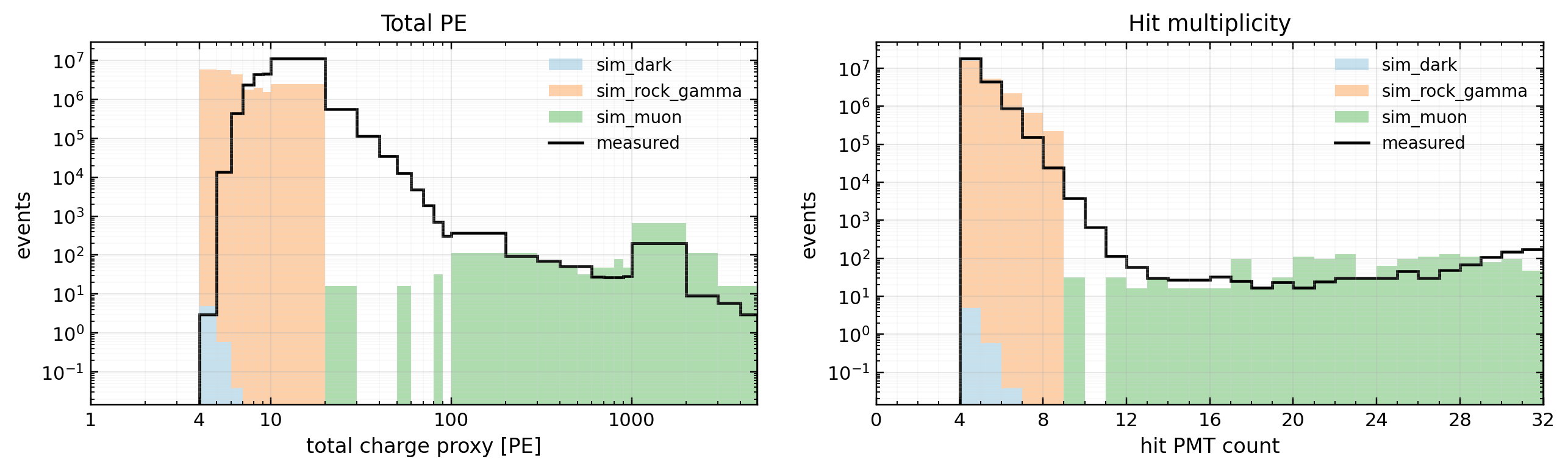}
\captionof{figure}
{
Comparison between the measured water-filled spectra and Geant4 source
templates. The black curves show the measured data, and the filled histograms
show the dark-noise, rock-gamma, and cosmic-muon simulation templates. The
left panel compares total PE, and the right panel compares hit multiplicity.
}
\label{fig:one_tonne_water_mc}
\end{center}

The next stage is LS preparation and LS filling. After the LAB purification and
LS-mixing checks are completed, the PMMA target tank will be filled with the
working LS. During this phase, additional water shielding around the
stainless-steel tank is planned to reduce external backgrounds. The LS-filled
data will be used to measure the scintillator light response, PMT stability in
the final optical configuration, radon-related activity, gamma backgrounds,
muon-induced events, and the agreement between measured data and detector
simulation. Thus the one-tonne prototype provides the commissioning records,
background measurements, detector-stability information, and liquid-handling
experience needed for the next stage of the $\nEYE$ detector program.

The data-taking plan is therefore organized as a sequence of commissioning
steps rather than as a final physics exposure. Each step has a specific
operational check and a defined output for the next detector stage, as
summarized in Table~\ref{tab:one_tonne_data_taking_plan}.

\begin{center}
\begingroup
\small
\setlength{\tabcolsep}{2.5pt}
\renewcommand{\arraystretch}{1.15}
\begin{tabular}{@{}p{0.15\linewidth}p{0.15\linewidth}p{0.31\linewidth}p{0.33\linewidth}@{}}
\hline
Stage & Status & Main measurement & Deliverable/use \\
\hline
Dry run & Completed &
Dark rate, gain map, channel mapping, trigger threshold, and waveform quality &
PMT operating point, channel-readiness check, and DAQ readiness before liquid
filling
\\
Water-filled run & Ongoing &
Trigger rate, PMT coincidence, timing alignment, cosmic-muon candidates, and
environmental-background samples &
Water-phase background baseline, DAQ stability check, and input for Geant4
background validation \\
LS preparation & Planned &
LAB purification, LS mixing, optical sampling, and transfer-line checks &
LS recipe confirmation, purification readiness, and filling-procedure check \\
LS-filled run & Planned &
LS light response, gamma and radon backgrounds, muon-induced events, and
data--simulation comparison &
Inputs to shielding, trigger, calibration, and full-detector design choices \\
\hline
\end{tabular}
\endgroup
\captionof{table}
{
Data-taking plan for the one-tonne prototype commissioning program.
}
\label{tab:one_tonne_data_taking_plan}
\end{center}

\subsection{\fontsize{12}{15}\selectfont International collaboration}
The Korean neutrino community has
accumulated 20 years of 
technological expertise from 
on-shore and offshore neutrino experiments, positioning the $\nEYE$ 
experiment for success. We may miss the 
state-of-the-art radio purification experience 
and the logistics on the high-dose radioactive source. 
\hfill \break
To address these, we have established a robust international 
collaboration with Borexino experts through  the
\href{https://indico.ibs.re.kr/event/709/}{IBS-INFN} framework of
2024.
\hfill \break
The $\nEYE$ project has been in regular discussions with the 
the US IsoDAR team every two months. 
Additionally, we 
have ongoing discussions with China for JUNO and the Jinping lab.
Japanese neutrino community has always maintained a close
partnership with 
Korean neutrino researchers, especially with the recently launched
Korea Neutrino Observatory project. In summary, a
significant part of the global neutrino physics community is 
already engaged with us, and this collaboration is expected
to expand rapidly upon receiving official approval for the project.

\section{Construction and Utilization Plan}
The construction timeline for the $\nEYE$ detector is estimated to be 
six years, followed by an additional two years dedicated to
commissioning.
This is consistent with the typical timeline for constructing
a $\mathcal{O}$(1) kilo-tonne neutrino experiment.
A major advantage is that the $\nEYE$ pit, the purification system area,
and the space for the potential IsoDAR accelerator are already
prepared, significantly reducing both time and budget during
the construction phase.
\hfill \break
To ensure the success for the project, we are currently designing a
prototype $\mathcal{O}$(1)-tonne
detector. This will help us:
\begin{itemize} \itemsep -2pt
\item Validate the  LS detector performance,
\item Understand radioactive background levels, and 
\item Refine the construction details for the full $\nEYE$ detector.
\end{itemize} 
Additionally, we maintain close collaboration with the Yemilab
operation center to finalize the detector construction
and installation procedure, 
including the purification and LS detector system infrastructure.
\section{\bf Plan to Organize Research Groups}
The $\nEYE$ center will be structured with two key committees:
\begin{itemize} \itemsep -2pt
\item {\bf Executive Committee}: Responsible for reviewing regular 
progress, overseeing the hiring process, and managing budget allocation, 
and spending.
\item {\bf International External Review Committee}: Tasked with  
evaluating annual progress and providing strategic guidance for
center's overall direction.
\end{itemize}
Within the center, there will also be two major research groups,
which will focus on different aspects of the $\nEYE$ experiment.
\begin{itemize}
\item {\bf Instrumentation Group}: 
Responsible for the detector design, electronics,
infrastructure construction, and radioactive source preparation.
\item {\bf Physics Group}: 
Focuses on simulation tool development,
detector optimization, and physics analyses.
\end{itemize}
These groups will work in close collaboration to ensure the
success of the $\nEYE$ experiment.

In the early phase of the experiment, 
the hardware group will take on a primary role, focusing on
detector construction, electronics, and infrastructure setup.
Once the detector is operational, 
the hardware group will shift its focus to maintenance,
while the physics group will take the lead in data analysis, simulation
improvements, and physics studies. 
Both groups expect to work closely to exchange
inputs and outputs to develop the experiment most
effectively. 
\hfill \break
\hfill \break
The Korean neutrino research community has its roots in the
RENO project, which played a crucial role in training and
developing a strong pool of neutrino experts over the past
20 years.  This proposal brings together experienced researchers
who have contributed significantly to past and ongoing neutrino 
experiments, ensuring that the $\nEYE$ project benefits from a 
highly skilled and knowledgeable team.
\hfill \break
\hfill \break
The installation of the $\nEYE$ experiment will
mark the completion of the full scientific program at Yemilab, 
complementing the experiments of Center for Underground Physics 
of IBS which 
doesn't detect neutrinos directly. 
With 
the $\nEYE$ experiment,
Yemilab will establish itself as one of the world’s leading 
multipurpose underground laboratories, contributing to 
cutting-edge research in neutrino physics and rare event searches. 
\section{Expected Research Outcome }
If the radioactive source experiment yields a positive signal,
it would provide strong evidence for the existence of a sterile
neutrino, aligning with the results from the BEST experiment.
Such a discovery would be profound implications for our 
understanding of neutrino physics, shedding light on the origin of 
neutrino mass and opening a new frontier in sterile neutrino research.
\hfill \break
Additionally, we aim to study CNO neutrinos from the Sun to 
address the ongoing tension in metallicity measurements. 
Precision measurements of  $^7$Be, $pep$, and $^8$B 
solar neutrinos will allow us to probe the vacuum-MSW transition 
region of the electron survival probability. This will serve 
as a critical test of the current neutrino oscillation framework, 
potentially confirming or challenging our existing theoretical models.
In addition to the studies on sterile and solar neutrinos,
it will serve the unique supernova burst alert detector in Korea. 
With the IsoDAR, it will be a unique facility in the world to 
search for dark sector and non-standard neutrino interactions coupling 
ultra-low background and high-power accelerator.
\hfill \break
The $\nEYE$ experiment would be an unique reactor neutrino experiment to
be able to validate results at 65 km baseline.
\section{Implementation Plan by Phase}
\subsection{Implementation plan}
The present implementation plan to operate the $\nEYE$ detector is described
below.
\hfill \break
\hfill \break
\indent \tikz\draw[black,fill=black] (1,0) circle (.4ex); 
{\bf Water Cherenkov} Phase (first 6 years)
\hfill \break
Our immediate focus is on constructing the detector while
finalizing the LS R\&D to ensure optimal
physics output. Once the detector is built, we will proceed
with filling the veto tank with water.
\hfill \break
By leveraging Cherenkov light detection, we aim to observe
$^8$B solar neutrinos and reactor neutrinos. This initial
phase will allow us to debug the detector system and generate the
first physics results from the
$\nEYE$ experiment. Additionally, we are exploring the
feasibility of adding elements to enable charged
current interactions, which could further enhance our ability
to probe neutrino properties.
\hfill \break
\hfill \break
\indent \tikz\draw[black,fill=black] (1,0) circle (.4ex); 
{\bf LS} Phase (6-8 years)
\hfill \break
Once the LS is installed in the 
target volume, we will prepare and deploy radioactive 
sources to initiate the sterile neutrino search. 
If IsoDAR is available, it will be the first underground accelerator-assisted
sterile neutrino search.
Simultaneously, 
we will continue detecting solar and reactor neutrinos 
to refine our understanding of neutrino oscillations.
\hfill \break
In parallel, we will conduct ongoing measurements of 
geo-neutrinos to study Earth's internal 
composition and supernova neutrinos to probe astrophysical 
phenomena. These efforts will provide a comprehensive 
neutrino detection program, maximizing the physics reach of the 
$\nEYE$ experiment.
\hfill \break
\hfill \break
\indent \tikz\draw[black,fill=black] (1,0) circle (.4ex); 
$\bm{0\upnu \upbeta \upbeta}$ Phase (8 years and beyond)
\hfill \break
To extend
the physics potential of the 
$\nEYE$ experiment, we will initiate the 
$0\nu \beta \beta$
search by loading an appropriate metal isotope into 
the liquid scintillator. This step will allow us to probe 
the Majorana nature of neutrinos and explore the neutrino mass hierarchy.
\hfill \break
A construction and implementation timeline for all 
phases of the experiment, including LS installation, detector 
commissioning, and physics data collection, is outlined in 
Fig.~\ref{fig:nEYE_Gantt}.
%
%
\begin{center}
\includegraphics[width=\linewidth]{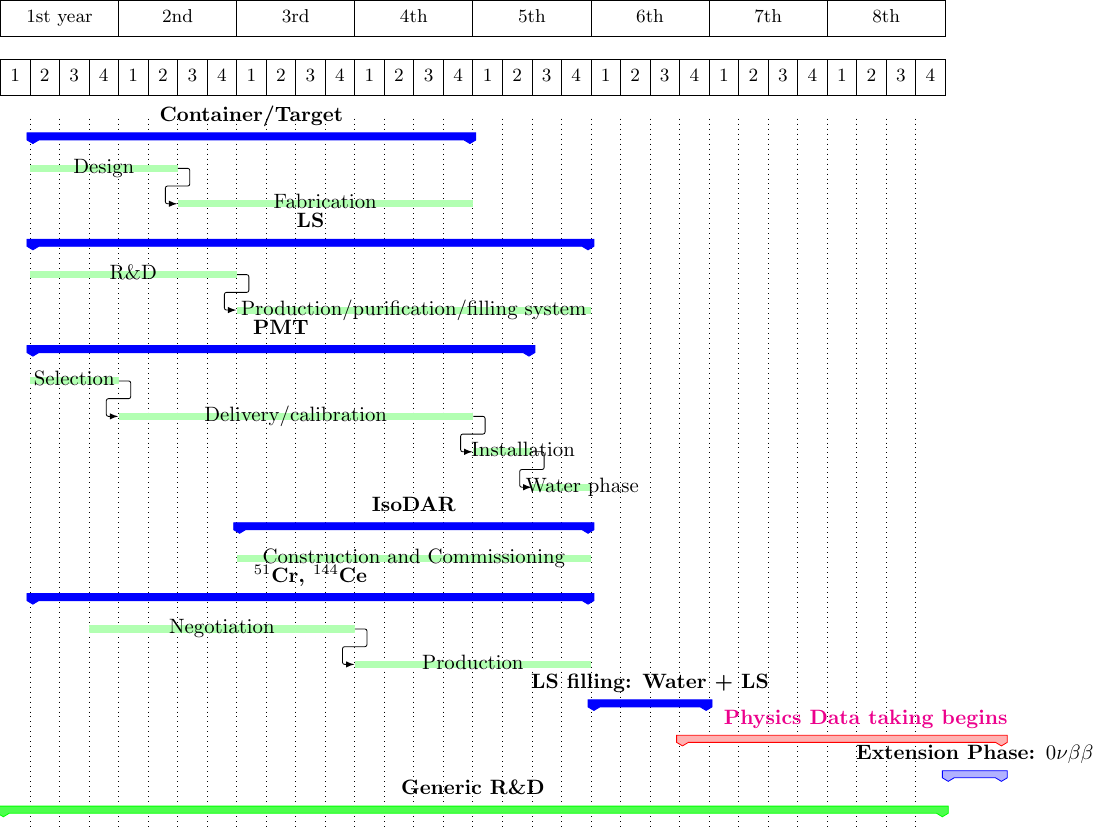}
\captionof{figure}
{
Our plan of the $\nEYE$ experiment assuming 8 years of
running the center. Note that as is explained
above, we plan to have three phases of Water Cherenkov, LS,
and $0\nu \beta \beta$ (beyond 8 years).
}
\label{fig:nEYE_Gantt}
\end{center}

}

\end{document}